\documentclass  [usegraphicx] {mn2e}
\newcommand\kms{$\rm{km\,s^{-1}}$}
\newcommand\HII{H\,{\sc ii}}

\title[OH polarization spectra, longitudes 240$^\circ$ to 350$^\circ$]
{Parkes full polarization spectra of OH masers - II. Galactic longitudes 
240$^\circ$ to 350$^\circ$}
\author[J.L.~Caswell, J.A.~Green \& C.J.~Phillips]
        {J.L.~Caswell\thanks{james.caswell@.csiro.au}, 
J.A.~Green,
and C.J.~Phillips
\\CSIRO Astronomy and Space Science, Australia Telescope National 
Facility, PO Box 76, Epping, NSW, Australia 2121 \\}

\date{Accepted .
      Received ;
      in original form 2013}

\pagerange{\pageref{firstpage}--\pageref{lastpage}}
\pubyear{2014}

\begin{document}

\maketitle

\label{firstpage}

\begin{abstract}

Full polarization measurements of 1665 and 1667-MHz OH masers at 261  
sites of massive star formation have been made with the Parkes 
radio telescope.  Here we present the resulting spectra for 
157 southern sources, complementing our previously published 
104 northerly sources. 
For most sites, these are the first measurements of linear polarization, 
with good spectral resolution and complete velocity coverage.  

Our spectra exhibit the well-known predominance of highly 
circularly polarized features, interpreted as $\sigma$ components 
of Zeeman patterns.  Focusing on the generally weaker and rarer 
linear polarization, we found three examples of likely full  
Zeeman triplets (a linearly polarized $\pi$ component, straddled 
in velocity by $\sigma$ components), adding to the solitary example 
previously reported.  We also identify 40 examples of likely isolated 
$\pi$ components, contradicting past beliefs that $\pi$ components 
might be extremely rare.  These were recognised at 20 sites where
a feature with high linear polarization on one transition is  
accompanied on the other transition by a matching feature, at the 
same velocity and also with significant linear polarization. 

%We also identify the rare occurrence of large velocity 
%ranges, eight cases exceeding 25 \kms, some of them indicative of high 
Large velocity ranges are rare, but we find eight exceeding 25 \kms, 
some of them indicating high velocity blue-shifted outflows.  
Variability was investigated on timescales of one year and over 
several decades.  More than 20 sites (of 200) show high variability 
(intensity changes by factors of four or more) in some prominent 
features.  Highly stable sites are extremely rare.

\end{abstract}


\section{Introduction}

This paper completes a project to obtain full polarization spectra, at 
high velocity resolution, of masers at the 1665- and 1667-MHz 
transitions of OH in the southern sky accessible to the Parkes 
telescope.  We focus exclusively on maser sites believed to be in star 
formation regions (SFRs), associated with high-mass young 
stellar objects (YSOs).  
We present the results for sources in the southern Galactic 
longitude range 240$^\circ$ to 350$^\circ$, thus complementing the 
previously published northern range  350$^\circ$ through the Galactic 
Centre to 40$^\circ$ (Caswell, Green \& Phillips 2013, hereafter Paper 
I).

OH masers play a very special role in star formation studies by virtue 
of their remarkable polarization properties.  The multiple narrow 
features commonly display prominent circular and linear polarization, 
sometimes approaching 100 per cent;  this has long been recognised as 
the signature of magnetic fields where the masers arise, 
primarily caused by the Zeeman effect in fields of 
several milliGauss.

%Several hundred OH masers in SFRs have been catalogued 
%(e.g. Caswell 1998).  Precise positions for the southerly sources 
Catalogued OH masers in star formation regions now exceed  
several hundred, with especially good coverage in the southern sky, 
Our target list for the current Parkes observations is 
primarily the catalogue of Caswell (1998), supplemented by a 
few southern sources discovered since 1998.  The catalogued positions, 
to arcsecond accuracy, were obtained with the Australia Telescope 
Compact Array (ATCA),  mostly more than a decade ago.  At that time, 
limited spectropolarimetric capability on the ATCA did not 
allow corresponding high quality spectra characterising each source.  
For some of our targets, known for several decades, Parkes spectra have 
been published, but limited to circular polarization (notably, with no 
linear polarization measurements), and dating from 25 years ago.  
Improved spectropolarimetric capabilities on the Parkes telescope 
now allow the present study with high spectral resolution and full 
polarization (simultaneous circular and linear) spectra at each 
catalogued OH position.  

Based on the large sample of full polarization spectra obtained from 
combining our present results for 157 sources  with the 104 northern 
sources from Paper I, we are able to make a good assessment of the 
incidence of linear polarization, enlightening us as to whether the 
occasional instances of extremely high linear polarization are 
explicable by the Zeeman effect.  
The results also cast light on more general puzzles of the physics 
of astronomical masers, including the relevance of magnetic beaming 
(Sections 6, 7) and the existence of co-propagation for the 1665 and 
1667-MHz transitions (Sections 4.1, 6.1, 6.3).

Our long term objectives are to conduct detailed study of maser 
emission at each site, which can reveal not only 
remarkable clues to the kinematics and magnetic field structure 
around each newly formed massive star, but also the physical properties 
of density, molecular abundances, and temperature implied by the masing 
of the OH molecules.  Such studies ultimately require the high spatial 
resolution provided by long baseline arrays (e.g. Wright et al. 2004a, 
b), and these will require large dedicated programs in the future.  
However, the present data allow selection of high priority sites that 
will most benefit from these studies.  
More readily achievable in the short term are measurements with shorter 
baseline arrays with superb spectropolarimeters that have 
recently become available.  Notably, the ATCA in 
combination with CABB (Compact Array Broadband Backend) has recently 
demonstrated its capability (Green et al. 2012b) and will be 
a major contributor to follow-up of the present observations of 
southern sources.  The reliability of magnetic field measurements from 
interpretation of Zeeman splitting will be improved relative to single 
dish measurements, and will be a focus of such studies.  
Complementing these with ATCA studies of 6-GHz transitions of 
excited-state OH will further strengthen these interpretations.  

\section[]{Observations and data reduction}

Paper I gave full details of the observing and reduction procedure.  
Here we summarize the key features.  The observations were made with 
the Parkes 64-m radio telescope in two observing periods, 2004
November 23-27 and  
%(five days 6am to 6.30pm)
2005 October 26-30 (project p484).   
The receiver accepted two orthogonal linear polarizations, followed by  
digital filters that restrict the processed signal to a 4-MHz 
bandpass before entering a correlator.  
The `Parkes multibeam correlator' was used in a non-standard   
configuration, with the 32 x 1024 spectral channels concatenated to 
provide single-beam full Stokes polarimetry yielding four polarization 
products:  auto-correlations from the orthogonal linear feeds, XX 
and YY, and real and imaginary parts of the cross-correlations, 
ReXY and ImXY, each with 8192 channels.

For each target, an observation of 10 min was made in most cases, 
but reduced to 4 min for some strong sources, and increased to 20 
min for several weak sources.  

%Note narrow RFI v +12 at 1665 in 2005 for 284.351 and.........?
%306.322-0.334 2004 could av d3 and d5 both 20m but rfi d5)}  

Improved position determinations for two sources (298.262+0.739  and 
326.780-0.241) were made with the 
ATCA 2005 March 28, conducted in a 6-km configuration, 
yielding position accuracy of 0.4 arcsec.

\subsection{Spectral line reduction in {\sc asap}}

The ATNF program {\sc asap} (ATNF Spectral line Analysis Package) was 
used for data reduction.  The intensity calibration is relative to the 
source 1934-638, for which a total intensity at 1666 MHz of 14.16 Jy has 
been adopted, essentially the same scale as used in earlier work 
(Caswell \& Haynes 1987a and references therein) where Hydra A was used 
as a calibrator with assumed flux density 36 Jy.  

%The recorded data comprised 8192-channel spectra across 4 MHz for each of 
%the autocorrelations from the orthogonal linear probes, XX, and YY, and 
%for the real and imaginary parts of their cross-correlation, ReXY and ImXY.  

As described in Paper I, the recorded correlator outputs were 
combined to generate 
the Stokes parameters I (total intensity), Q and U (components of linear 
polarization), and V (net circular polarization).  For some purposes, 
different representations of polarization are more illuminating, so we 
also generated the two individual right and left hand circular 
polarization components, RHCP and LHCP,  (I+V)/2 and 
(I$-$V)/2 respectively; total linearly polarized intensity, subsequently 
abbreviated to LINP when used quantitatively, where
LINP =$ \sqrt{Q^2 + U^2}$;  and linear polarization position angle
(ppa), where  ppa = $1/2\tan^{-1}{(U/Q)}$.
\\
For graphical display of full polarization spectral data we present  
the results as two panels of spectra for each transition, showing: 
\\
1.  	Spectra of  RHCP and LHCP, overlaid with LINP. 
\\
2.  	Overlaid spectra of I with Q and U.
\\
The presence of linear polarization is most readily seen on the Q and U 
spectra.  The value of LINP (derived from Q 
and U as noted above) has a positive noise bias;  for display 
purposes, we have chosen to clip it at five times the rms 
noise level, so that features with significant linear polarization are 
clearly evident from comparing this plot with the plot of total 
intensity. 
The polarization position angle is a noisy quantity which we have chosen 
not to show, but the plots of Q and U indicate it qualitatively, 
noting that ppa = $1/2\tan^{-1}{(U/Q)}$.

\section[]{Results}

\begin {table*}

\caption{Polarization measurements of OH masers at 1665 and 1667 MHz. 
References to positions are C98 (Caswell 1998);  
%CVF95 (Caswell et al. 1995c) FC89?? (Forster \& Caswell 1989); 
A00 (Argon et al. 2000); C01 (Caswell 2001); C04 (Caswell 2004c); 
G12 (Green et al. 2012b) and `text' refers to the notes of Section 3.3.
Peak intensities refer to the stronger circular polarization; the 
intensity values shown in boldface correspond to the epoch for which the 
spectra are shown in Fig. 1. Lin(5,7) 
summarizes  linear polarization at 1665 and 1667 MHz with  `P' denoting  
more than 50 per cent in at least one feature, and `p', detectable 
but weaker polarization.   Abbreviated references to earlier 
polarization observations are M (MAGMO from Green et al. 2012b); L (LBA 
measurements from Caswell, Kramer \& Reynolds 2011b and refs therein) 
c (Caswell \& Haynes 1987a and refs therein) and 
v (vla from Argon et al. 2000).  Column 13 heading `m/OH' refers to the 
intensity ratio of the peak of an 
associated 6.6-GHz methanol maser to the highest OH peak. }

\begin{tabular}{llrlcccccclll}
%\begin{tabular}{lcrcccrcrl}

\hline

\multicolumn{1}{c}{Source Name} & \multicolumn{2}{c}{Equatorial 
Coordinates} & \multicolumn{1}{l}{Refpos}  & \multicolumn{2}{c}{Vel. range}& 
\multicolumn{2}{c}{$\rm S_{peak}(2004)$} & \multicolumn{2}{c}{$\rm 
S_{peak}(2005)$} 
& \multicolumn{1}{l}{Lin(5,7)}  & \multicolumn{1}{l}{Refpol} 
& \multicolumn{1}{l}{m/OH}   \\

\ (~~~l,~~~~~~~b~~~)    &       RA(2000)        &       Dec(2000)       
&  &     $ \rm V_{L}$&$ \rm V_{H}$ &  $\rm S_{1665}$    &  $\rm 
S_{1667}$  &  $\rm S_{1665}$ & $\rm S_{1667}$     &       \\

\ (~~~$^\circ$~~~~~~~$^\circ$~~~) & (h~~m~~~s) & (~$^\circ$~~ '~~~~") & 
& \multicolumn{2}{c}{(\kms)} & (Jy) &  (Jy) & (Jy) & (Jy) \\

\hline

%350.011$-$1.342&17 25 06.50& $-$38 04 
%00.7&C98&$-$26.5&$-$17.5&\bf{11.0}&\bf{4.3}&4.5&4.4&5P,7p&- & C98\\

%350.011$-$1.342  &	17 25 06.50	&	 $-$38 04 00.7	&	C98	&	$-$26.5	&	$-$17.5	&	\bf{4.3}	&	4.4	&	11.0	&	4.5	&	5P;  7p	&	v  &  1/1.9	\\
%350.015+0.433	&	17 17 45.44	&	 $-$37 03 12.9	&	C98	&	$-$35	&	$-$32	&	1.0	&	$<0.15$	&	\bf{1.1}	&	0.15	&		&	&  1/6.5	\\

240.316+0.071	&	07 44 51.98	&	 $-$24 07 42.4	&	C98	&	62.5	&	69	&	1.0	&	0.27	&	\bf{1.0}  &	\bf{0.27}	&		&		&  $<1/3.3$	\\
263.250+0.514	&	08 48 47.80	&	 $-$42 54 28.8	&	C98	&	6	&	16	&	\bf{0.5}  &  \bf{0.5}	&	0.5	&	0.55	&		&		&	176	\\
284.351$-$0.418	&	10 24 10.73	&	 $-$57 52 33.3	&	C98, G12  &	4	&	9	&	\bf{1.5}  &  \bf{1.2}	&	2.0	&	1.3	&	5p	&	M	&	$<1/10$	\\
285.263$-$0.050	&	10 31 29.87	&	 $-$58 02 18.0	&	C98, G12  &	2	&	15	&	\bf{34}	&  \bf{1.7}	&  80	&	2.9	&		&	Mc	&	$<1/400$	\\
287.371+0.644	&	10 48 04.45 	&	 $-$58 27 01.2	&	C98, G12  &	$-$5	&	1	&	\bf{1.5}  &  $\bf{<0.1}$ &	2.3	&	$<0.1$	&		&	M	&	45	\\
290.374+1.661	&	11 12 18.10	&	 $-$58 46 20.3	&	C98, G12  &	$-$24.5	&	$-$12	&	3.4	&	0.3	&	\bf{3.7}  &  \bf{0.4}	&	5p	&	M	&	1/1.8	\\
291.274$-$0.709	&	11 11 53.44	&	 $-$61 18 23.4	&	C04, G12  &	$-$25	&	$-$23.5	&	\bf{0.45}  &  $\bf{<0.2}$	&	 $-$	&	 $-$	&		&	M	&	155	\\
291.579$-$0.431	&	11 15 05.74	&	 $-$61 09 40.7	&	C98, G12  &	10.5	&	16.5	&	\bf{1.4}  &  $\bf{<0.2}$	&	1.2	&	$<0.2$	&		&	M	&	1/1.4	\\
291.610$-$0.529	&	11 15 02.59	&	 $-$61 15 49.5	&	C98, G12  &	16	&	22	&	\bf{10}   &  \bf{2.7}	&	10	&	2.7	&		&	Mc	&	$<1/25$	\\
291.654$-$0.596	&	11 15 10.74	&	 $-$61 20 32.3	&	G12	&	14	&	23	&	\bf{2.0}  &  $\bf{<0.8}$  &	2.0	&	$<0.8$	&		&	M	&  $<1/2.9$	\\
294.511$-$1.621	&	11 35 32.22	&	 $-$63 14 42.6	&	C98, G12  &	$-$20.5	&	$-$9	&	37	&	0.85	&	\bf{32}	&  \bf{0.85}	&	5p; 7p	&	M	&	1/4.5	\\
297.660$-$0.973	&	12 04 08.99	&	 $-$63 21 36.0	&	C98	&	22	&	33	&	4.5	&	0.6	&	\bf{4.2} &  \bf{0.55}	&	5P	&	c	&	$<1/15$	\\
298.262+0.739	&	12 11 47.74	&	 $-$61 46 20.6	&	text	&	$-$29	&	$-$26.5	&	1.2	&	$<0.2$	&	\bf{1.3}  & \bf{0.15}	&		&		&	11.2	\\
299.013+0.128	&	12 17 24.66	&	 $-$62 29 03.7	&	C98	&	19.5	&	24	&	\bf{1.0}  & $\bf{<0.1}$	&	1.0	&	$<0.1$	&	5p	&	c	&	8.0	\\
300.504$-$0.176	&	12 30 03.49	&	 $-$62 56 49.8	&	C98	&	2	&	24	&	0.45	&	$<0.1$	&	\bf{0.4}  & $\bf{<0.1}$	&		&	c	&	11.1	\\
300.969+1.147	&	12 34 53.24	&	 $-$61 39 40.3	&	C98	&	$-$44.5	&	$-$34.5	&	13.5	&	3.3	&	\bf{16}	&  \bf{3.9}	&		&	Lc	&	1/3.3	\\
301.136$-$0.226	&	12 35 35.07	&	 $-$63 02 32.4	&	C98	&	$-$64	&	$-$33	&	17	&	6.2	&	\bf{17}	&  \bf{6.5}	&	5p; 7p	&	c	&	1/11	\\
305.200+0.019	&	13 11 16.90	&	 $-$62 45 54.7	&	C98	&	$-$35	&	$-$29	&	1.5	&	$<0.2$	&	\bf{1.6}  & $\bf{<0.2}$	&		&		&	31	\\
305.202+0.208	&	13 11 10.61	&	 $-$62 34 37.8	&	C98	&	$-$43	&	$-$42	&	1.5	&	$<0.2$	&	\bf{1.5}  & $\bf{<0.2}$	&		&	c	&	60	\\
305.208+0.206	&	13 11 13.78	&	 $-$62 34 41.1	&	C98	&	$-$44.5	&	$-$32	&	21	&	33	&	\bf{30}	&  \bf{27}	&		&	v	&	13.3	\\
305.362+0.150	&	13 12 35.87	&	 $-$62 37 17.9	&	C98	&	$-$42	&	$-$32	&	15	&	8.5	&	\bf{17}	&  \bf{8.5}	&	5p; 7p	&	c	&	1/3.4	\\
305.799$-$0.245	&	13 16 43.32	&	 $-$62 58 32.9	&	C98	&	$-$37	&	$-$32	&	2.7	&	1.2	&	\bf{3.0}  & \bf{1.3}	&	5p; 7p	&	c	&	1/5.4	\\
306.322$-$0.334	&	13 21 23.00	&	 $-$63 00 30.4	&	C98	&	$-$25	&	$-$20	&	0.15	&	$<0.1$	&	\bf{0.15} & $\bf{<0.1}$	&		&		&	3.3	\\
307.805$-$0.456	&	13 34 27.40	&	 $-$62 55 13.8	&	C98	&	$-$19	&	$-$13	&	4.5	&	0.55	&	\bf{5.0}  & \bf{0.57}	&	5P	&		&	$<1/50$	\\
308.754+0.549	&	13 40 57.56	&	 $-$61 45 42.9	&	C04	&	$-$51	&	$-$43	&	0.5	&	$<0.2$	&	\bf{0.5}  & $\bf{<0.2}$	&		&		&	24	\\
308.918+0.123	&	13 43 01.67	&	 $-$62 08 51.9	&	C98	&	$-$70	&	$-$47.5	&	\bf{42}	&	\bf{8}	&	51	&	8.8	&	5p; 7p	&	c	&	1/1.2	\\
309.384$-$0.135	&	13 47 24.10	&	 $-$62 18 12.5	&	C98	&	$-$55.5	&	$-$49	&	1.1	&	$<0.1$	&	\bf{0.95} & $\bf{<0.1}$	&	5P	&	c	&	1.3	\\
309.921+0.479	&	13 50 41.73	&	 $-$61 35 09.8	&	C98	&	$-$64	&	$-$58	&	83	&	8.8	&	\bf{75}	&  \bf{10}	&	5p; 7P	&	c	&	12	\\
310.144+0.760	&	13 51 58.26	&	 $-$61 15 39.5	&	C98	&	$-$60	&	$-$50	&	\bf{0.3}  &  \bf{0.25}	&	0.28	&	0.22	&		&		&	267	\\
311.643$-$0.380	&	14 06 38.76	&	 $-$61 58 24.0	&	C98	&	15	&	50	&	8	&	4.4	&	\bf{8.9}  & \bf{4.7}	&	5P; 7p	&	c	&	1.2	\\
311.94+0.14	&	14 07 48.7	&	 $-$61 23 22	&	text	&	$-$42	&	$-$40	&	$\bf{<0.2}$  &	$\bf{<0.2}$  &	$<0.2$	&	$<0.2$	&		&	c	&	1.0	\\
312.598+0.045	&	14 13 15.01	&	 $-$61 16 53.6	&	C98	&	$-$72	&	$-$58	&	2.4	&	1.8	&	\bf{2.5}  &	\bf{1.6} &	5p	&	c	&	8.4	\\
313.469+0.190	&	14 19 40.98	&	 $-$60 51 47.1	&	C98	&	$-$11	&	$-$6	&	2.7	&	0.35	&	\bf{2.7}  &	\bf{0.3} &	5p	&	c	&	11	\\
313.577+0.325	&	14 20 08.70	&	 $-$60 41 59.9	&	C98	&	$-$48	&	$-$41	&	0.55	&	0.27	&	\bf{0.45}  &	\bf{0.2} &	5P	&		&	200	\\
313.705$-$0.190	&	14 22 34.72	&	 $-$61 08 27.4	&	C98	&	$-$46	&	$-$35.5	&	2.6	&	0.25	&	\bf{2.9}  &	\bf{0.2} &	5p	&		&	1/1.7	\\
313.767$-$0.863	&	14 25 01.63	&	 $-$61 44 58.1	&	C98	&	$-$57	&	$-$49	&	\bf{7}	& \bf{0.27}	&	7.5	&	0.25	&	5P	&		&	1.4	\\
314.320+0.112	&	14 26 26.34	&	 $-$60 38 31.6	&	C98	&	$-$75	&	$-$44	&	0.5	&	0.35	&	\bf{0.65}  &	\bf{0.3} &	5P	&		&	57	\\
316.359$-$0.362	&	14 43 11.00	&	 $-$60 17 15.3	&	C98	&	$-$4	&	5	&	\bf{0.95}  &	$\bf{<0.1}$ &	0.75	&	$<0.1$	&		&		&	111	\\
316.412$-$0.308	&	14 43 23.34	&	 $-$60 13 00.0	&	C98	&	$-$9	&	7	&	\bf{0.4}  &	$\bf{<0.1}$ &	0.5	&	$<0.1$	&		&	c	&	20	\\
316.640$-$0.087	&	14 44 18.35	&	 $-$59 55 11.3	&	C98	&	$-$35	&	$-$15	&	\bf{2.4}  &	\bf{1.3} &	2.0	&	1.5	&	5p	&	c	&	40	\\
316.763$-$0.012	&	14 44 56.34	&	 $-$59 48 00.8	&	C98	&	$-$40	&	$-$35	&	0.8	&	$<0.2$	&	\bf{0.8}  &	$\bf{<0.2}$ &		&	c	&  $<1/2.7$	\\
316.811$-$0.057	&	14 45 26.34	&	 $-$59 49 15.4	&	C98	&	$-$46	&	$-$40	&	27	&	0.7	&	\bf{30}	&	\bf{0.7} &		&	c	&	1.7	\\
317.429$-$0.561	&	14 51 37.62	&	 $-$60 00 20.2	&	C98	&	22	&	26.5	&	6.7	&	3.7	&	\bf{7}	&	\bf{3.9} &		&		&	$<1/35$	\\
318.044$-$1.405	&	14 59 08.74	&	 $-$60 28 25.7	&	C98	&	35	&	46	&	0.3	&	$<0.1$	&	\bf{0.21}  &	$\bf{<0.1}$ &		&		&	23	\\
318.050+0.087	&	14 53 42.69	&	 $-$59 08 52.6	&	C98	&	$-$54.5	&	$-$49.5	&	7.8	&	5.6	&	\bf{6.6}  &	\bf{10.5}	&	5P; 7p	&	c	&	1.7	\\
318.948$-$0.196	&	15 00 55.36	&	 $-$58 58 52.8	&	C98	&	$-$41.5	&	$-$23	&	38	&	14	&	\bf{40}	&	\bf{13}	&	5p	&	c	&	15	\\
319.398$-$0.012	&	15 03 17.41	&	 $-$58 36 12.2	&	C98	&	$-$14.5	&	$-$1	&	0.45	&	0.2	&	\bf{0.4}  &	\bf{0.25} &		&	c	&  $<1/2.2$	\\
319.836$-$0.196	&	15 06 54.57	&	 $-$58 32 57.1	&	C98	&	$-$13	&	$-$5	&	0.9	&	1.3	&	\bf{1.0}  &	\bf{0.9} &		&	c	&	1/2.0	\\
320.120$-$0.440	&	15 09 43.85	&	 $-$58 37 05.8	&	C98	&	$-$58	&	$-$54	&	$<0.1$	&	$<0.1$	&	$\bf{<0.1}$  &	\bf{0.2}	&		&		&	$<1.0$	\\
320.232$-$0.284	&	15 09 51.96	&	 $-$58 25 38.3	&	C98	&	$-$70	&	$-$59	&	9.7	&	6.8	&	\bf{10.8}  &	\bf{7.0} &	5p; 7p	&	c	&	4.6	\\
321.030$-$0.485	&	15 15 51.67	&	 $-$58 11 18.0	&	C98	&	$-$77	&	$-$65	&	0.6	&	0.85	&	\bf{0.6}  &	\bf{0.9} &		&		&	33	\\
321.148$-$0.529	&	15 16 48.39	&	 $-$58 09 50.2	&	C98	&	$-$67	&	$-$60.5	&	1.7	&	$<0.1$	&	\bf{1.7}  &	$\bf{<0.1}$	&	5P	&	c	&	5.3	\\
322.158+0.636	&	15 18 34.62	&	 $-$56 38 25.6	&	C98	&	$-$65	&	$-$43	&	35	&	$<0.15$	&	\bf{36}	&	\bf{0.4}	&	5p	&	c	&	8.3	\\
323.459$-$0.079	&	15 29 19.36	&	 $-$56 31 21.4	&	C98	&	$-$72	&	$-$64	&	12	&	5.8	&	\bf{12}	&	\bf{5.5}	&	5p	&	c	&	1.4	\\

\hline

\end{tabular}
\label{}
\end{table*}

\begin {table*}

\addtocounter{table}{-1}

\caption{\textit{- continued p2 of 3}}

\begin{tabular}{llrlcccccclll}
%\begin{tabular}{lcrcccrcrl}

\hline

\multicolumn{1}{c}{Source Name} & \multicolumn{2}{c}{Equatorial 
Coordinates} & \multicolumn{1}{l}{Refpos}  & \multicolumn{2}{c}{Vel. range}& 
\multicolumn{2}{c}{$\rm S_{peak}(2004)$} & \multicolumn{2}{c}{$\rm 
S_{peak}(2005)$} 
& \multicolumn{1}{l}{Lin(5,7)}  & \multicolumn{1}{l}{Refpol} 
& \multicolumn{1}{l}{m/OH}   \\

\ (~~~l,~~~~~~~b~~~)    &       RA(2000)        &       Dec(2000)       
&  &     $ \rm V_{L}$&$ \rm V_{H}$ &  $\rm S_{1665}$    &  $\rm 
S_{1667}$  &  $\rm S_{1665}$ & $\rm S_{1667}$     &       \\

\ (~~~$^\circ$~~~~~~~$^\circ$~~~) & (h~~m~~~s) & (~$^\circ$~~ '~~~~") & 
& \multicolumn{2}{c}{(\kms)} & (Jy) &  (Jy) & (Jy) & (Jy) \\

\hline

323.740$-$0.263	&	15 31 45.49	&	 $-$56 30 50.7	&	C98	&	$-$80.5	&	$-$37	&	\bf{1.9}  &	\bf{3.3} &	\bf{4.2}  &	\bf{3.5} &	5P; 7p	&		&	762	\\
324.200+0.121	&	15 32 52.92	&	 $-$55 56 07.5	&	C98	&	$-$94	&	$-$83.5	&	12.5	&	12.5	&	\bf{15}	&	\bf{14}	&	5P; 7p	&	c	&	$<1/75$	\\
324.716+0.342	&	15 34 57.42	&	 $-$55 27 24.0	&	C98	&	$-$56	&	$-$44	&	\bf{2.7}  &	$\bf{<0.15}$	&	2.2 & $<0.15$	&	5P	&		&	3.7	\\
326.670+0.554	&	15 44 57.21	&	 $-$54 07 12.7	&	C98	&	$-$41	&	$-$40	&	1.4	&	$<0.1$	&	\bf{1.8}  &	$\bf{<0.1}$ &	5p	&		&  $<1/5.7$	\\
326.780$-$0.241	&	15 48 55.16	&	 $-$54 40 37.9	&	text	&	$-$66	&	$-$57	&	3.0	&	$<0.1$	&	\bf{2.5}  &	$\bf{<0.2}$	& 5p	&	c	&	$<1.8$	\\
327.120+0.511	&	15 47 32.82	&	 $-$53 52 39.4	&	C98	&	$-$88	&	$-$80	&	14	&	2.8	&	\bf{14}	&	\bf{3.1} &	5P; 7p	&	c	&	3.9	\\
327.291$-$0.578	&	15 53 07.78	&	 $-$54 37 06.8	&	C98	&	$-$72.5	&	$-$37.5	&	\bf{7.5}  & \bf{2.0}	&	8	&	1.8	&	5P	&	c	&	1/2.7	\\
327.402+0.444	&	15 49 19.68	&	 $-$53 45 14.3	&	C98	&	$-$86	&	$-$73	&	2.9	&	0.45	&	\bf{1.7}  &	\bf{0.45} &	5P; 7P	&	c	&	34	\\
328.237$-$0.547	&	15 57 58.26	&	 $-$53 59 22.5	&	C98	&	$-$47	&	$-$23	&	7.5	&	0.4	&	\bf{8}	&	\bf{0.4} &	5p	&	c	&	175	\\
328.254$-$0.532	&	15 57 59.80	&	 $-$53 57 59.6	&	C98	&	$-$58	&	$-$31	&	15	&	9.5	&	\bf{15}	&	\bf{9.2} &	5p	&	c	&	27	\\
328.307+0.430	&	15 54 06.48	&	 $-$53 11 40.3	&	C98	&	$-$94	&	$-$88.5	&	22	&	0.35	&	\bf{22}	&	\bf{0.35} &	5p	&	c	&  $<1/147$	\\
328.809+0.633	&	15 55 48.55	&	 $-$52 43 05.6	&	C98	&	$-$49.5	&	$-$33.5	&	110	&	8.5	&	\bf{110} &	\bf{7.8} &	5p; 7p	&	c  &	3.2	\\
329.029$-$0.205	&	16 00 31.79	&	 $-$53 12 49.8	&	C98	&	$-$42	&	$-$36	&	18.5	&	5.0	&	\bf{19}	&	\bf{5.9} &	5p; 7p	&	c	&	6.3	\\
329.029$-$0.200	&	16 00 30.38	&	 $-$53 12 35.5	&	C98	&	$-$40	&	$-$37	&	3.3	&	$<0.3$	&	\bf{3.3}  &	$\bf{<0.3}$ &		&	c	&  $<1/4.7$	\\
329.031$-$0.198	&	16 00 30.32	&	 $-$53 12 27.8	&	C98	&	$-$48	&	$-$45	&	1.0	&	$<0.2$	&	\bf{1.0}  &	$\bf{<0.2}$ &		&	c	&	30	\\
329.066$-$0.308	&	16 01 09.94	&	 $-$53 16 02.9	&	C98	&	$-$44	&	$-$42.5	&	5.3	&	1.7	&	\bf{4.3}  &	\bf{1.7} &	"5p, 7p"	&	c  &	3.5	\\
329.183$-$0.314	&	16 01 47.02	&	 $-$53 11 43.7	&	C98	&	$-$58	&	$-$46.5	&	5.9	&	3.0	&	\bf{6.5}  &	\bf{3.4} &	5p	&	c	&	1.7	\\
329.339+0.148	&	16 00 33.15	&	 $-$52 44 39.8	&	C01	&	$-$108	&	$-$97	&	0.25	&	$<0.1$	&	\bf{0.25}  &	$\bf{<0.1}$ &		&		&	100	\\
329.405$-$0.459	&	16 03 32.15	&	 $-$53 09 31.0	&	C98	&	$-$76	&	$-$67	&	9.4	&	14.2	&	\bf{8.8}  &	\bf{14.2} &	5p; 7p	&	c	&	3.6	\\
330.878$-$0.367	&	16 10 20.01	&	 $-$52 06 07.7	&	C98	&	$-$74.5	&	$-$50	&	440	&	82	&	\bf{530}  &	\bf{88}	&	5p; 7p	&	c	&	1/500	\\
330.953$-$0.182	&	16 09 52.38	&	 $-$51 54 57.3	&	C98, text  &	$-$90	&	$-$87	&	7.3	&	$<1$	&	\bf{7.3}  &	$\bf{<1}$ &	5P	&	Lc	&	1.0	\\
330.954$-$0.182	&	16 09 52.60	&	 $-$51 54 53.7	&	C98, text  &	$-$102	&	$-$80	&	43	&	22	&	\bf{45}  &	\bf{23}	&	5p; 7p	&	Lc	&	1/64	\\
331.132$-$0.244	&	16 10 59.72	&	 $-$51 50 22.7	&	C98	&	$-$95	&	$-$88	&	15	&	2.3	&	\bf{15}	&	\bf{3.4} &	5p; 7p	&	c	&	2.0	\\
331.278$-$0.188	&	16 11 26.57	&	 $-$51 41 56.5	&	C98	&	$-$92	&	$-$78.5	&	\bf{75}	&	\bf{5.0} &	100	&	4.8	&	5p; 7p	&	c	&	1.3	\\
331.342$-$0.346	&	16 12 26.46	&	 $-$51 46 16.3	&	C98	&	$-$67.5	&	$-$64	&	\bf{8.0}  &	\bf{4.6} &	10.5	&	5.5	&	5p; 7P	&	c	&	10	\\
331.442$-$0.186	&	16 12 12.41	&	 $-$51 35 09.5	&	C98	&	$-$84	&	$-$82	&	\bf{0.8}  &	 $-$	&	0.8	&	 $-$	&		&		&	87	\\
331.512$-$0.103	&	16 12 10.12	&	 $-$51 28 37.7	&	C98	&	$-$94	&	$-$85	&	82	&	17	&	\bf{98}	&	\bf{17}	&	5p	&	c	&  $<1/140$	\\
331.542$-$0.066	&	16 12 09.05	&	 $-$51 25 47.2	&	C98	&	$-$87	&	$-$81.5	&	4	&	 $-$	&	\bf{4}	&	 $-$	&		&	c	&	1.75	\\
331.543$-$0.066	&	16 12 09.16	&	 $-$51 25 45.3	&	C98	&	$-$86	&	$-$85	&	$<0.5$	&	$<0.5$	&	$\bf{<0.5}$  &	$\bf{<0.5}$ &		&	c	&	$>30$	\\
331.556$-$0.121	&	16 12 27.19	&	 $-$51 27 38.1	&	C98	&	$-$103	&	$-$96	&	\bf{0.9}  &	$\bf{<0.4}$ &	 $-$	&	 $-$	&		&		&	67	\\
332.295+2.280	&	16 05 41.72	&	 $-$49 11 30.5	&	C98	&	$-$41.5	&	6	&	\bf{1.6}  &	\bf{2.4} &	1.6	&	2.2	&	5p; 7P	&		&	61	\\
332.352$-$0.117	&	16 16 07.16	&	 $-$50 54 30.4	&	C98	&	$-$52	&	$-$42	&	1.7	&	$<0.2$	&	\bf{1.3}  & $\bf{<0.2}$	&	5p	&		&	3.9	\\
332.726$-$0.621	&	16 20 03.02	&	 $-$51 00 32.1	&	C98	&	$-$50	&	$-$44	&	\bf{1.2}  &	\bf{2.3} &	0.9	&	2.1	&	7p	&	c	&	2.6	\\
332.824$-$0.548	&	16 20 10.23	&	 $-$50 53 18.1	&	C98	&	$-$59	&	$-$52.5	&	\bf{1.6}  &	$\bf{<0.2}$ &	1.7	&	$<0.2$	&		&		&  $<1/8.0$	\\
333.135$-$0.431	&	16 21 02.97	&	 $-$50 35 10.1	&	C98	&	$-$60	&	$-$44	&	\bf{29}	&	\bf{12.5} &	30	&	13	&	5p	&	c	&	1/30	\\
333.234$-$0.060	&	16 19 50.90	&	 $-$50 15 10.0	&	C98	&	$-$96	&	$-$80.5	&	\bf{3.0}  &	\bf{0.4} &	3.3	&	0.4	&	5P	&	c	&	1/3.0	\\
333.315+0.105	&	16 19 29.00	&	 $-$50 04 41.2	&	C98	&	$-$48	&	$-$45	&	\bf{0.9}  &	\bf{0.2} &	0.9	&	0.2	&	5P	&		&	20	\\
333.387+0.032	&	16 20 07.58	&	 $-$50 04 47.4	&	C98	&	$-$75	&	$-$70.5	&	1.0	&	$<0.15$	&	\bf{1.0}  & $\bf{<0.15}$ &	5P	&		&	3.0	\\
333.466$-$0.164	&	16 21 20.19	&	 $-$50 09 48.2	&	C98	&	$-$46	&	$-$36.5	&	\bf{2.8}  & \bf{1.5}	&	2.7	&	1.5	&	5p	&	c	&	12	\\
333.608$-$0.215	&	16 22 11.06	&	 $-$50 05 56.3	&	C98	&	$-$57.5	&	$-$46	&	\bf{11}	&	\bf{5}	&	14	&	6	&		&	c	&	$<1/47$	\\
335.060$-$0.427	&	16 29 23.13	&	 $-$49 12 26.9	&	C98	&	$-$48	&	$-$32.5	&	\bf{0.7}  &	\bf{0.5} &	0.8	&	0.3	&	5P; 7P	&		&	62	\\
335.556$-$0.307	&	16 30 56.00	&	 $-$48 45 51.0	&	C98	&	$-$115	&	$-$113	&	\bf{0.2}  &	\bf{0.4} &	0.25	&	0.4	&		&		&	62	\\
335.585$-$0.285	&	16 30 57.33	&	 $-$48 43 39.9	&	C98	&	$-$49	&	$-$40.5	&	\bf{5.2}  &	\bf{0.7} &	5.5	&	0.7	&	5P;  7P	&	c	&	7.3	\\
335.585$-$0.289	&	16 30 58.63	&	 $-$48 43 50.8	&	C98	&	$-$59	&	$-$49	&	\bf{2.6}  &	\bf{0.5} &	2.6	&	0.5	&	5p	&	c	&	27	\\
335.789+0.174	&	16 29 47.37	&	 $-$48 15 51.1	&	C98	&	$-$55	&	$-$45	&	1.2	&	10.0	&	\bf{1.1}  &	\bf{10.1} &	5P; 7P	&	c	&	20	\\
336.018$-$0.827	&	16 35 09.35	&	 $-$48 46 47.1	&	C98	&	$-$49	&	$-$36	&	2.0	&	2.5	&	\bf{2.4}  &	\bf{2.4} &	7p	&		&	42	\\
336.358$-$0.137	&	16 33 29.16	&	 $-$48 03 43.7	&	C98	&	$-$84	&	$-$80	&	\bf{0.8}  &	\bf{0.6} &	0.7	&	0.55	&		&	c	&	16	\\
336.822+0.028	&	16 34 38.26	&	 $-$47 36 33.0	&	C98	&	$-$79	&	$-$74	&	\bf{1.2}  &	$\bf{<0.3}$ &	0.8	&	$<0.2$	&		&	c	&	21	\\
336.864+0.005	&	16 34 54.39	&	 $-$47 35 37.2	&	C98	&	$-$90	&	$-$80	&	\bf{1.9}  &	$\bf{<0.3}$ &	1.6	&	$<0.2$	&	5p	&		&	32	\\
336.941$-$0.156	&	16 35 55.22	&	 $-$47 38 45.7	&	C98	&	$-$71	&	$-$65	&	0.6	&	0.5	&	\bf{0.7}  &	\bf{0.4} &		&		&	33	\\
336.984$-$0.183	&	16 36 12.46	&	 $-$47 37 55.0	&	C98	&	$-$82	&	$-$76	&	0.7	&	0.2	&	\bf{0.7}  &	\bf{0.2} &		&		&	20	\\
336.994$-$0.027	&	16 35 33.95	&	 $-$47 31 11.4	&	C98	&	$-$125	&	$-$114	&	1.7	&	1.3	&	\bf{2.0}  &	\bf{1.3} &	5p	&	c	&	15	\\
337.258$-$0.101	&	16 36 56.36	&	 $-$47 22 26.8	&	C98	&	$-$72	&	$-$55	&	\bf{2.0}  &	\bf{1.0} &	1.9	&	1.1	&		&		&	4.0	\\
337.405$-$0.402	&	16 38 50.57	&	 $-$47 27 59.3	&	text	&	$-$58.5	&	$-$33	&	\bf{140}  &	\bf{55}	&	140	&	58	&	5P; 7P	&	c	&	1/1.9	\\
337.613$-$0.060	&	16 38 09.50	&	 $-$47 04 59.8	&	C98	&	$-$45	&	$-$39	&	\bf{0.8}  & $\bf{<0.2}$	&	0.8	&	$<0.2$	&		&	c	&	25	\\
337.705$-$0.053	&	16 38 29.68	&	 $-$47 00 35.2	&	C98	&	$-$55	&	$-$47.5	&	20	&	5.6	&	\bf{21}	& \bf{5.6}	&		&	Lcv	&	8.1	\\
337.916$-$0.477	&	16 41 10.43	&	 $-$47 08 03.1	&	C98	&	$-$63.5	&	$-$31	&	36	&	4.0	&	\bf{43}	& \bf{4.0}	&	5p	&	c	&	$<1/61$	\\
337.920$-$0.456	&	16 41 06.07	&	 $-$47 07 02.4	&	C98	&	$-$40	&	$-$38.5	&	3.0	&	$<0.15$	&	\bf{5}	& $\bf{<0.2}$	&		&	c	&	6.0	\\
337.997+0.136	&	16 38 48.48	&	 $-$46 39 57.6	&	C98	&	$-$41	&	$-$32.5	&	7.0	&	13.0	&	\bf{7.5}  & \bf{12.5}	&	5P; 7p	&	c	&	1/2.2	\\
338.075+0.012	&	16 39 39.05	&	 $-$46 41 28.0	&	C98	&	$-$54	&	$-$43	&	2.9	&	0.9	&	\bf{2.9}  & \bf{0.9}	&		&	c	&	4.5	\\

\hline

\end{tabular}
\label{}
\end{table*}

\begin {table*}

\addtocounter{table}{-1}

\caption{\textit{- continued p3 of 3}}

\begin{tabular}{llrlcccccclll}
%\begin{tabular}{lcrcccrcrl}

\hline

\multicolumn{1}{c}{Source Name} & \multicolumn{2}{c}{Equatorial 
Coordinates} & \multicolumn{1}{l}{Refpos}  & \multicolumn{2}{c}{Vel. range}& 
\multicolumn{2}{c}{$\rm S_{peak}(2004)$} & \multicolumn{2}{c}{$\rm 
S_{peak}(2005)$} 
& \multicolumn{1}{l}{Lin(5,7)}  & \multicolumn{1}{l}{Refpol} 
& \multicolumn{1}{l}{m/OH}   \\

\ (~~~l,~~~~~~~b~~~)    &       RA(2000)        &       Dec(2000)       
&  &     $ \rm V_{L}$&$ \rm V_{H}$ &  $\rm S_{1665}$    &  $\rm 
S_{1667}$  &  $\rm S_{1665}$ & $\rm S_{1667}$     &       \\

\ (~~~$^\circ$~~~~~~~$^\circ$~~~) & (h~~m~~~s) & (~$^\circ$~~ '~~~~") & 
& \multicolumn{2}{c}{(\kms)} & (Jy) &  (Jy) & (Jy) & (Jy) \\

\hline

338.280+0.542	&	16 38 09.05	&	 $-$46 11 03.1	&	C98	&	$-$63.5	&	$-$60	&	1.5	&	0.65	&	\bf{1.5}  & \bf{0.75}	&	5P; 7P	&	c	&	4.0	\\
338.461$-$0.245	&	16 42 15.53	&	 $-$46 34 18.7	&	C98	&	$-$61	&	$-$53	&	7.4	&	2.1	&	\bf{8}  & \bf{2.1}	&	5p; 7p	&	c	&	8.7	\\
338.472+0.289	&	16 39 58.88	&	 $-$46 12 35.7	&	C98	&	$-$41.5	&	$-$31	&	3.0	&	0.9	&	\bf{2.9}  & \bf{0.75}	&	5P	&	c	&	1/3.3	\\
338.681$-$0.084	&	16 42 23.99	&	 $-$46 17 59.4	&	C98	&	$-$23.5	&	$-$19	&	\bf{5.4}  & $\bf{<0.15}$ &	5.4	&	$<0.15$	&	5P	&	c	&	$<1/36$	\\
338.875$-$0.084	&	16 43 08.23	&	 $-$46 09 12.8	&	C98	&	$-$43	&	$-$33.5	&	2.3	&	4.5	&	\bf{2.4}  & \bf{5.3}	&	5P; 7p	&	c	&	3.8	\\
338.925+0.557	&	16 40 33.57	&	 $-$45 41 37.2	&	C98	&	$-$65	&	$-$55.5	&	17.5	&	0.9	&	\bf{19}	& \bf{1.0}	&	5P; 7p	&	c	&	1/2.4	\\
339.053$-$0.315	&	16 44 49.16	&	 $-$46 10 14.4	&	C98	&	$-$122	&	$-$109	&	0.85	&	0.22	&	\bf{0.95} & \bf{0.25}	&		&	&	137	\\
339.282+0.136	&	16 43 43.12	&	 $-$45 42 08.4	&	C98	&	$-$74	&	$-$66.5	&	\bf{1.5}  & \bf{1.3}	&	1.0	&	1.6	&	5p; 7P	&		&	4.0	\\
339.622$-$0.121	&	16 46 06.03	&	 $-$45 36 43.7	&	C98	&	$-$41	&	$-$30	&	28	&	6.0	&	\bf{30}	& \bf{7.5}	&	5P; 7P	&	c	&	2.9	\\
339.682$-$1.207	&	16 51 06.21	&	 $-$46 15 57.8	&	C98	&	$-$27.5	&	$-$22	&	2.9	&	3.8	&	\bf{2.6}  & \bf{3.9}	&	7p	&	c	&	18	\\
339.884$-$1.259	&	16 52 04.67	&	 $-$46 08 34.7	&	C98	&	$-$40.5	&	$-$26.5	&	\bf{110}  & \bf{160}	&	\bf{59}	& \bf{140}	&	5P; 7p	&	c	&	11	\\
340.054$-$0.244	&	16 48 13.88	&	 $-$45 21 45.1	&	C98	&	$-$59	&	$-$48	&	\bf{37}	& \bf{30}	&	38	&	47	&	5P; 7P	&	c	&	1.0	\\
340.785$-$0.096	&	16 50 14.81	&	 $-$44 42 26.9	&	C98	&	$-$108.5  &	$-$88.5	&	17	&	4.7	&	\bf{19}	& \bf{5}	&	5p; 7p	&	cv	&	7.9	\\
341.218$-$0.212	&	16 52 17.84	&	 $-$44 26 52.5	&	C98	&	$-$41.5	&	$-$35.5	&	12	&	2.0	&	\bf{14}	& \bf{2.8}	&	5p; 7p	&	cv	&	14	\\
341.276+0.062	&	16 51 19.43	&	 $-$44 13 44.5	&	C98	&	$-$75	&	$-$69	&	1.75	&	0.2	&	\bf{2.0}  & \bf{0.3}	&		&	c	&	3.5	\\
343.127$-$0.063	&	16 58 17.19	&	 $-$42 52 08.4	&	C98	&	$-$37	&	$-$22	&	105	&	2.7	&	\bf{112}  & \bf{2.9}	&	5p	&	cv	&  $<1/373$	\\
343.930+0.125	&	17 00 10.92	&	 $-$42 07 18.7	&	C98	&	8	&	17	&	0.35	&	$<0.1$	&	\bf{0.35}  & $\bf{<0.1}$ &		&	c	&	26	\\
344.227$-$0.569	&	17 04 07.81	&	 $-$42 18 40.2	&	C98	&	$-$32	&	$-$18	&	1.8	&	8.8	&	\bf{2.7}  & \bf{12.5}	&	5p; 7p	&	cv	&	7.2	\\
344.419+0.044	&	17 02 08.67	&	 $-$41 47 08.6	&	C98	&	$-$66	&	$-$62	&	\bf{0.55}  & \bf{0.2}	&	0.55	&	$<0.15$	&		&	c	&	4.2	\\
344.421+0.045	&	17 02 08.77	&	 $-$41 46 58.5	&	text	&	$-$74.5	&	$-$73.5	&	\bf{0.2}  & $\bf{<0.1}$	&	$<0.1$	&	$<0.1$	&		&	c	&	84	\\
344.582$-$0.024	&	17 02 57.75	&	 $-$41 41 53.4	&	C98	&	$-$13	&	2	&	\bf{24}	& \bf{9.5}	&	28	&	9.5	&	5p; 7p	&	cv	&	1/8	\\
345.003$-$0.224	&	17 05 11.26	&	 $-$41 29 06.7	&	C98	&	$-$32.5	&	$-$20.5	&	7.5	&	2.4	&	\bf{5.5}  & \bf{2.2}	&	5p; 7P	&	cv	&	27	\\
345.010+1.792	&	16 56 47.58	&	 $-$40 14 25.2	&	C98	&	$-$31	&	$-$15	&	16	&	8	&	\bf{17}	& \bf{9}	&	5p; 7p	&	cv	&	15	\\
345.407$-$0.952	&	17 09 35.45	&	 $-$41 35 57.3	&	C98	&	$-$20	&	$-$16.5	&	\bf{0.7}  & \bf{0.25}	&	0.7	&	0.3	&		&	c	&	2.9	\\
345.437$-$0.074	&	17 05 56.59	&	 $-$41 02 55.6	&	C98	&	$-$35	&	$-$19	&	0.65	&	0.5	&	\bf{1.1}  & \bf{0.6}	&	5P; 7P	&	&	$<1/2.2$	\\
345.494+1.469	&	16 59 41.61	&	 $-$40 03 43.3	&	C98	&	$-$22	&	$-$7	&	10	&	$<0.2$	&	\bf{8}	& $\bf{<0.2}$	&	5p	&		&	$<1/67$	\\
345.498+1.467	&	16 59 42.81	&	 $-$40 03 36.2	&	C98	&	$-$17	&	$-$13	&	12	&	0.75	&	\bf{12}	& \bf{0.8}	&	5p	&		&	1/10	\\
345.487+0.314	&	17 04 28.13	&	 $-$40 44 25.5	&	A00	&	$-$23.5	&	$-$22.5	&	\bf{0.5}  & $\bf{<0.1}$	&	0.5	&	$<0.1$	&		&	v	&	$<1.4$	\\
345.504+0.348	&	17 04 22.87	&	 $-$40 44 22.9	&	C98	&	$-$24.5	&	$-$8	&	\bf{16}	& \bf{12.5}	&	14	&	10.5	&	5P; 7P	&	cv	&	19	\\
345.698$-$0.090	&	17 06 50.62	&	 $-$40 50 59.4	&	C98	&	$-$11	&	$-$3	&	9	&	12	&	\bf{9}	& \bf{12.5}	&	5p; 7p	&	cv	&	$<1/31$	\\
346.480+0.221	&	17 08 00.11	&	 $-$40 02 15.9	&	text	&	$-$18	&	$-$16	&	\bf{0.4}  & \bf{0.2}	&	0.4	&	0.2	&		&		&	75	\\
346.481+0.132	&	17 08 22.76	&	 $-$40 05 25.8	&	C98	&	$-$9	&	$-$1.5	&	\bf{1.2}  & $\bf{<0.2}$	&	1.4	&	$<0.2$	&		&	c	&	1.4	\\
347.628+0.148	&	17 11 51.02	&	 $-$39 09 29.3	&	C98	&	$-$99	&	$-$92.5	&	19	&	7	&	\bf{19}	& \bf{7.5}	&	5P; 7p	&	cv	&	1.0	\\
347.870+0.014	&	17 13 08.80	&	 $-$39 02 29.5	&	C98	&	$-$34	&	$-$30	&	\bf{5.2}  & $\bf{<0.15}$ &	4.9	&	$<0.15$	&		&	c	&	$<1/17$	\\
348.550$-$0.979	&	17 19 20.39	&	 $-$39 03 51.8	&	C98	&	$-$22	&	$-$10	&	\bf{4.5}  & \bf{0.3}	&	4.9	&	0.25	&	5p	&	cv	&	8.2	\\
348.579$-$0.920	&	17 19 10.56	&	 $-$39 00 24.5	&	C98	&	$-$28	&	$-$26	&	$\bf{<0.15}$  &	$\bf{<0.15}$ &	$<0.15$	&	$<0.15$	&	&		&	1/2.8	\\
348.698$-$1.027	&	17 19 58.91	&	 $-$38 58 14.1	&	C98	&	$-$16	&	$-$15	&	2.0	&	$<0.2$	&	\bf{2.0}  & $\bf{<0.2}$	&	&	cv	&	$<1/6.7$	\\
348.703$-$1.043	&	17 20 03.96	&	 $-$38 58 31.3	&	C98	&	$-$18	&	$-$11	&	0.6	&	$<0.1$	&	\bf{0.6}  & $\bf{<0.1}$	&		&		&	108	\\
348.727$-$1.037	&	17 20 06.55	&	 $-$38 57 08.2	&	C98	&	$-$12	&	$-$2	&	3.5	&	0.5	&	\bf{3.3}  & \bf{0.6}	&	5p	&	c	&	23	\\
348.884+0.096	&	17 15 50.15	&	 $-$38 10 12.5	&	C98	&	$-$75	&	$-$70	&	11.5	&	2.6	&	\bf{9.6}  & \bf{2.4}	&	5p; 7p	&	c	&	1.1	\\
348.892$-$0.180	&	17 17 00.21	&	 $-$38 19 27.9	&	C98	&	$-$0.5	&	14	&	1.45	&	0.2	&	\bf{1.6}  & \bf{0.35}	&	5P	&	c	&	1.6	\\
349.067$-$0.017	&	17 16 50.74	&	 $-$38 05 14.4	&	C98	&	8.5	&	16	&	\bf{1.4}  & \bf{0.3}	&	1.8	&	0.3	&		&	c	&	1.3	\\
349.092+0.106	&	17 16 24.59	&	 $-$37 59 45.5	&	C98	&	$-$90.5	&	$-$72.5	&	2.1	&	1.6	&	\bf{2.0}  & \bf{2.0}	&	5p; 7p	&	c	&	4.8	\\

\hline

\end{tabular}
\label{}
\end{table*}

Source parameters are listed in Table 1.  
Column 1 gives the Galactic coordinates, used as a source name, and 
derived from the more precise equatorial coordinates given in columns 2 
and 3.  Column 4 gives a reference to a position measurement for the OH 
emission, with `text' referring to the text of Section 3.3.  
The velocity range of emission is given in columns 5 and 6, and in a few 
cases is larger than currently detectable since it encompasses features 
at outlying velocities that have been prominent in 
the past but subsequently weakened.   
The values of peak intensity of emission, for epochs 2004 and 2005, at 
both 1665 and 1667, are given in columns 7-10, listing the highest peak 
seen in the circular polarization spectra;  non-detections are given as 
upper limits, for example, if no feature exceeds 0.2 Jy, we list as 
$< 0.2 $; a dash indicates no measurement available.  Boldface 
font identifies the epoch of the spectra selected for display in Fig. 1.  

%maybe `lt 0.2' 

Linear polarization detectability from the present spectra is 
summarised in column 11, with 5P and 7P (upper case P) indicating 
the presence of a feature with more than 50 per cent 
at 1665 and 1667 MHz respectively, and 5p and 7p (lower case p) 
indicating our clear detection of linear polarization, but not above 50 
per cent in any feature.  References to past published polarization 
spectra with comparable sensitivity are given in column 12.  

Column 13 refers to the relative prominence of maser 
emission at the 6.6-GHz methanol transition and the stronger 
of the ground-state 1665 or 1667-MHz OH transitions.  
The ratio has been evaluated from the highest peak spectral intensity 
from a methanol 
spectrum and the highest peak of OH emission (generally taken from 
the circularly polarized spectrum displayed here).   
Methanol values for these comparisons were taken from the Methanol 
Multibeam survey (Caswell et al. 2010b, 2011c; Green et al. 2012c;  
see also Caswell 2009).  
The comparison of methanol to OH intensity is superior to earlier 
investigations (e.g. Caswell 1998), owing to many improved methanol 
positions and some improved positions of OH in the present paper, 
allowing confirmation or rejection of some earlier apparent 
associations.  The ratio is believed to be an indicator of the 
evolutionary stage of the high-mass star formation maser site 
(Caswell 1997; 1998) and is discussed further in Section 4.2.

There are two sources tentatively listed in Table 1 that have not 
previously been reported as OH maser sites.  They correspond to features 
newly detected towards known targets, but which we suggest arise 
from an offset position that we think likely to be at a methanol maser 
site.  These sites (344.421+0.045 346.480+0.221) will require future 
confirmation.  We also note, at a less confident level (and thus 
with a remark in the source notes, but with no entry in Table 1) that 
some OH emission seen towards  333.234-0.060 may be from the location 
of methanol maser  333.234-0.062.

In a few cases, we list an OH source, with a previously measured precise 
position, which in the present observations was close to our detection 
limit or not detected.  331.543-0.066 and 348.579-0.920 are confused, 
but the positive OH detections of the past are supported by 
the fact that the measured precise position in each case coincides with 
a methanol maser; 320.120-0.440 is an OH site barely detectable, with no 
corresponding methanol maser,

Our listing of 311.94+0.14 deserves special attention;  not only 
does it have a large position uncertainty of several arcminutes, but we 
were unable to detect it in either 2004 or 2005.  We regard it as an 
interesting OH maser which has been variable, as detailed in the source 
notes.  With the inclusion of this source, Table 1 is a complete list 
of main-line ground-state OH SFR masers, in this region of sky, known up 
until 2005.

%The epoch of the spectrum chosen for display is given in column 12.  

\subsection{Spectra}

Spectra of the 157 maser sites are presented in Figure 1, and these are 
displayed in just 140 panels since, in some instances, plots at a single 
position are sufficient for several adjacent sites that are in closely 
spaced clusters.  
The ordering of the panels follows that of the Table, with minor 
deviations to allow nearby sources that are confused by each other to be 
shown on the same page, usually aligned in velocity.   
The remarkable source 330.878-0.367 is the persistently strongest 
source known (although flares on other sources have surpassed it 
briefly).  For this source we show a second set of plots at an 
expanded intensity scale to reveal the weak features extending over the 
wide velocity range from -50 to -74 \kms, approximately symmetrical 
about the most prominent emission.  
For two other sources we also show a second set of plots, spectra at 
both the 2004 and 2005 epochs, as a demonstration of strong 
variability: 323.740-0.263 where a highly blue-shifted feature 
has flared more than a factor of five to 4 Jy, with more than 50 
per cent polarization; and 339.884-1.259, which includes a feature that 
has flared from 12 Jy to 28 Jy, with more than 50 per cent linear 
polarization.

%aside:  meth at 330.878-0.367 is -60 to -58; companion has meth 
%-73.5 to -55.5;  so systemic seems likely to be near -63
%In the printed article. only the sample first page of Fig. 1 
%is given, with the full set of spectra available in the on-line version 
%of the article as Supporting Information.  
%cf paper I max plot range was 45, OR 30 plus 30 side by side.  
For the majority (128) of the 140 plots, a velocity range of 30 \kms\ 
is sufficient to show all detected features, and display the fine 
detail present.  Larger ranges are used for a few sources with large 
velocity extents, the most extreme being  323.740-0.263 
(55 \kms) and 332.295+2.280 (60 \kms).   

Spectra have a channel separation of 0.488 kHz 
(equivalent to 0.088 \kms) and have not been smoothed, so the 
`resolution', full-width to half-maximum (FWHM), is 1.21 times the 
channel separation.  

For a typical source observed with integration time of 10 min, our rms 
noise level on a spectrum at full spectral resolution is 0.05 Jy.  
%of total intensity 
Longer integration times of up to 20 min were used towards some targets 
where the background sky noise is high.  The most extreme background 
noise was towards 291.610-0.529 (system noise 110 Jy compared with a 
typical value of less than 20 Jy), and the rms noise is 0.25 Jy despite 
the slightly longer integration time.  Towards 291.274-0.709 with 
system noise of 55 Jy, the integration time was increased to 120 
min so as to reduce the rms noise sufficiently below the typical value 
to allow measurements on the weak maser known to be present.  
There is no  detectable interference on any of the spectra 
presented here.

\subsection{Other OH data sets consulted for comparison}

Compilation of source notes has included comparisons with 
several earlier data sets.  Most sources studied here lie south of 
Declination -38$^\circ$ and thus very few can be observed with northern 
hemisphere instruments.

\subsubsection{VLA comparisons}

%note conventions

The VLA dataset of Argon et al. (2000) provides good spectral 
resolution for northerly sources.  Although limited to the two 
circular polarizations, the VLA observations allow comparisons 
at both 1665 and 1667 MHz with the Parkes data for five sources.  
Comparisons for a further nine sources are useful, but limited, 
owing to either the absence from the VLA dataset of 1667-MHz 
observations, or incomplete velocity coverage.  
VLA observations of these southern sources are necessarily made at low 
elevation near the VLA southern horizon.  As remarked by Caswell, Vaile 
and Forster (1995b), the VLA declination errors can then be quite large, 
and Argon et al. (2000) note that in the most extreme example of their 
dataset, the error reached 8 arcsec.  

%limited v coverage for:  and 1667 not shown for:

\subsubsection{Earlier Parkes data}

Many of the subsequent comparisons regarding variability relate to 
earlier Parkes data, with spectra from 1978 onwards (Caswell, Haynes  \& 
Goss 1980 and later citations) displaying good sensitivity and spectral 
resolution in the two circular polarizations especially useful.

\subsubsection{LBA}

The southern Long Baseline Array (LBA) has been used in studies 
of the 1665 and 1667-MHz transitions towards four of the sources studied 
here (Caswell, Kramer \& Reynolds 2011b and references therein).  The 
beamsize was approximately 100 mas, but only the two circular 
polarizations were computed.  Successful pilot observations computing all 
four Stokes parameters were made more recently, of 340.054-0.244  (Bains 
et al. 2007), and we are now using this mode in newer LBA observations.

\subsubsection{ATCA data from the MAGMO survey}

The installation of CABB on the ATCA now allows excellent 
spectropolarimetry suitable for OH masers, as first demonstrated 
by Caswell \& Green (2011).  Using this new capability, a large-scale 
study of OH masers (the `MAGMO' survey) is now underway, with 
observations of a pilot region already available (Green et al. 2012b).  
There are nine sources in the pilot region (longitude range 280$^\circ$ 
to 295$^\circ$) with excellent ATCA data suitable for comparison 
with the corresponding Parkes spectra in the present survey.

\subsection{Source notes}

The following source notes discuss comparisons  
with earlier data, and variability, and draw attention to unusual 
features such as large velocity widths, unusual ratio of 1665 
to 1667-MHz intensity, and exceptionally high linear 
polarization.  We also remark on the upper limit of methanol emission in 
those cases where methanol is absent.  
Thus the notes are a foundation and foretaste of future detailed studies 
of individual sources, and a guide to those which are most urgent.

For an evaluation of ppa from our displayed 
values of Q and U, it is useful to recall that the relative values of Q 
and U as a function of ppa are: ppa 0$^\circ$ (Q=+1); 45$^\circ$ (U=+1); 
90$^\circ$ (Q=$-1$);  135$^\circ$ (U=$-1$).  

The notes include remarks on a few sources that 
appear to be currently in a quiescent mode (not detected in the 
present observations) but have been listed in Table 1.

\subparagraph{240.316+0.071 }  
The current peak of 1 Jy is comparable to epoch 1993 (Parkes 
archival data); we see for the first time 1667-MHz emission of 0.27 Jy.   

Unusually, OH emission in the excited-state at 6035 MHz is stronger 
than the 1665-MHz ground-state maser, and there is no associated 
methanol maser (Caswell et al. 1995).

\subparagraph{263.250+0.514 }  
1665-MHz emission remains similar to 1993, and comparable emission 
is present at 1667 MHz (as previously noted with the ATCA), with 
more features.

\subparagraph{284.351-0.418 }  

This is the first of nine sources for which new ATCA polarimetric 
data have been obtained in the pilot phase of the MAGMO project (Green 
et al. 2012b).  Continuing the variability that occurred since 1992 
and 1995, we now also see significant changes between 2005 and 
2011, despite a superficial similarity.  Most notably,   
MAGMO (epoch 2011) shows weakening of 1665-MHz emission at 6.1 
\kms, predominantly RHCP, (from 1.6 Jy 2004; 2.0 Jy 2005 to 0.25 Jy), 
but a flare of RHCP at velocity +5.36 \kms\ to 1.26 Jy, and of a 
likely LHCP Zeeman partner at +5.71 \kms\ to 0.55 Jy.

\subparagraph{285.263-0.050 }  
Many features seen in 1982 are still recognisable, despite some 
variability.  Between 2004 and 2005 there was a doubling of the peak 
1665-MHz feature.  Green et al. (2012b) show that differences 
continue with the 2011 epoch observations of MAGMO, but there is 
generally clear recognition and persistence of many features.     
Zeeman patterns corresponding to magnetic fields of approximately 
+10 mG, for emission centred near velocity +9 \kms, were most evident
from past spectra at both 6035 MHz and 1665 MHz (Caswell \& Vaile 
1995).  Green et al. (2012b) demonstrate that some 1665-MHz and 
1667-MHz features centred near velocity +3.3 \kms\ reveal Zeeman 
pattern fields of approximately +3 mG.  Our 2004 spectra displayed
here also support the +3 mG field, based on features between 
+2 and +5 \kms\ at 1667-MHz (the main features),
and  1665-MHz (secondary features). However, our 1665-MHz main LHCP
features between +5 and +7 \kms\ have matching RHCP features
that are weaker and less obvious, and at a much larger shift of
+5.5 \kms, to between  +10 and +13 \kms, and correspond to a field
of approximately +9 mG (as was more clearly evident in the 1665-MHz
spectra from 1993 shown by Caswell \& Vaile 1995).

%JAG*****remarks above and for following source to be checked by JAG; 
%and note required discussion of 285.263-0.050 in 6035 and 6030 paper****

\subparagraph{287.371+0.644  }  
Emission only at 1665 MHz, with no detection at 1667 MHz, is similar 
to 1993 (LHCP at -4, RHCP at 0 \kms) but LHCP is now twice 
as strong;  the weak RHCP feature of 0.2 Jy (displayed for 
2004) faded to 0.15 Jy (2005) and 0.1 Jy in 2011 (MAGMO).  
The features appear to be a likely Zeeman pair, indicating a field 
of +7 mG.

\subparagraph{290.374+1.661 }  
The secondary 1665-MHz feature at -19.4 \kms\ flared in 2005 
(as displayed) by a factor of three but decayed again in MAGMO 
observations (2011);  MAGMO spectra also  
showed a similar decrease at -24 \kms, but a continuing increase of 
1667-MHz emission, at just the peak, at -23.6 \kms.

\subparagraph{291.274-0.709 }
This source, at 1665 MHz, was first reported by Bourke et al. (2001) 
from Parkes observations in 1995 and 1996.  Its precise position was 
determined with the ATCA  (Caswell 2004c) and confirmed by 
MAGMO observations (Green et al. 2012b). 
Bourke et al. showed the total intensity and V spectra.  Our 
observations with similar sensitivity and spectral resolution 
(with a 2-hour integration taken in 2004, and not observed 2005) yielded 
similar spectra for I and V but, as with our other spectra, we choose to 
show RHCP and LHCP; these emphasise that the double 
peaked RHCP polarized emission (evident from close comparison of RHCP 
and LHCP spectra) is dwarfed  by  the absorption of strong continuum 
background emission.  Our total intensity spectrum shows the amplitude 
of the narrow features to be the same as seen in RHCP-LHCP (i.e. V) and 
thus no evidence of LHCP emission.  The ATCA 
spectra from MAGMO show the maser very clearly, and also 
show that the absorption minimum is offset from the maser, 
approximately located at the peak of extended continuum background 
emission.
 
All three data sets indicate that the maser emission remains similar 
at epochs 1995, 2004 and 2011, with peak flux density approximately 
0.4 Jy.

\subparagraph{291.579-0.431, (291.579-0.434), 291.610-0.529 and 
 291.654-0.596 }  
Targeted observations were made towards both 291.579-0.431 and 
291.610-0.529 and, since their separation is only 6 arcmin, each 
is seen with half intensity at site of the other.  The aligned 
spectra demonstrate this.  

Following detections with peak flux density of 8.5 Jy in 
1970 and 1976, 291.579-0.431 was not detectable in 1982 (Caswell \& 
Haynes 1987a) but in 1988 was dominated by RHCP emission of 2 and 1.5 
Jy at 14.5 and 16.2 \kms. LHCP emission was a single feature of 1 
Jy at 13.45 \kms.  The latter was still present 
in 2004 and 2005, at about 0.7 Jy, with a stronger feature of 1.3 Jy  at 
12.2 \kms;  both features are also present in 2011 (MAGMO). RHCP 
emission  has weakened since 1988, with a single feature at 16.2 \kms\ 
clearly detected and a likely feature at 15.7 \kms.  The 2005 
appearance suggests two Zeeman pairs with RHCP at velocity more 
positive than LHCP by 3.1 \kms.  
No significant 1667-MHz emission was seen in 2004 or 2005 (apparent 
1667-MHz emission is 291.610-0.529).

A nearby source 291.579-0.434 (offset 10 arcsec from 291.579-0.431) 
was recognised from the MAGMO 2011 data as a 1665-MHz RHCP feature of 
0.7 Jy at 18.8 \kms;  MAGMO shows that a 1667-MHz RHCP 0.9-Jy feature at 
13.6 \kms\ also originates from this site.  291.579-0.434 was not 
detected in 1982 (Caswell \& Haynes 1987a), but showed strong emission of 
4 Jy at 1665 MHz in 1988, although not then recognised as distinct from 
291.579-0.431.  We do not list the source in Table 1 since it is not  
detectable on our spectra of 2004 or 2005, or earlier archival spectra 
(e.g. 1982).  
Both  291.579-0.431 and 291.579-0.434 are sites of water maser emission 
(Breen et al. 2010b), and 291.579-0.431 is a site of methanol maser 
emission.  

291.610-0.529 has remained remarkably stable over several decades, as 
noted by Caswell \& Haynes (1987a) from their 1982 spectra and 
earlier comparisons, and continuing through 1988 and 2004, 2005 into 
2011 (MAGMO).  1665-MHz emission is dominated by LHCP emission but 
accompanied by weaker RHCP emission;  1667-MHz emission shows a 
weaker, equal amplitude, RHCP and LHCP pair of features. 

MAGMO shows an additional source, 291.654-0.596, offset 
by nearly 6 arcmin from 291.610-0.529 (and 11 arcmin from 291.579-0.431 
and thus absent from that spectrum), with 
strongest emission of 0.8 Jy at 1665 MHz, velocity 16.1 \kms, RHCP.  
With hindsight, 291.654-0.596 is recognisable in the 1982 spectrum  
(Caswell \&  Haynes 1987a) and, more clearly, in an unpublished Parkes 
1988 spectrum, with peak flux density of 2 Jy. It is also seen in our 
2004 data, with peak of nearly 2 Jy again (after correction accounting 
for its offset to the halfpower point of the targeted position).  

No methanol maser has been detected at either site, but 291.610-0.529 
has associated water maser emission.

%assume Chris plotted Im and labelled it V, whereas V=2xIm;  
%then maser intensity same as Tyler; file with ngc3576 paper.

%\clearpage

%\clearpage

\subparagraph{294.511-1.621 }  
The peak  1665-MHz feature, RHCP, remained between 32 and 37 Jy from 
1989 through 2004 to 2005;  the 2011 spectrum showed a decrease 
of this feature to 20 Jy, and the reduction of a feature at 
-9.5 \kms\ to 0.2 Jy.  But increases occurred in blue-shifted features 
at -19.5 \kms\ (0.4, 1.0, 4.0 Jy), and -15.84 \kms\ (0.3, 0.3,  0.8 Jy), 
both with strong linearly polarized emission.     
Close Zeeman pairs centred near -12 \kms\ are evident at both 1665 and 
1667 MHz as reported from MAGMO.  
This is the ninth (final) of our targets that were later 
observed in the MAGMO pilot survey.  
MAGMO demonstrated that, for sources between Galactic longitudes 
284.5$^\circ$ and 295$^\circ$,  Zeeman patterns indicate a 
persistent positive magnetic field (RHCP at velocity more 
positive than LHCP), consistent with a coherent field over many kpc 
of the Carina-Sagittarius arm, where we view it along a tangent.

\subparagraph{297.660-0.973 }  
Some 1665-MHz  features are comparable with 1982, especially at  
+27.6 \kms, now seen to display linear polarization greater than 
90 per cent (consistent with 1982 no net circular).  
In Section 6.1 and 6.2 we suggest that it is part of a previously 
unrecognised Zeeman triplet.  

\subparagraph{298.262+0.739 } 
The source was first reported (as 298.262+0.740) by MacLeod et al. 
(1998) and a precise position was measured in new ATCA observations. 
Current emission is not distinguishably different from the 
discovery observation pre-1998.

\subparagraph{299.013+0.128 }  
The major feature, 1665-MHz  LHCP (but with 30 per cent linear 
polarization) at +20.1 \kms, although similar to its epoch 1982 
appearance, flared by a factor of three in 1990 and 1992 
archival observations. Our current good sensitivity now allows 
weak  RHCP 0.1 Jy at +23.3 \kms\ to be seen.

\subparagraph{300.504-0.176 }  
The original detection of this source in 1982 was merely 1665-MHz 
RHCP emission of 0.6 Jy near +22.5 \kms\ (Caswell \& Haynes 1987a).   
By 1990, a possible LHCP feature of less than 0.4 Jy was all that 
remained, but in 1992 a flare of 1.4 Jy in both RHCP and LHCP 
emission (separated by 1 \kms\ in velocity) had occurred near 
+22.5 \kms, and weak emission of 0.2 Jy was detectable between 
+3 and +7 \kms.  The position determined with the ATCA in 1995 
was based on the +22.5-\kms\ feature.  
Emission at epoch 2005 (displayed), and epoch 2004, is now 
strongest between +2 and +4 \kms, but with weak LHCP emission seen 
at +22.5 \kms.  OH absorption at 1667 MHz is centred at +8 \kms, 
with a hint of emission LHCP at +7.7 and RHCP at +8.8 \kms.  
The associated methanol maser has a peak at +7.5 \kms, and range +2.5 
to +10 \kms, and associated water has features between -37 and +14 
\kms;  a nearby water maser (300.491-0.190, offset 1 arc min) has 
a peak at +23 \kms\ (Breen et al. 2010b).  We regard the systemic 
velocity of these masers as not reliably determined.

\subparagraph{300.969+1.147   }  
The spectra remain similar to epoch 1982 for both 1665 and 1667 MHz.  
Observations with high spatial resolution are presented by Caswell, 
Kramer \& Reynolds (2009), and confirm clear Zeeman patterns and a 
consistent magnetic field direction.  
From this Galactic longitude onwards, it may be the first of 
several sources showing a magnetic field opposite to the 
persistent field seen between longitudes 284.5$^\circ$ 
and 295$^\circ$ in the MAGMO pilot sample.

\subparagraph{301.136-0.226 }  
Emission has remained strong at 1665 and 1667 MHz over several 
decades.  Its very wide velocity  range extends from -64 \kms\ 
where there used to be (1982) 1665-MHz  RHCP emission, 
and certainly -51 \kms\ where there is currently a 1667-MHz 
feature (and formerly a 1665-MHz feature), to at least -33  
\kms\ (currently at 1665 and 1667 MHz)
An accompanying methanol maser has a velocity range of only 4 
\kms, mid-range -39 \kms\ (the likely systemic velocity), and thus 
indicative of the OH showing an extra blue-shifted outflow.

\subparagraph{305.200+0.019 }  
There is some confusion from 305.362+0.150 and 305.208+0.206, 
causing 7 and 10 per cent sidelobe responses respectively.  
Genuine features are only at 1665 MHz: 1.5 Jy RHCP at -33 \kms\ 
and probably 0.4 Jy LHCP at -30 \kms.  

%The LHCP features at -34.5 and -33.8 \kms\ may also be here.  

\subparagraph{305.202+0.208, 305.208+0.206 and 305.362+0.150 }  
We display a shared spectrum for the first two (separation less than 30 
arcsec), with an additional spectrum for the third, offset nearly 10 
arcmin, but strong enough to be seen as a 10 per cent sidelobe at the 
other position.  
The last of these, 305.362+0.150, is strongest and, with the exception 
of 6-Jy RHCP features at -38.1 (1665 MHz) and -37.1 \kms\ (1667 MHz),  
all features seen when targeting 305.362+0.150 do indeed arise here:  
mainly RHCP, but with linear polarization exceeding 50 per cent for 
the 3-Jy 1665-MHz feature at -41.6 \kms. 
1667 and 1665-MHz features are all recognisably similar to their 1982 
appearance, with amplitude variations by a factor of two.  

305.202+0.208 may be merely a 1665-MHz 6-Jy feature near velocity 
-40.8  \kms.  

The remainder of the first spectrum, with the exception of the 
weak sidelobes from 305.362+0.150, arises from 305.208+0.206.  RHCP 
features dominate at both 1665 and 1667 MHz, at -38.1 and 
-37.1 \kms\ respectively, and a LHCP 1665-MHz feature at -32.5 \kms.  
Linear polarization exceeds 50 per cent at the isolated 2-Jy 
1667-MHz feature at -44 \kms.  

\subparagraph{305.799-0.245 }  
The spectrum remains the same as in 1982, with plausible Zeeman 
pairs at 1665 and  1667 MHz, with LHCP at more positive velocity.  
There is a coincident very weak methanol maser, and weak (1 mJy) 
compact continuum emission, weaker at 4.8 than 8.6 GHz, and thus 
most likely an ultra- or hyper-compact \HII\ region (Guzman 
et al. 2012).  
  
\subparagraph{306.322-0.334 }  
An emission peak of 1.0 Jy LHCP was present in 1992, but is now 
very weak, 0.15 Jy,  mainly RHCP.    
Weaker 1667-MHz emission may be present.    

\subparagraph{307.805-0.456 }  
Generally similar to 1993.
In addition to highly circularly polarized features, there is 
also strong linear polarization:  80 per cent for 1665-MHz 1 Jy at 
-18.1 \kms, 60 per cent -16.2 \kms; 40 per cent at 1667 MHz -14.8 \kms.

\subparagraph{308.754+0.549 }  
First reported by Caswell (2004c) as a maser site towards the optical 
HII nebula RCW 79.  Our 2005 polarization spectrum, displayed here, 
shows just 1665-MHz emission, mainly RHCP for several features.  

\subparagraph{308.918+0.123 }  
1665-MHz LHCP features are similar to 1982, but RHCP emission is 
now weaker. 1667-MHz emission remains generally similar.  
Note a quite wide velocity range of 22.5 \kms, with 1665-MHz 
emission extending to -47.5 \kms, and 1667-MHz emission to -70 \kms, 
beyond the range of archival data, and also quite weak.    
Since the accompanying methanol maser velocity range is -56 to 
-52.5 \kms, the wide range seems to be caused by a blue-shifted 
outflow.

\subparagraph{309.384-0.135}  
Emission at 1665 MHz remains similar to 1990;  linear polarization 
of 80 per cent is now seen at -50.1 \kms.  

\subparagraph{309.921+0.479 }  
A strong source with long history, remaining generally similar to 
spectra in 1982, but the LHCP 1665-MHz feature at -60 \kms\ has 
flared from 28 Jy to 80 Jy (now the strongest feature).  At 1665 MHz 
there is only weak linear polarization.   A new 1667-MHz feature, 
the strongest in 2004 and 2005, with total intensity peak of 10 Jy 
at  -61.8 \kms\ displays essentially  100 per cent linear 
polarization.  In Section 6.1 and 
6.2 we suggest that it is part of a Zeeman triplet.  

\subparagraph{310.144+0.760 }  
The features of this weak source at 1665 and 1667 MHz are 
mainly LHCP, similar in 2005 and 2004.

\subparagraph{311.643-0.380 }  
Still fairly similar to 1982 for both 1665 and 1667 MHz.  
Linear polarization is especially high at 1665 MHz for 1.5-Jy 
features at 26.9 and 28.1 \kms.  
We note that weak features over a large velocity range were reported 
by Caswell (1998), and are confirmed in the present spectra.

\subparagraph{311.94+0.14 }  
We note that the position of 311.94+0.14, has a large uncertainty of 
several arminutes;  we were unable to detect it in either 2004 or 2005 
and a detailed discussion follows.    

The site was first reported by Caswell \& Haynes (1987a) from a 
Parkes observation in 1982, citing the target position where it 
was discovered but with no refinement except to note its 
undetectability from positions 6 arcmin away.   
The 1982 spectrum is poor, but archival spectra from Parkes in 1990 
January and 1992 August with much lower noise level gave clear  
detections confirming a LHCP feature with peak of 0.5 Jy and width 
nearly 1 \kms.  However no detection was present in our current 
spectra in 2004 (as displayed) or 2005, with upper limit 0.2 Jy, and 
clearly indicating variability since 1992.  
A weak methanol maser (Caswell et al. 1995a, Caswell 2009, Green et 
al. 2012c) is at the precise position 311.947+0.142 
%(14 07 49.72, -61 23 08.3) 
($14^h07^m49.72^s,~-61^{\circ}23{\arcmin}08.3{\arcsec}$),
and this may be the 
location of the OH site.  We report the current OH non-detection as a 
record of its variability should it be successfully measured in 
future.  We note that likely associated \HII\ region emission is 
estimated to be at the far kinematic distance (Caswell \& Haynes 1987b; 
Caswell et al. 1975), 8.5 kpc (using current Galactic size parameters).  
The report of a water maser at this site prompted the first detection 
of the OH, and the OH and water reports prompted the detection of 
the methanol.  It is ironic that the water and OH have faded and 
prevented precise position measurements, and only the methanol has 
a precisely known position, but 
it seems likely that the other species are at this same position.

\subparagraph{312.598+0.045 }  
The weak features seen in 1982 at both 1665 and 1667 MHz remain 
similar, but the strongest  1665-MHz feature at -61.5 \kms\ 
has fallen dramatically, from 12 Jy to less than 1 Jy.  1665-MHz 
weak linear polarization is seen, corroborated 2004 and 2005.  

\subparagraph{313.469+0.190 }  
Very little change since 1982 or 1989.  Weak 1667-MHz emission is 
now seen, and linear polarization at 1665 MHz.  

\subparagraph{313.577+0.325 }  
Weak emission is present at both 1665 and 1667 MHz, and linear 
polarization is seen at 1665 MHz.

\subparagraph{313.705-0.190 }  
LHCP emission dominates both the strong 1665-MHz emission, and the 
weak 1667-MHz emission, with some linear polarization now seen at 
1665 MHz.

%previous location of figures

\subparagraph{313.767-0.863 }  
Significant variability at 1665 MHz has occurred since 1993, and 
linear polarization is seen in weak features;  weak emission is 
seen at 1667 MHz.

\subparagraph{314.320+0.112 }  
A 1665-MHz feature at -45 \kms\ remains the same as in 1993,  
and is now seen to display high linear polarization.  The 1667-MHz 
Q and U spectra show a strong hint of similar linear polarization 
at the same velocity.  If future more sensitive observations confirm 
this, it will be another candidate for a Zeeman pattern isolated 
$\pi$ component of the variety discussed in Sections 6.1 and 6.3.  
A prominent blue-shifted emission peak at -72 \kms\ (seen with the 
ATCA epoch 1996 and shown by Caswell (1998)) is not evident in 2004 
or 2005, but weak broad emission extends from -45 to -60 \kms\ 
at both 1665 and 1667 MHz, as noticed in 1993 spectra.  
Spectra of methanol and water (Breen et al. 2010b) similarly 
suggest a blue-shifted outflow in all three maser species, 
with the strongest methanol peak (rather than the mid-range) 
perhaps indicative of the systemic velocity.  

\subparagraph{316.359-0.362 and 316.412-0.308 }  
The sources are separated by 4 arcmin and their spectra have been 
aligned to better distinguish them.  The second source, 
316.412-0.308, has been known since 1987, remaining similar, with 
0.3 Jy RHCP at -8 \kms,  and 0.4 Jy LHCP at -2 \kms. 
316.359-0.362 emission was at +4.5 \kms\ (1993) and -0.5 \kms\ (ATCA 
in 1996), but is now mainly a new LHCP feature at -3.1 \kms, of  
about 1 Jy.     A +6-\kms\ feature is 
present at both positions, with similar low amplitude at both epochs and 
might arise from the location of methanol maser 316.381-0.379 
(velocity range -6 to +1.5 \kms) which lies between the OH target pair.

\subparagraph{316.640-0.087 }  
Rich spectra at 1665 and 1667 MHz extend  from $-35$ to $-15$ \kms.  
There has been much variation since 1982 and 1989 and, notably, 
a 1665-MHz feature at $-29.8$ \kms\ in 2004 (displayed spectrum) 
disappeared 2005.  At 1667 MHz there are persistent features 
from $-35$ to $-32$ \kms, blue-shifted relative to the OH peak, 
and the methanol peak and to the methanol median velocity near 
$-20$ \kms.

\subparagraph{316.763-0.012 and 316.811-0.057 }  
Features are seen between -47 and -35 \kms, within deep absorption.  
ATCA observations (Caswell 1998) show the two sources separated by 
only 4 arcmin, A single spectra is shown, at 316.763-0.012, and thus 
emission from 316.811-0.057 is reduced by the offset, and its 
tabulated peaks have been corrected from the measured peaks by 
a factor of 1.3.   
Most of the emission between -47 and -40 \kms\ is 
from the second position.  Only the  1665-MHz emission -37 to -35 \kms\  
LHCP is from the first position.   
Amongst the features ascribed to 316.811-0.057, we note that the 
strongest current feature of 30 Jy at -43.2 \kms\ was only 2 
Jy in 1982,

\subparagraph{317.429-0.561 }  
The positive velocity indicates an unambiguous large kinematic distance 
outside the solar circle, beyond 15 kpc.  Coincident with the OH maser, 
there is a uc\HII\ region with flux density nearly 20 mJy at 8.6 GHz, 
and weaker at 4.8 GHz (Guzman et al. 2012).  
There is no methanol maser reported at this location.  
The OH intensity at both 1665 and 1667 MHz remains comparable to 
values  measured with the ATCA, epoch 1996, and the corresponding 
luminosity at the large distance is notably high.

\subparagraph{318.044-1.405 }  
The spectral shape is similar to 1993 but weakened to half the 
intensity.

\subparagraph{318.050+0.087 }  
Since 1982, the 1665-MHz peak at $-52.8$ \kms\ has  
decreased from 57 (30 in 1990) to 7 Jy, but at 1667 MHz, increased from 
1 to 5.6 Jy (2004) and to 10.5 Jy (2005), and thus now exceeds the 
1665-MHz peak intensity.  

At both 1665 and 1667 MHz, RHCP is stronger but with significant 
linear polarization in some features.

\subparagraph{318.948-0.196 }  
Hugely variable since 1982 and less so since 1989.  At 1665 MHz, the 
peak has increased from 1 Jy to 40 Jy;  at  1667 MHz, increased from 1.5 
Jy to 13 Jy.  
The main features at 1665 and 1667 MHz resemble matching Zeeman 
pairs, with LHCP at higher velocity.  
Also, 1667-MHz emission now has an outlying feature at -24 
\kms, with peak of 3 Jy (2004) and 1 Jy (2005).

\subparagraph{319.398-0.012 }  
The peak in 1982 exceeded 1 Jy at both 1665 and 1667 MHz, and has 
subsequently weakened, although the higher sensitivity of current 
observations reveal additional features over a larger velocity range.

\subparagraph{319.836-0.196 }  
1665-MHz features remain similar to epoch 1982, although a flare of 
5 Jy occurred at -10.1 \kms\ in 1989.  1667-MHz emission is now stronger 
and comparable to 1665 MHz.

\subparagraph{320.120-0.440 }  
In 2004 and 2005,  the only distinguishable feature here is a very 
weak 1667-MHz 0.2-Jy LHCP feature near -55.5 \kms.  All remaining 
emission on the displayed spectra is a sidelobe response to 
320.232-0.284 (see following source note).

\subparagraph{320.232-0.284 }  
1665-MHz emission has shown persistent features over many decades, but 
with strong intensity variations since 1982 and  1989.  In particular, 
the  strongest 1665-MHz feature in 1982 exceeded 20 Jy at -67.9 
\kms\ (now fallen to 9 Jy 2004 and 5 Jy 2005), and the strongest feature 
in 2005 and 2004 is 10.5 Jy at -64.2 \kms, previously 2 Jy.  1667-MHz 
emission has shown no obvious change since 1982. 
At each transition, there is significant linear polarization, mostly  
in features that are predominantly circularly polarized.  

\subparagraph{321.030-0.485 }  
Strongest at 1667 MHz.  A 0.5-Jy RHCP feature at -55 \kms\ 
was present in 1993 at 1665 MHz, and at this velocity there is now 
a weak 1667-MHz feature.  
At 1665 MHz there is some confusion from the following source, 
321.148-0.529, whose spectrum is aligned beneath it to make this 
clear.   Features in the range -77 to -65 \kms\ are not affected, 
apart from the 0.9-Jy (total intensity) peak at  -66.5 \kms\ 
which is blended with a 0.45-Jy sidelobe contribution from a narrow 
linearly polarized feature of 321.148-0.529. 

\subparagraph{321.148-0.529 }  
Note that, at 1667 MHz, only the narrow feature at -65.9 \kms\ is 
from this site, and the remainder from the previous 
source.  The 1665-MHz emission remains remarkably similar to 1982 with 
total velocity range -67 to -60.5 \kms.  Note that the -66.5 \kms\ 
feature of 1.45-Jy peak total intensity includes  a small 0.15-Jy 
contribution of unpolarized emission from the previous source;  the 
residual emission of 1.3 Jy is thus essentially 100 per cent linearly 
polarized.   

\subparagraph{322.158+0.636 }  
In 1982, 1989, 1996,   1665-MHz emission comprised a single feature 
near -61 \kms\ varying between 3 and 1 Jy; it has  now flared to more 
than 30 Jy, with linear polarization, but ppa changed from 30$^\circ$ 
to 70$^\circ$ 2004 to 2005.  
Methanol  at this site is also strongly  variable.  
Weak 1667-MHz emission is seen, for the first time, between 
-45 to -43 \kms, apparently red-shifted relative to 
systemic, but no check has yet been made to verify its precise 
position, and we suspect that it arises from water 
maser site 322.165+0.625 (Breen et al. 2010b), offset nearly 
one arcminute, close to a compact \HII\ region but with no methanol 
maser.  

\subparagraph{323.459-0.079 }  
In LBA 1998 measurements (Caswell \& Reynolds 2001), 
only  circular polarization could be analysed, but features could 
be distinguished spatially as well as by frequency.  Most features 
were highly circularly polarized, many 
of them in identifiable Zeeman pairs.  The present single dish spectra 
still closely resemble the 1998 spectra (and also earlier Parkes 
spectra of 1982 and 1990), and reveal that linear polarization is, 
indeed, low.

\subparagraph{323.740-0.263 }  
OH multiple features from -60 to -37 \kms\ at both 1665 and 1667 MHz 
are present with comparable strength.  
Accompanying methanol is strong, and confined to the range -59 to 
-42 \kms,  i.e. midrange -50.5 \kms, which is also the peak of 
emission, and the likely systemic velocity.  
However 1665-MHz emission, with a flaring feature at -79 \kms, 
extends to -80 \kms, blue-shifted -29.5 \kms\ from the systemic 
velocity.  
For this source we show spectra at both the 2004 and 2005 epochs, as a 
demonstration of strong variability:  the highly blue-shifted 
1665-MHz feature at -79.5 \kms\ has flared from 0.7 to 4 Jy, with more 
than 50 per cent linear  polarization.  Other features also show 
strong linear polarization.

\subparagraph{324.200+0.121 }  
Most features at 1665 and 1667 MHz remain similar to 1982.  

The strongest feature is at -91.6 \kms\ for both 1665 and 1667-MHz 
transitions;  it is highly linearly polarized, more than 50 per 
cent at 1665 MHz and almost 50 per cent at 1667 MHz, with 
comparable ppa.  The feature is discussed later (Sections 6.1 and 
6.3), along with several other sources, in the context of Zeeman 
pattern isolated $\pi$ components.

\subparagraph{324.716+0.342 }  
Only 1665-MHz emission is confidently detected; it is similar to 
1982, but twice as strong, and very high linear polarization is 
present.

\subparagraph{326.670+0.554 }  
A single 1665-MHz feature remains similar to the earliest 
known spectrum (1993, Parkes archival), and is  now (2004, 2005) 
seen to display 40 per cent linear polarization as well as LHCP.

\subparagraph{326.780-0.241  }  
The earliest spectrum in 1978 (at 326.77-0.26) showed  1665-MHz 
emission LHCP of 2 Jy at -58 \kms; no emission was subsequently 
detectable for several years until 2004 November, as 2.5 Jy 1665-MHz 
RHCP at  -65.2 \kms.  
It was again detected, 2005 March with the ATCA, near -65 \kms, and 
an accurate position measured.  

It coincides with a water maser (with slightly worse position 
uncertainty, 0.04 s and 1 arcsec), 16 Jy at -64 \kms\ (Breen et al. 
2010b).  The site  is assumed to be the `lost' OH site 
326.77-0.26.  No methanol maser has been detected here.

\subparagraph{327.120+0.511 }  
Slowly varying, with little change from 2004 to 2005, and still generally 
similar to 1982 and 1989, but with intensity changes greater than 
factors of 2.   Linear polarization of more than 50 per cent is present 
at 1665 MHz in 3 features, and is accompanied also at 1667 MHz by nearly 
50 per cent polarization in one of them, at -84.80 \kms\ (see 
Sections 6.1 and 6.3).   

\subparagraph{327.291-0.578 } 
Highly variable by factors of more than 5 between successive 
observations 1976 and 1978 (Caswell et al. 1980), 1993 and subsequently, 
and with large velocity range from -72.5 to -37.5 \kms.  
Emission near the extremities of the velocity range are currently 
seen at both 1667 MHz and 1665 MHz.  
Accompanying methanol maser emission over the range -49 to -36 \kms\ 
indicates a systemic velocity near -42.5 \kms. 
Thus the wide OH range arises predominantly from highly blue-shifted 
emission.

\subparagraph{327.402+0.444 }  
Highly variable  1978 through 1989 to 2004 and 2005, but still with  
features over the range -86 to -73 \kms.  
2004 and 2005 data concur on the high linear polarization, at both 
1667 and 1665 MHz, for two features;  notably 
at -82.5 \kms, where a 0.5-Jy feature at both 1665 and 1667 MHz 
is more than 50 per cent linearly polarized (see Sections 6.1 and 6.3).

\subparagraph{328.237-0.547 and 328.254-0.532 }  
A single spectrum is used to display emission from both sites which are 
separated by only 80 arcsec.   The main features remain recognisable 
from 1978 to our 2004 and 2005 measurements, but with amplitude 
variability, both decreases and decreases, greater than factors of two 
or three.  
Current emission extends beyond the velocity range of ATCA measurements 
(-54 to -30 \kms); the  total range is now -58 to -23 \kms.  It is 
difficult to distinguish which features arise from each site, 
thus our estimated velocity ranges in the Table are largely guided by 
previous ATCA data.  
A new 1667-MHz feature, LHCP 1.3-Jy at -56 \kms\ was present 2005 (see 
displayed spectrum) but not 2004.  

New emission at 1667 MHz LHCP, 0.8 Jy from -27 to -23 \kms, and 
1665-MHz at -29 \kms\ (detected 2004 but not 2005), is most 
likely from 328.237-0.547.

\subparagraph{328.307+0.430 }  
1665-MHz emission remains very similar to 1978 and 1989, and the 
detection of 1667-MHz emission  is confirmed at the better 
sensitivity now achieved.  

\subparagraph{328.809+0.633 }  
Spectra at both 1665 and 1667 MHz remain similar to 1978 and 1989, with 
velocity range -49 to -33.5 \kms.  

\subparagraph{329.029-0.205, 329.029-0.200 and 329.031-0.198 }  
The total extent of this complex is less than 30 arcsec, so sites 
are distinguishable only with the ATCA (Caswell 1998) from which we  
identify as the  first source, 1665-MHz emission of 19 Jy LHCP at 
-38.5 \kms;  we use the ATCA data for the confused  
second source, and the third source which has apparently weakened.  
The location of 1667-MHz emission between velocities -31 and 
-28 \kms\ is not clear.   
Note that sidelobes from 329.183-0.314 are present beyond -50 \kms, and 
from 329.066-0.308 at -43.5 \kms.

\subparagraph{329.066-0.308 }  
At this target position, there is some confusion from both the previous 
and following sources,  but clearly located at 329.066-0.308 is 
the RHCP feature at -43 to -44 \kms, both 1665 
MHz (4 Jy) and 1667 MHz (1.5 Jy) showing weak linear polarization.

\subparagraph{329.183-0.314 }  
Recognisably similar to 1978 and 1989 at both 1665 and 1667 MHz 
in the range -55 to -47 \kms, as previously listed,  and also 
extending now to -58 \kms\ with a weak 0.5 Jy LHCP 1665-MHz  feature.  
There is confusion at -39 and -37 \kms\ from 329.029-0.205;  and 
at -43.5 \kms\ from 329.066-0.308, which is plotted alongside.

\subparagraph{329.339+0.148 }

At this  remarkable site, weak emissions from the excited states, 
13441 GHz, 6035 GHz and 6030 GHz, all show clear Zeeman patterns 
interpreted as positive magnetic fields (Caswell 2004b).  
The 1665 and 1667-MHz emissions studied here are comparably weak, 
with 1665-MHz spectra showing LHCP emission at -107.3 \kms\
and weaker RHCP emission at -104.0 \kms, and thus a Zeeman pattern 
similar to stronger emission seen at 1720 MHz (Caswell 2004a).

\subparagraph{329.405-0.459 }  
Generally similar to 1978 and 1989, but at 1665 MHz generally 
weaker (halved) and the 3.5-Jy feature at -75.5 \kms\ in 1978 is now 
absent, whereas at 1667 MHz now stronger, typically double.  
Linear polarization exceeds 30 per cent at 1665 MHz in one feature.

\subparagraph{330.878-0.367 }  
At 1665 MHz, the observed flux density over more than three decades has  
been the highest persistent emission of any source in the sky.  And in 
comparison  with W3(OH), which has comparable flux density at some 
epochs (Wright et al. 2004a, 2004b), we note that 330.878-0.367 is 
more distant than W3(OH), implying that its luminosity considerably 
surpasses W3(OH).    
Spectra at 1667 MHz remain very similar to the 1978 appearance, 
with most variations less than factors of two.  1665-MHz 
spectra have also stayed similar, but LHCP emission at -61.5 \kms\ has 
nearly doubled from 1978 to 2005.  Because of the 
high flux density, we show an additional spectrum at expanded 
scale to reveal the weaker features, clearly exhibiting a large velocity 
range extending from  -74.5 to -50 \kms.  There is modest linear 
polarization, consistent between the 2004 and 2005 measurements.  
Methanol emission at this location, with velocity -59.2 \kms, is quite 
weak, suggesting that this is an example of an evolved site where 
the OH may have progressed to its ultimate peak, 
and the methanol emission has begun to fade.  An additional weak 
methanol maser lies 30 arcsec away (Caswell et al. 2011c) and there 
is no known OH emission at that site.

\subparagraph{330.953-0.182 and 330.954-0.182 }  
Emission from this direction has been generally stable since 1978, 
with most variations less than a factor of two.  
The presence of two distinct sites (Caswell 1998) was recognised 
from ATCA data, but the current distinction between them is based 
on higher resolution LBA data (Caswell et al. 2010a), with 
330.954-0.182 encompassing most of the emission, whereas 330.953-0.182 
refers to a discrete weaker site to its south.

At the southern site 330.953-0.182, 1665-MHz features near the 
three velocities, -89.4 \kms\ (RHCP), -87.2 \kms\ (LHCP) and 
-88.2 \kms\ (strong at both R and LHCP) are located at 
$16^h09^m52.38^s,~-51^{\circ}54{\arcmin}57.3{\arcsec}$,
within 0.3 arcsec of associated methanol and 6035-MHz excited OH.  
The LBA data (limited to circular polarization) of the 
-88.2 \kms\ feature (with near equal R and LHCP of more than 5 Jy) 
was suggested as having linear polarization, and our spectrum 
indeed shows LINP of 70 per cent.  
Its velocity is essentially midway between the components of a 
Zeeman pair and we argue in Sections 6.1 and 6.2 that the 
combination is a Zeeman triplet.  

The centroid of other emission, the majority, to the north east, is 
at $16^h09^m52.60^s,~-51^{\circ}54{\arcmin}53.7{\arcsec}$,
%16 09 52.60 -51 54 53.7, 
i.e. 330.954-0.182 in rounded Galactic coordinates.  
A feature at 1665 MHz exceeding 4 Jy (4.65 Jy R and 12.58 Jy L) in 
the LBA observations and suggested as having elliptical polarization 
with significant elliptical fraction was near  velocity -87 \kms\ 
and indeed shows LINP of 4 Jy in our 2004/5 data.  The only other 
significant linear polarization in our 2004/5 observations is at 
-85.9 \kms, where there are multiple confusing features, and not 
recognised as likely elliptical in the LBA data.  1667-MHz emission 
at -85.4 \kms\ also shows 
significant linear polarization in 2004 and 2005.  

Note that the LBA lower sensitivity was unable to map the weak 
1667-MHz emission seen on our spectra from -91 to -86.8 \kms.

\subparagraph{331.132-0.244 }  
Note that the spectra include a 10 per cent sidelobe contribution 
from the previous source, and a 20 per cent sidelobe from  
331.278-0.188 (notably 1665-MHz LHCP at -89.5 \kms\ and 1667 RHCP at 
-88.8 \kms). 

The main feature RHCP 1665-MHz at -88.8 \kms\ has persisted from 
1978 to 2004 and 2005. 1667-MHz emission has shown a flare of LHCP 
at -92.8 \kms\ from 0.5 Jy (1978)  to 2.3 and 3.4 Jy (2004, 2005).

\subparagraph{331.278-0.188 }  

Generally similar to 1978 except: in 1990, a feature at -86.8 \kms\ 
flared at RHCP 1665-MHz from 5 to 26 Jy, and is now less than 5 Jy.   
Even more dramatic is the recent flare of a 1665-MHz LHCP feature 
at -89.5 \kms\ from below 2 Jy 
(1978) to 9 Jy (1990), to 75 Jy (2004), and 100 Jy (2005);  at 1667 MHz 
there has been a doubling to 5 Jy of RHCP emission at -89 \kms.
  
Note that the 5-Jy RHCP 1665-MHz feature at -88.5 \kms\ is a 
sidelobe of 331.132-0.244.

\subparagraph{331.342-0.346 }  
Spectra at 1665 and 1667 MHz remain recognisably similar to epoch 1978.  
High linear polarization of 90 per cent is present at the strongest 
1667-MHz feature;  prominent linear polarization is also present at 
the strongest 1665-MHz feature which, notably, is at the same 
velocity, $-66.95$ \kms\  (see Sections 6.1 and 6.3).

\subparagraph{331.442-0.186  }  

Most emission seen in the spectrum at this site is from 
sidelobes of 331.512-0.103 (offset 6.5 arcmin), with weaker 
contribution at 1665 MHz from 331.542-0.066 (offset 9.4 arcmin) plus a 
strong LHCP spike at -89.5 \kms\ from 331.278-0.188 (offset 10 arcmin).  
A weak 1665-MHz feature of 1 Jy from 331.442-0.186 at 
$16^h12^m12.41^s$, $-51^{\circ}35{\arcmin}09.5{\arcsec}$,
%RA 16 12 12.41, Dec -51 35 09.5 
with velocity -83 \kms\ was recognisable in 1994 (Caswell 1998), 
but has faded below 0.2 Jy on our spectra of 2004 and 2005.  
However, 331.442-0.186 is probably now seen as a 
weak 1665-MHz LHCP 0.8-Jy feature at -85.2 \kms.  
The spectrum aligned with 331.512-0.103 and 331.542-0.066 shows 
that this feature at -85.2 \kms\ has no strong counterpart 
in the other spectra (unlike, for example, the similar strength RHCP 
feature at -84 \kms\ which corresponds to a 4 Jy feature of 
331.542-0.066).  Thus the feature is not from 331.512-0.103 (where 1.6 
Jy would be expected), nor from 331.542-0.066 (where 4 Jy would be 
expected), and thus seems to be a genuine feature of 331.442-0.186.

\subparagraph{331.512-0.103 and 331.542-0.066 with close companion 
331.543-0.066 }  
The displayed spectrum is taken at the second pair of sites, 
$16^h12^m09.05^s,~-51^{\circ}25{\arcmin}47.2{\arcsec}$,
%16 12 09.05 -51 25 47.2, 
offset  3 arcmin from the first site 
(at $16^h12^m10.12^s,~-51^{\circ}28{\arcmin}37.7{\arcsec}$),
%(at 16 12 10.12, -51 28 37.7) 
and thus an intensity correction factor 
of 1.2 is needed for features from the first site.  

In more detail, Caswell (1997, 1998) remarks that 331.512-0.103, may  
be double, since it has associated 6035-MHz emission 
that appears to have a companion offset by 1.7 arcsec (distinguished by 
the names 331.511-0.102 and 331.512-0.103);    
however,  here we discuss it as a single site with name 331.512-0.103.   
This site accounts for nearly all of the observed emission seen in our 
spectrum in the range -94 to -85 \kms.  

Caswell (1998) lists both 331.542-0.066 and the weaker source  
331.543-0.066,  offset 3 arsec, north and at later RA.    
At 331.542-0.066, there is strong continuum, 191 mJy, with weak 
methanol  and 6035-MHz masers; the second, weaker, OH source  
331.543-0.066 has stronger accompanying methanol, but no 
continuum or 6035-MHz 
emission, suggesting a  clear physical distinction between these 
two sites, with perhaps the second site younger.  From the present 
confused single-dish spectrum, the isolated RHCP feature of 4 Jy  
at -84 \kms, and probably the 10-Jy feature at -86.1 \kms, and 
1-Jy feature at -82 \kms, are recognisably from the 
very close pair of sites, most likely from 331.542-0.066.  We can place 
only an upper limit to emission from 331.543-0.066.

Despite very little change between 2004 and 2005, overall, we note that 
the spectra at these sites are amongst the most variable recorded since 
observations in 1978 (Caswell et al. 1980):  most features are barely 
recognisable as a result of these changes (some increases, some 
decreases) commonly by factors of 4.   For example, 1665-MHz 
RHCP at -92.8 \kms\ has decreased 
from 43 to 15 Jy, and LHCP at -88.3 \kms\ has increased from 18 Jy to 80 
Jy and RHCP at   -87 \kms\ from 10 to 35 Jy.  At 1667 MHz, the RHCP and 
LHCP features near -93 \kms\ have decreased from 32 and 30 Jy to less 
than 1 Jy.  
We note that the weak (0.5-Jy) emission seen near 
-101 to -100 \kms\ in 1978, although probably from the 331.512-0.103 
site, is now barely seen, and not included in our estimated velocity 
range.

\subparagraph{331.556-0.121 }   
Weak emission is confined to the velocity range -103 to -96 \kms, 
as first reported (Caswell 1998), with peak emission of nearly 1 Jy 
at -100 \kms\ in 1994.  
Note that there is a sidelobe response to the AGB star 331.594-0.135 
(offset nearly 2.5 arcmin, 1994 peak of 4 Jy from -108 to -106 \kms, 
and weaker from -86 to -82 \kms) as reported by Caswell (1998).

\subparagraph{332.295+2.280  }  
The OH emission was discovered in 1990 (te Lintel Hekkert \& Chapman 
1996) in a search towards possible AGB stars.  It was further studied by 
Caswell (1998) where it was conclusively shown that the emission 
coincided with a methanol counterpart and thus associated with a 
high mass star formation region (or YSO). The small 
velocity range of methanol emission (Caswell et al. 2011c)  
indicates a likely systemic velocity of -23.5 \kms.  
The ATCA OH data confirm this 
single site as the source of all the OH features over a wide velocity 
range as displayed from ATCA spectra (Caswell 1998).  
1667-MHz features include emission of 0.4 Jy 
at -40 \kms\ and 0.3 Jy from -3 to +6 \kms, as well as the 
stronger emission, -31 to -15 \kms, matching the 1665-MHz emission, 
and symmetric about the likely systemic velocity.  
The 1667-MHz  outlying features 
(displayed for 2004) are corroborated by our  2005 spectra and 
represent OH outflows highly redshifted (as large as 29.5 \kms), 
as well as blue-shifted (as large as 18 \kms).  Outflow 
sources are discussed further in Section 4.3.    

The earliest known spectra from 1990 (te Lintel Hekkert \& Chapman 1996), 
despite lower spectral resolution of 0.9 \kms, are very similar to the 
present ones, and there are no significant changes from 2004 and 2005.  

Our new spectra are the first to reveal high linear polarization, 
especially at 1667 MHz, -24.8 \kms.

\subparagraph{332.352-0.117 }  
From 2004 to 2005 we note the decay of a strong 1.7-Jy feature at -43.5 
\kms\ to 0.4 Jy, and of a weak (0.3-Jy) -52-\kms\ feature 
(not seen 2005), but an increase of a broad feature 0.8 to 1.3 Jy.

\subparagraph{332.726-0.621 }  
Emission is stronger at 1667 than 1665 MHz, and remains similar to 
epoch 1978.  
Note that the velocity range for this source is -50 to -44 \kms; 
outside this range, the 1665-MHz features arise from the 
following source.

\subparagraph{332.824-0.548 }  
Apparent emission at 1667 MHz is wholly from the previous source.  The 
velocity range of 1665-MHz emission is -59 to -52,5 \kms.  
Note that the OH maser is offset 7 arcsec from any methanol maser 
(see Caswell 1998, Caswell et al, 2011c), so this is an OH maser 
site without an accompanying methanol maser.

\subparagraph{333.135-0.431 }  
Strong OH emission has remained remarkably similar to 1978, apart 
from a doubling of the strongest 1667-MHz feature in 2004 and 2005.  
The spread of positions over nearly 3 arcsec, similar to that 
of excited-state OH maser emission, and approximately matching  
the extent of a strong compact \HII\ region (1.4 Jy, and 
approximately 4 arcsec:  Caswell 1997;  Guzman et al. 2012), also 
encompasses a methanol maser.  The continuum emission suggests 
that it is most likely to be a single extended site rather than 
two close sites (Caswell 1997, 1998).

\subparagraph{333.234-0.060 }  
The 1665-MHz spectra retain a general resemblance to 1978 and 
1990 spectra, but with a new maser feature in 2004 and 2005 of 
LHCP at -87 \kms.  A new 
RHCP feature at -95.6 \kms, 1.2-Jy, was present only in 2004 
(see displayed figure); it disappeared in 2005.  
Most emission coincides with methanol maser 333.234-0.060, but 
some may arise from the nearby methanol maser site 333.234-0.062 
(with velocity range -92.5 to -80 \kms).  
The 1667-MHz spectra show a single weak emission feature, and are 
dominated by broad absorption -96 to -83 \kms.  

\subparagraph{333.315+0.105 } 
The strongest feature, 1665 MHz 1 Jy at -46 \kms, shows strong 
linear polarization, comparable to RHCP emission.  Weak 
1667-MHz emission is present  at -47.5 and -46.9 \kms\ (in 2005 
as well as the displayed spectrum 2004).

\subparagraph{333.387+0.032 }  
Emission is present only at 1665 MHz, with peak at -73.5 \kms\ 
(displaying near 100 per cent linear polarization), which extends 
to -70.5 \kms.  Note that apparent emission near -85 \kms\ is from 
333.234-0.060 (plotted alongside).

\subparagraph{333.466-0.164 }  
Note that emission -58 to -46 \kms\ arises from 333.608-0.215 (as 
is clear from the aligned plot beneath), 1665-MHz emission is still 
similar to epoch 1978, and 1667-MHz emission is seen clearly for the 
first time, with a 1-Jy LHCP feature near -38.1 \kms.

\subparagraph{333.608-0.215 }  
Features lie between  -58 and -46 \kms\  at both 1665 and 1667 MHz; 
they mostly resemble emission seen in 1978.

\subparagraph{335.060-0.427 }  
Several features are present at 1665 and 1667 MHz and, notably, at 
-42.7 \kms\ there is more than 50 per cent linear polarization 
of matching 1665 and 1667 MHz emission (see Sections 6.1 and 6.3).

\subparagraph{335.556-0.307  }  
Weak emission is seen at 1665 MHz, and stronger emission at 1667 MHz 
is now observed for the first time.

\subparagraph{335.585-0.285 and 335.585-0.289 }  
For this pair of sources (separated by 15 arcsec), the velocity 
range is wide.   

ATCA data from 1996 (Caswell 1998) for 335.585-0.289 show at 1665 MHz 
a weak feature at -58.5 \kms\ and stronger features at -53.5 and -50 
\kms, and a velocity range -60 to -49 \kms.  In  this velocity 
range, we note from our 2004 and 2005 data that the stronger 
features are persistent, with weak, but repeatable, linear 
polarization.   Weak 1667-MHz LHCP  
emission is also now seen at -50.2 \kms.  

The ATCA data for 335.585-0.285 showed the main 1665-MHz feature near  
-48.0 \kms\ and a weak feature at -40 \kms; although the weak feature  
is now absent, at similar velocity we do see emission at 1667-MHz. We 
also see 1667-MHz emission matching the 1665-MHz emission near -48 \kms, 
and note that the feature has very high  linear polarization  
at both transitions (see Sections 6.1 and 6.3).  

At 1667 MHz there is an additional weak feature with high linear 
polarization at -49.1 \kms.  Although we cannot be sure that it is not 
a feature of 335.585-0.285, we tentatively attribute it to 
335.585-0.289, implying a current velocity range -56 to -49 \kms.  
We note that the velocities of methanol emission from the two sources 
slightly overlap, with ranges  -51 to -43 \kms\ and -56 to -50 \kms.  
An additional  methanol maser site 3 arcsec further south at 
335.585-0.290 (velocity range -48 to -45 \kms) has  no reported OH 
maser emission.  

Variability has been high at both OH maser sites since 1978 (listed 
by Caswell Haynes \& Goss (1980) as 335.61-0.31 and at that time showing 
a single RHCP 1665-MHz feature of nearly 3 Jy at -53.5 \kms), through 
1989,  1993, to 2004, 2005, with the persistent feature of 
335.585-0.289 showing both increases and decreases by factors of 2. 
In contrast, 335.585-0.285 increased from a non-detection in 1976;  
the 1665-MHz feature at -48 \kms\  flared from 
2.3 Jy (ATCA in 1996) to its current total intensity of 9 Jy total 
intensity (LHCP and RHCP both approximately 4.5 Jy).  We recall that 
this flaring feature has high linear polarization, and is matched by a 
similar high linear polarization 1667-MHz feature at the same 
velocity, and thus both interpreted as isolated Zeeman $\pi$ components 
(Sections 6.1 and 6.3).

\subparagraph{335.789+0.174  }  
Emission remains similar to 1978, but most features at 1667 and 
1665 MHz are now 3 times stronger, and 1665-MHz emission shows  
a wider velocity range, from -54.5 to -47.5 \kms.  
Peaks at 1665 and 1667 MHz are at different velocity, but both 
exhibit strong linear polarization, 1667 MHz 70 per cent, and 
1665 MHz more than 80 per cent.

\subparagraph{336.018-0.827 }  
Isolated, similar at 1665 and 1667 MHz, and with no earlier spectra 
of high quality.

\subparagraph{336.358-0.137 }  
Isolated, with weak 1665-MHz emission remaining similar since 1978 
(and 1988), and weaker 1667-MHz emission now seen.

\subparagraph{336.822+0.028 and 336.864+0.005  }  
The sources are separated by 2 arcmin.  Broad absorption prevents 
any confident detection of 1667-MHz emission.  At 1665 MHz, 
features between -79 and  -74 arise from 
336.822+0.028 (cf.  methanol  -78 to -76.5 \kms); and in the range   
-90 to -86 \kms\ from 336.864+0.005 (cf. methanol  -82.5 to -74 \kms).  
Features at -81 and -84 \kms\ were absent from ATCA spectra (Caswell 
1998) at these positions but seem more likely to be at 336.864+0.005.  
 
%or, perhaps,  a new OH source at methanol  336.809+0.119 (-86 
%to -80 \kms), 2 arcmin north, and RA smaller by 2 and 4 arcmin 
%from 336.822+0.028 and 336.864+0.005, respectively.

\subparagraph{336.941-0.156 and 336.984-0.183  }  
Weak, separated 2 arcmin, and features distinguishable only from 
ATCA information, which suggests a clear separation in velocity: 
336.941-0.156 from -71 to -65 \kms, and 336.984-0.183 from 
-82 to -76 \kms.  
Both have weak 1667-MHz emission, and both have matching methanol 
masers.

\subparagraph{336.994-0.027 }  
1665 and 1667-MHz features span 10 \kms, and straddle deep 
absorption.  Weak linear polarization seen from the displayed 
2005 Q and U spectra was corroborated in 2004.

\subparagraph{337.258-0.101  }  
In an absorption dip, this maser is quite weak, with 1667 and 1665-MHz 
emission near -70 \kms, and with additional weaker 1665-MHz features 
extending from -64 to nearly -55 \kms.   

%whereas the velocity range noted by Caswell (1998) with 
%the ATCA was only -72 to -68 
%\kms.    We note methanol not only at 337.258-0.101 with velocity 
%similar to the OH main features, but also at 337.388-0.210 (offset 10 
%arcmin) with emission peak at -56 \kms\ (range -67.5 to -52 \kms), which 
%may be the location of the OH emission in the vrange -64 to -55 \kms\ (a 
%position where the OH Parkes primary beam correction would be 6).  

\subparagraph{337.405-0.402 }  
The 1665-MHz peak now exceeds 130 Jy (66 Jy 1978, 100 Jy 1989, 120 Jy 
1993). A 1665-MHz LHCP feature at -39.9 \kms\ now has a peak of 
80 Jy, but was less than 5 Jy in 1978 (Caswell, Haynes \& Goss 1980).  

The current OH velocity range is -58.5 to -33 \kms\ whereas  
the mid-range of methanol emission (the likely systemic velocity) 
is  near -38 \kms;  thus there are highly blue-shifted OH features, 
(especially prominent at 1667 MHz), first highlighted several years 
ago  (Caswell 2007).  
There is linear polarization of more than 50 per cent at both 1667 
and 1665 MHz near -36 \kms,  but notably not at exactly the same 
velocities.
This ground-state OH  site displays several other interesting 
properties.  Although methanol emission is very close in position 
and almost certainly shares the same source of excitation, the 
nearby excited-state OH emission is offset 
3 arcsec to the south (Caswell 2001, 2003) and straddles a strong 
compact \HII\ region.  Further study of the \HII\ region by Guzman  
et al. (2012) confirms its position, and indicates a spectrum 
increasing with frequency to 139.5 mJy at 8.6 GHz, indicative of 
being ultra- or hyper-compact, or perhaps a thermal jet at a 
high mass YSO.

\subparagraph{337.613-0.060 }  
Most features in the displayed spectrum are from 337.705-0.053, 
the next source, only 4.5 arcmin away (and thus attenuated by  
0.7 at this offset position), and displayed beneath this one.  
Unique at the target location 337.613-0.060 is weak ($< 1$-Jy peak) 
1665-MHz emission between -45 and -39 \kms.

\subparagraph{337.705-0.053 }  
This site has previously been observed with high spatial resolution 
and good spectral resolution with the LBA (Caswell, Kramer \& Reynolds 
2011b).  The spectra remain similar, and the apparent multiple 
Zeeman pairs are confirmed by the LBA spatial coincidences.  
Overall, the spectra remain similar to 1978.  
There is no pronounced linear polarization, demonstrated for 
the first time from the present spectra, and consistent with the 
absence from the LBA of any strong emission of both RHCP and LHCP,  
coincident spatially and in velocity.  

1667-MHz weak emission, offset in velocity between  -61 to -57 \kms, 
has not previously been noted; it is weak, and in a region where 
the baseline is affected by broad absorption, but the reality of the 
features is in no doubt, since there is some circular polarization, 
the same in both  2004 and 2005.   Its location is uncertain, but 
is most likely a weak `outflow' from 337.705-0.053.

\subparagraph{337.916-0.477 and 337.920-0.456 }  

The secondary  source, 337.920-0.456, offset by 75 arcsec, 
is a single 5-Jy LHCP 1665-MHz feature near -39 \kms, with 
methanol counterpart at  -38.5 \kms. 

The main source (337.916-0.477) has prominent features over the wide 
velocity  range -56 to -34 \kms\ at 1665 MHz; weak 0.5-Jy features (at 
both 2004 and 2005 epochs) extend this range at 1665 MHz RHCP to 
-63.5 \kms,  and at 1667 MHz  to -31 \kms, a total span of 32.5 \kms.  
There is no accompanying methanol maser, but if the systemic 
velocity is near -39 \kms\ (similar 
to 337.920-0.456, as seems likely), it would imply the most negative 
velocity emission is a blue-shifted outflow (see Section 4.3).    

The current peak of LHC 1665-MHz emission, at -51 \kms, in the 
suggested flaring blue-shifted outflow, is 43 Jy, but 
was less than 6 Jy in 1978.
Moderate linear polarization is present, chiefly in the 
blue-shifted outflow.

\subparagraph{337.997+0.136 }  
1667-MHz emission is slightly stronger than 1665 MHz,  but all 
features have been quite variable. 
The velocity span is -41 to -32.5 \kms, with weak features outside 
this range attributed to 338.075+0.012, the next source, offset   
8 armin, and displayed in the aligned spectra beneath.  
Linear polarization is more than 50 per cent in the 1665-MHz 
feature at -33.8 \kms.

\subparagraph{338.075+0.012  }  
The velocity span is  -54 to -43 \kms\ (disregarding emission 
from the previous source).  Emission is present at both 1665 MHz 
and, weaker, at 1667 MHz.

\subparagraph{338.280+0.542 }  
1665-MHz emission flared from 4 Jy in 1978 to 10 Jy in 1989 and has 
now fallen to 1.5 Jy. The  1665-MHz feature at -63 \kms\ is 100 per 
cent linearly polarized.  1667-MHz emission 
is currently seen as a single feature at -61.6 \kms\ with 60 per cent 
linear polarization and 50 per cent RHCP.  Near this velocity, 
1665-MHz features also show high linear polarization (see Sections 
6.1 and 6.3).

\subparagraph{338.461-0.245 }  
Emission remains similar to 1978.  Modest linear polarization at 
1665 MHz is now seen.  

\subparagraph{338.472+0.289 }  
Spectra remain very similar to 1978 and  1989.  The 1665-MHz 
emission peak shows  60 per cent linear polarization, and at 1667 
MHz a pure RHCP  feature is present at -31.5 \kms.  Weak 1667-MHz 
emission is also seen at -41 \kms.    

\subparagraph{338.681-0.084 }  
1665-MHz emission is similar to 1978 and the strongest feature 
shows 70 per cent linear polarization.  
No 1667-MHz emission is seen, and there is no accompanying 
methanol maser.

\subparagraph{338.875-0.084 }  
More features are now evident than in 1978, and the strongest 
feature is now a RHCP feature at 1667 MHz.  
A weak 1665-MHz feature at -36.5 \kms\ has linear polarization 
exceeding 50 per cent.

\subparagraph{338.925+0.557 }
At this site, all four ground-state transitions have been seen 
(and are variable),  and accompanied by maser emission from the 
6035-MHz OH excited state, and methanol.  Especially notable 
is the 1665-MHz LHC peak at -63.5 \kms, which was 38 Jy in 
1978 at -63.5 \kms\ but now near 4 Jy, whereas the current peak 
of 20 Jy near -61.5 \kms\ has shown little change.    
Some 1665-MHz features show  linear polarization over 50 per cent.

\subparagraph{339.053-0.315 }  
Weak emission was first discovered near -110 \kms\ towards a methanol 
maser at -110 \kms, and our spectra suggest that it is a likely 
Zeeman pair.  A feature 
now seen offset 10 \kms\ at  -121 \kms\ is one of the narrowest 
ever recorded;  It is pure LHCP,  and accompanied by weaker  
emission at 1667 MHz, also pure LHCP.  

\subparagraph{339.282+0.136 }  
The highest 1665-MHz peak, at -71 \kms, decreased from 1.5 
to 0.7 Jy between 2004 and 2005.  
At 1667 MHz, the single feature, at -73.5 \kms, displays about 50 
per cent linear polarization.  At this same velocity there is a 
1665-MHz feature which, although weak, is clearly seen from the Q and 
U spectra to have with similar linear polarization (see Section 6.1 
and 6.3).

\subparagraph{339.622-0.121 }  
Considerable variability has occurred since 1978 (when the  
largest velocity range was seen), but the overall spectral 
shape is still readily recognisable at 1665 and 1667 MHz.  
1720-MHz emission is present, but note that nearby 1612-MHz 
emission is clearly offset by 20 arcsec (Caswell 1999).  
Linear polarization at both 1665 MHz (at -35 \kms) and 1667 MHz is 
prominent,  with some features exceeding 70 per cent, especially 
at 1667 MHz. The high linear polarization of the 
strongest 1667-MHz feature at -36.3 \kms\ is matched by significant 
linear polarization at 1665 MHz (see Sections 6.1 and 6.3), although 
the coincidence is somewhat confused in the complex 1665-MHz 
spectra.

\subparagraph{339.682-1.207 }  
Note that there is a clear separation, in space and velocity,  
from the stronger, better-known, companion (339.884-1.259, the 
next source, seen as a 5 per cent sidelobe at velocities more 
negative than -33 \kms). 339.682-1.207 is now weaker than in 
1982,  but 1667-MHz emission has not changed as much.

\subparagraph{339.884-1.259 }  
The cited velocity range from -40 to -26 \kms\ is taken from 
1980 sensitive observations displayed by Caswell \& Haynes (1983) 
and subsequent ATCA observations.  
The prominent features are now stronger than in 1980 or 1989.  
The 1665-MHz LHCP peak at -36.1 \kms\ of 110 Jy in 2004 was less 
than 4 Jy in 1980;  however, the RHCP peak of 30.5 Jy at -29.2 \kms\ 
in 1980 dropped to 6 Jy in 2005.  
Similarly extreme changes occurred at 1667 MHz.  Some remarkable 
changes occurred even between 2004 to 2005, 
demonstrated by our plots displayed for both these epochs.  
Several of the strongest peaks at both transitions 
approximately halved, whereas other features flared, for example, 
the 1665-MHz feature at -33.7 \kms\ doubled from 7 to 14 Jy, 
with more than 50 per cent linear polarization.

\subparagraph{340.054-0.244 }  
There are similarities to spectra from 1978 and 1980, but some big 
changes at both 1665 and 1667 MHz, with the strongest features  now 
three times stronger than at early epochs.  

2004 and 2005 spectra show stability of linear polarization at both 
1665 and 1667 MHz and, in particular, for the strongest feature at 
1665 MHz, indicate more than 50 per cent linear polarization with 
stable ppa.

\subparagraph{340.785-0.096 }  
Spectra remain very similar to 1993 and fairly similar to 1980, with 
wide velocity range at both 1665 and 1667 MHz from -108.5 to -88.5 \kms, 
comparable to matching maser emission of methanol and 6035-MHz 
excited OH.

\subparagraph{341.218-0.212 }  
There is excellent spectral agreement between 2004, 2005, and generally 
remarkable similarity to epochs 1980 and 1989 at both 1665 and 1667 MHz.

\subparagraph{341.276+0.062 }  
The current 1665-MHz spectrum is similar to 1980, and weak 
1667-MHz is now detectable.

\subparagraph{343.127-0.063 }  
Still strong as in 1980, with LHCP at 1665 MHz approximately 100 Jy, 
and generally unchanged at both 1665 and 1667 MHz, 
We note that there is no known associated 6.6-GHz methanol maser. 
However, there is coincident continuum emission which is very compact 
at 25 GHz, with flux density 16 mJy, and rising at higher frequencies, 
and thus with characteristics of a uc\HII\ region, or perhaps a 
thermal jet (Brooks et al. 2007)

\subparagraph{343.930+0.125 }  
The strong 1665-MHz LHCP feature at +12 \kms\ of 1980 doubled to 
8 Jy in 1989, but is now an order of magnitude weaker.  A number of 
weak features span the range +8.5 to +17.5 \kms.  
There is no significant 1667-MHz emission, but both transitions 
show prominent absorption centred near +5.8 \kms.  

\subparagraph{344.227-0.569 }  
1665 and 1667-MHz  major features remain similar in strength to 1980, 
and emission is now seen to extend over the wider velocity range 
-32 to -18 \kms\ (with absorption in between) for both transitions.  
The strongest RHCP 1665 and 1667-MHz peaks increased more than 50 
per cent from 2004 to 2005.

The systemic velocity of this source seems likely to be near -20 \kms, 
based on the methanol maser peak here, and weaker methanol emission 
extending to both red and blue shifts (Caswell et al. 2011c).  If so, 
then the dominant OH emission  near -31 \kms, which is persistent 
yet variable, and stronger at 1667 MHz, seems likely to be strong 
blue-shifted emission, with only weak features present near -20 \kms\ 
at the systemic velocity (and stronger at 1665 than 1667 MHz as is most 
common for emission near the systemic velocity). The closely 
associated water maser emission, 344.228-0.569, shows a group 
of features from -28 to -8 \kms, also suggestive of the systemic 
velocity being near -18 \kms, and an additional feature at 
-51 \kms\ (Breen et al. 2010b) would then be  blue-shifted.  
Apparently in the same stellar cluster, offset by about 30 arcsec, 
is water maser 344.226-0.576 which shows a single weak feature 
at -19 \kms, further corroborating a systemic 
velocity near -20 \kms\ for objects near here.   
Thus although the OH maser is 
not one of the largest total velocity spreads, the above evidence 
indicates a distinctive  blue-shifted outflow, which shows flaring 
activity,  and shows significant linear polarization at 1667 MHz in 
the -31.2-\kms\ feature, and also in a nearby weaker 1665-MHz feature.  

The methanol spectrum likewise shows a strong asymmetry, with much 
stronger emission at blue-shifted rather than red-shifted velocity.

\subparagraph{344.419+0.044 and 344.421+0.045 }  
The spectrum (displayed for 2004) towards 344.419+0.044 is weaker 
than in 1980 or 1989, but the current improved sensitivity again 
shows dominant LHCP of the main 1665-MHz feature at $-65$ \kms, and 
similar in 2005.  The 1667-MHz spectrum shows deep absorption matching 
1665-MHz absorption, and a weak emission feature at -62.1 \kms.  
Methanol emission is also present at 344.419+0.044 from -65.5 to 
-63.0 \kms.  
There is a further, quite distinct, stronger, methanol maser at 
344.421+0.045 
($17^h02^m08.77^s,~-41^{\circ}46{\arcmin}58.5{\arcsec}$)
%(17 02 08.77, -41 46 58.5) 
with features from -72.5 to 
-70.0 \kms.  We suggest that an OH 1665-MHz feature at -74 \kms\ 
newly detected on our 2004 spectrum is probably from this site 
(offset 10 arcsec from 344.419+0.044).  We accordingly list it 
tentatively as a new OH maser site, adopting the methanol coordinates.

\subparagraph{344.582-0.024 }  
At 344.582-0.024 there are multiple features which are quite strong 
at 1667 as well as 1665 MHz, generally remaining similar from 1980 
and 1989 to the present. The most notably varying feature 
is 1665-MHz LHCP at -6.6 \kms, increasing from 4 Jy 1980 (Caswell 
\& Haynes 1983), 12 Jy 1989 (Parkes archival data), 15 Jy 1993 
(Argon et al. 2000), and declining to 10 Jy 2004 and 2005. 
Many features show significant linear polarization.
The velocity span is from -13 \kms\ (at 1667 MHz) to +2 \kms.

\subparagraph{345.003-0.224 }  
At our two epochs, intensities have shown 25 per cent changes.  None 
the less, most features over the whole range -32.5 to -20.5 \kms\ 
are clearly recognisable from 1980, 1989, and from 1998 (Argon 
et al. 2000), but with somewhat 
larger changes over the long timespan.  In particular, near -31 \kms, 
the 2-Jy 1665-MHz feature from 1980 has halved, its 1667-MHz 
near-counterpart has doubled (with even larger peak in 1998).  Our 
linear polarization measurements are the first for this source and show 
more than 50 per cent linear polarization for the 1667-MHz emission 
near -31 \kms, but barely significant polarization otherwise.

\subparagraph{345.010+1.792  }  
Variations since 1980 exceed factors of two, for example at -31 
\kms\ (1665 MHz), the 1980 peak of 14 Jy changed to 36 Jy (1989),  
to 30 Jy (1991),  to 6 Jy in 2004 and 2005.  
The velocity range has been -31 to -14 \kms\ in the past, but 
is currently -24 to -14 \kms.

\subparagraph{345.407-0.952 }  
A weak source that is somewhat confused by absorption at both 
1665 and 1667 MHz.  It remains strongest (as in 1980 and 1989) 
at 1665 MHz, LHCP at $-17.8$ \kms.  
The 1667-MHz transition has weak but persistent (2005 and 2004) LHCP 
emission at -17.8 and RHCP at -16.8 \kms.

\subparagraph{345.437-0.074 }  
Total intensity spectra in 1995 (from the ATCA, displayed by Caswell 
1998) best reveal the quite wide velocity range and similar 
intensity at 1665 and 1667 MHz.  
The current polarization spectra show considerable differences 
between the transitions, and between epochs, with the displayed  
(2005) 1665-MHz spectrum flaring at $-25.3$ \kms\  from 0.5 (2004) 
to 1.1 Jy (chiefly LHCP but also with strong linear polarization).   

There has been no associated  methanol maser detection, but the site 
is, none the less, likely to be a SFR maser (rather than an AGB variety)  
since the site is very close to the Galactic plane where nearby OH masers  
have similar velocities, and an OH absorbing cloud is also seen at the 
same velocity.

\subparagraph{345.487+0.314 and 345.504+0.348 }  
345.487+0.314 is a quite distinct source offset by 2 arcmin from 
345.504+0.348, as seen from Argon et al. (2000);  their  observations 
reveal it as a solitary weak 1665-MHz LHCP peak near -22.9 \kms;  it 
is seen to be 0.5 Jy at -22.9 \kms\ in our data from both 2004 and
2005, similar to their 1993 measurement.
Note that it coincides with 6035-MHz emission (Caswell 2001,
2003), and is clearly offset by nearly 4 arcsec from methanol maser
345.487+0.314, despite the same value of their rounded Galactic
coordinates.

Apart from the weak LHCP 1665-MHz feature of 345.487+0.314, all other 
emission seen in our spectra arises from 345.504+0.348.  
It remains generally similar to 1980 at both 1665 and 1667 MHz, with 
multiple features over the wide velocity range from -24.5 to -8 \kms.  
Note that a 1667-MHz LHCP feature  near -23.1 
\kms\ was 1 Jy in the Argon et al. (2000) data of 1993 and below 0.5 Jy 
in 1980 (CH83), but has flared to 4.5 Jy in 2004 and 9.8 Jy in 2005.  

Linear polarization of approximately 50 per cent is present in the 
feature near -19.5 \kms\ at both 1665 and 1667 MHz (see Sections 
6.1 and 6.3).  

It is worth noting that HI self absorption (Green \& McClure-Griffiths 
2011) indicates that both sites probably lie just beyond 1 kpc, 
in the Carina-Sagittarius arm, contrary to the suggestion of 
Caswell et al. (2010a) that 345.504+0.348 might be at 10.8 kpc, in 
the far side of the 3-kpc ring.

\subparagraph{345.494+1.469 and 345.498+1.467  }  
The two sites, with separation 18 arcsec, 
are blended on our single dish spectra but are distinguishable 
from ATCA measurements made in 1996 (Caswell 1998), with total 
intensity spectra yielding the individual velocity ranges 
which we cite in Table 1. From these ATCA spectra (resolution 0.7 
\kms), we see for the first source, 345.494+1.469, main features 
near -21, -16.3 and -12.7 \kms;  our spectra from 2004 and 2005 
(displayed) now reveal the first and last features as predominantly 
RHCP, and the second one a more complex blend.  
The -21-\kms\ feature was 6 Jy in 1993 (Parkes archival) but now 
only 0.5 Jy.  There is no methanol at this first site, but there 
is coincident continuum emission (Caswell 1998; Guzman, Garay \& 
Brooks 2010) with flux density 12 mJy at 8.6 GHz and weaker at 
lower frequency;  it is likely to be an ultra- or hyper- compact 
\HII\ region, and also shows weak jet emission.  
1665-MHz emission attributed (from ATCA data) to 345.498+1.467 
spanned -16.5 to -13.5 \kms;  
for our spectra, this region is very much blended with  
345.494+1.469, but we suggest that the strongest current peak at 
-16.0 \kms\ is from 345.498+1.467, and is accompanied by 1667-MHz 
emission which is now seen to resemble a Zeeman pair.  This second 
site, where there is also methanol, at -15 to -13.2 \kms, has no 
detected continuum emission (Guzman et al. 2010).  

The region also possesses further interesting maser emission: 
a pair of 1720-MHz emitting sites (Caswell 2004a) are offset 
nearly 30 arcsec south-east from the 1665-MHz emission and \HII\ 
region, but straddle 
weak (7-mJy, flat spectrum) extended (5 arcsec) continuum emission 
that might represent an earlier outflow from the compact \HII\ 
region (Guzman et al. 2010).

\subparagraph{345.698-0.090 }  
Multiple features at both 1665 and 1667 MHz from -11 to -3 
\kms\ remain generally similar to those recorded in 1980, with  
comparable intensities at the two transitions.  An 
additional 1665-MHz weak feature at +5.7 \kms\ is seen in 2005 
but not 2004 and may be real; it lies at the edge of an absorption 
dip seen at both 1665 and 1667, at both epochs.   
This is an OH maser site with no detected methanol maser.

\subparagraph{346.481+0.132 and 346.480+0.221 }  
The principal source 346.481+0.132 is now weaker than in 1980 or 1989, 
but still dominated by a LHCP 1665-MHz feature at -8 \kms,  and with 
additional emission between -3 and  -1.5 \kms.  
The velocity range of associated methanol, -11.6 to -4.9 \kms\ 
matches this well.  

A feature at -17 \kms\ is also evident in 2004 and 2005 at 1665 MHz, 
with a weak 1667 counterpart.  
Possibly the OH emission at this velocity arises from the methanol maser  
site 346.480+0.221 (velocity range -21 to -14 \kms) which is offset  3' 
north and RA 22s smaller (ie 4.3 arcmin smaller), total offset 
5 arcmin, and we list it as such, using the methanol position, pending 
confirmation of the OH location.  

\subparagraph{347.628+0.148 }  
Strong at 1665 and 1667 MHz, with most features showing an increase  
from 1980 1989 1993 to 2004 (similar to  2005), and the strongest 
feature more than doubled.  
Linear polarization exceeds 50 per cent for 1665 MHz  at -95.8 \kms, 
and for both 1665 and 1667 MHz,  at -97 \kms\ (see Sections 6.1 and 6.3).

\subparagraph{347.870+0.014 }  
Emission (only at 1665 MHz) has remained almost identical from 1980 
through 1989 to the present.

\subparagraph{348.550-0.979 (and 348.579-0.920) }  
The first source has many features at 1665 MHz between -22 and 
-10 \kms\ (the feature at -8.5 \kms\ is a weak sidelobe from the 
following source cluster), and has remained quite stable from 
1980 through 1989 to the present.   1667-MHz emission features 
are very weak, but confirmed in the 2005 data. 
The source 348.579-0.920, discovered 1993 at an offset of 4 arcmin 
by Caswell (1998) at 1665 MHz as a 1.4-Jy feature at -27 \kms, 
is no longer detectable.  
Methanol masers are present at both sites.

\subparagraph{348.698-1.027, 348.703-1.043 and 348.727-1.037 }  
348.550-0.979 is 9 arcmin away from the targeted present cluster 
of sources, and is seen as weak features in the velocity ranges 
-21 to -19 and -15 to -12 \kms, as is evident from the aligned 
spectra.   Features in the present cluster (spread over 2 arcmin) 
are assigned to positions determined from ATCA measurements 
(Caswell 1998), specifically: a 0.8-Jy LHCP 1665-MHz feature at 
-17.0 \kms\ may be from 348.703-1.043;  a solitary 2-Jy RHCP 
1665-MHz feature at -15.4 \kms\ is from 348.698-1.027 (see also 
Argon et al. 2000), and is probably a long-term stable feature 
(reported as 348.73-1.06 by Caswell \& Haynes 1983).    
Emission of up to 3-Jy peak between -10.5 and -2 \kms\ is from 
348.727-1.037, and was not detectable (below 0.5 Jy) in 1980 
(Caswell \& Haynes 1983), but had increased to 1 Jy in 1988 
(Parkes archival spectra);  it has weak 
accompanying emission from the 1667-MHz transition.  
Methanol masers are present at 348.703-1.043 and 348.727-1.037, 
but none at 348.698-1.027.

\subparagraph{348.884+0.096 }  
The currently seen features at 1665 MHz are more than three times 
stronger than in 1980 or 1989, but still dominated by a RHCP feature 
near -73 \kms, and the 1667-MHz feature is also stronger (seen 
for the first time using the ATCA in 1995).

\subparagraph{348.892-0.180 }   
There are multiple weak 1665-MHz features from -0.5 to +14 \kms, 
with appearance similar to 1980 and 1989 except for the new 
LHCP 1.2-Jy feature near 0 \kms.  Linear polarization exceeds 
50 per cent at +7.8 \kms.  Weak 1667-MHz emission near +4.5 \kms\ 
has persisted since the earliest discovery.

\subparagraph{349.067-0.017 }  
The same single 1665-MHz feature remains, as seen 1980 and 1989, 
predominantly RHCP.  High sensitivity in 2004 and 2005 reveals weak 
1667-MHz emission at +9.0 and +14.0 \kms\ at both epochs.

\subparagraph{349.092+0.106 }  
The bulk of the emission features, both 1665 and 1667 MHz, lie 
between -86 and -79 \kms.  The strongest feature in 2004 and 
2005, at both transitions, LHCP at -79 \kms,  was absent in 1980;  
an  outlier near -90 \kms\ seen in 1980 has weakened below our 
detection limit but a new outlier is present near -73 \kms,  
so the velocity range is now quite large, 18 \kms.

\section[]{Statistics from the combined data}

The data from this Paper II can be combined with data from Paper I to 
more than double the sample size.  We note that statistics within the 
two sub-samples are likely to differ in some respects because there is a 
larger proportion of weak sources in the present sample (partly a 
consequence of more thorough surveys, of higher sensitivity, in the 
southern sky).  For example, 12 per cent of sources in Paper 
I had peak amplitudes below 1 Jy, compared to 23 per cent for Paper II.   
Consequences include impacts on variability statistics, where past 
variability of the sources is not as well documented for weak sources;  
and on the fraction of sources showing linear polarization, which is 
smaller since it is limited for 
weak sources by the lower signal-to-noise ratio.

\subsection[]{Relative intensity  of emission at 1665 and 1667 MHz}

Among OH masers of the SFR variety studied here, the ratio of 1665 to 
1667-MHz intensity (peak or integrated) has long been known to favour 
1665-MHz emission (but with a small proportion 
of counter-examples).  Caswell \& 
Haynes (1987a) noted that, in their sample of 120 OH masers, only 13 
sources had peak intensity greater at 1667 MHz than at the 1665-MHz 
transition.  Our new measurements now yield improved quantitative 
statistics;  using the combined data of Paper I and the present 
Paper II, we find from our sample of 257 sources that for 85.5 
per cent (220 sources) the strongest feature is at 1665 MHz, and for 
14.5 per cent (37 sources) the strongest feature is at 1667 MHz.  This 
is now the best estimate of this statistic in view of our large sample 
observed with comparable sensitivity at both transitions.  
We also determined the ratio of peak intensity at 1665 MHz to 
that at 1667 MHz for each source;  we find that this ratio has a 
median value of 4.0.   The value of this ratio is a yardstick for 
recognising interesting departures that may reveal systematic 
differences in the physical conditions in different sources.  
The strong source W3(OH) has been the subject of many detailed studies,  
and the data of Wright et al. (2004b) are representative, showing an 
intensity ratio for the strongest peak at each transition (1665 to 
1667-MHz) of 12.  This is larger than we find for a typical source, but 
not exceptional, since we find 18 per cent of our source sample have 
more extreme ratios than 12.   
% in similar fashion to our result for an ensemble of sources, 
Wright et al. (2004b) also report that most of the ensemble of features 
in this one source are stronger at 1665 MHz, and for the summed flux 
density of all features, find a ratio of 15 (1665 to 1667-MHz), similar 
to that of the strongest peak.   However, they also investigated 
extensively (their section 3.1) a small subset of maser spots where 
the 1665- and 1667-MHz emission appears to co-propagate, i.e. 
the excitation of both transitions is in the same volume of gas.  
They were surprised to note that, at the rare locations where 
the 1665 and 1667-MHz transitions co-propagate, the summed 
1667-MHz intensity slightly surpasses the 1665-MHz intensity, 
and their data show a typical ratio for individual peaks close 
to 1.  We return to this issue in Section 6.3.

\subsection[]{Comparisons with masers of water and methanol, and 
with uc\HII\ regions}

With regard to associations of OH with water, we refer to Paper I and 
the analysis of a search for water towards a larger sample of OH 
masers (Breen, Caswell, Ellingsen \& Phillips 2010b).  The conclusion  
that 79 per cent of the OH masers have an associated water maser is 
expected to apply to the present sample since the majority of OH 
masers studied here are contained in these earlier water studies.  

We now concentrate on associations between OH and methanol.  
Previous statistics (for maser sites of the SFR or massive YSO 
variety) suggested that 80 per cent of OH  masers have an accompanying 
methanol maser (Caswell 1998).  Retaining the methodology of 
that study (as we also did in Paper I),  a simple comparison of 
methanol to OH is made in Table 1, listing the ratio:  peak 
methanol intensity to 
peak OH intensity.  Paper I showed that 87 of the 104 OH masers 
presented there had closely associated methanol masers, and 17 did not.  
In the present sample of 157 OH masers, 128 have methanol and 29 do not 
(essentially the same proportion) and thus the best statistics, from 
the combined sample of 261, show 215 (82.4 per cent) with, and 46 (17.6  
per cent) without associated methanol masers.  

We recall that our present source sample has been limited to 
maser sites believed to be associated with regions of young massive 
star formation.  Although many sources of the sample were discovered 
by extensive unbiased OH surveys, there are  others discovered  from 
targeted observations, notably some cases targeting known methanol 
masers.   
We might therefore expect when a full unbiased OH survey is conducted 
e.g. the planned GASKAP survey (Dickey et al. 2013), 
that the percentage without methanol maser emission will be larger than 
17.6 per cent, and that the additional OH masers without methanol would 
be more likely  associated with uc\HII\ (ultracompact \HII) 
regions (Caswell 1996, 
1997, 1998), in line with a previously noted trend.   The trend has 
been interpreted as evidence that maser sites where OH outshines the 
methanol are in the later stages of evolution.  
A likely progression (e.g. Breen et al. 2010a) is that the volume of gas 
with conditions favourable to molecular maser emission in the 
surroundings of a young high mass star increases with time, until a 
point during the development of a uc\HII\ region when  
rapid quenching of the maser emission begins.  It is suggested that 
maser emission excited by a massive YSO initially favours methanol, 
followed by OH main-line emission, and that the final quenching first 
affects the methanol, and subsequently the OH.  In this scenario, 
sources with only methanol 
emission represent the earliest phase;  sources with both prominent 
methanol and prominent OH represent a later phase; these then progress 
to sources with OH but very little  methanol (or none), at which stage 
there may be a clearly detectable 
uc\HII\ region.  Individual  exceptions to this simplistic 
pattern appear to be in a minority and we suggest that the 
46 OH sources with no detected methanol are prime targets for 
sensitive continuum measurements to see how well they fit this paradigm.  

In the source notes of 3.3, we remarked on some new continuum data 
that have become available, more clearly revealing uc\HII\ regions 
at several OH sites where methanol appears to be weaker than OH or 
absent;  specifically, 305.799-0.245. 317.429-0.561, 333.135-0.431, 
337.405-0.402, 343.127-0.063 and 345.494+1.469.																										

\subsection{Maser site velocity ranges and outflows}

Summarising the background information of Paper I, the velocity range of 
emission shown by masers in star formation regions can be a useful 
diagnostic in several ways.  Methanol masers most commonly have small 
velocity ranges, with a median value for the distribution of velocity 
range near 6 \kms\ (Caswell 2009), and the mid-range
velocity appears to be a good estimate of the systemic velocity.  
Water masers, in contrast, show a median velocity range of 15 \kms, 
with the significantly higher value partly accounted for by the common 
occurrence of features offset from the systemic velocity by more than 30 
\kms, interpreted as high velocity outflows. 

For OH masers studied in Paper I, the median velocity spread for a 
sample of 101 sources  was 8.3 \kms.  But a velocity spread 
exceeding 25 \kms\ was found for seven sources, whose detailed 
investigation suggested that there were two varieties, arising 
in quite different situations.   The 
first variety arise in a late stage of the maser evolution, with a 
general expansion over a wide angle, possibly driven by the enclosed 
\HII\ region; the archetype discussed in Paper I was the  well-studied 
evolved site 5.885-0.392, with 
a strong \HII\ region, displaying  OH outflows both blue- and redshifted. 
The second variety appear to be the result of collimated outflows 
similar to water masers, with indications of a preponderance of 
blue-shifted outflows;  the  possible OH archetype discussed in Paper 
I was 24.329+0.145, with 351.775-0.536 and 19.609-0.234 suggested 
as further examples.

The present study has shown eight additional examples of velocity 
spreads exceeding 25 \kms\ (see Table I, spectra, and notes 
of Section 3.3;  we omit 328.254-0.532 from consideration here because 
of possible confusion with 327.237-0.547).  Of these eight, the 
classification is not clear for candidates 301.136-0.226, 
311.643-0.380, and 332.295+0.280, whereas sources with 
blueshifted outflows that most likely resemble 24.329+0.145 are 
323.740-0.263, 327.291-0.578, 337.405-0.402, and 337.916-0.477.  
In particular, the blue-shifted features of 323.740-0.263 and 
337.916-0.477 are, at some epochs, stronger than emission near 
the systemic velocity, and the outflow from 337.405-0.402 is 
stronger at 1667 MHz than at 1665 MHz, in contrast to emission 
near the systemic velocity.  We also suggest that 314.320+0.112 
belongs to this group, with systemic velocity near -45 \kms\ 
where the strongest peak occurs for OH, methanol and water.  
Weak blue-shifted water emission extends to -70 \kms.  Weak methanol 
emission in the range -59 to -55 \kms\ also seems likely to be 
outflowing material, although such outflows seem rare amongst 
methanol masers.  The OH 1665-MHz spectrum in 1996 (Caswell 1998) 
showed blue-shifted  emission near -73 \kms\ stronger than 
at the systemic velocity.  It is no longer present, but our 2005 
spectra (displayed) show a broad weak plateau of blue-shifted 
emission strongest at 1667 MHz.

Our present extended study thus supports the 
argument for the existence of a distinct group of OH 
masers  with blue-shifted outflows.  They have been recognised from 
our statistics as sources with wide velocity range.  No doubt others  
occur (for example 309.918+0.123) but do not exceed our velocity 
range threshold, and it is generally more difficult to recognise these.  
However, two striking examples that we argue should be added to 
this class of blue-shifted outflows are 
12.908-0.260 (details in Paper I) and 344.227-0.569 
(see details in Section 3.3 of present paper).   In both instances, 
the velocity range has a modest value of 14 \kms, but is chiefly  
caused by the offset from the systemic velocity of 
dominant and variable blue-shifted features (which are strongest 
at 1667 MHz and, in the case of 12.908-0.260, have exceptionally 
high linear polarization).  

We also noted in Section 3.3 the rejection of an apparent red-shift 
outflow candidate (322.158+0.636), which seems  
likely to be a separate offset source with different systemic 
velocity; and uncertainty in the systemic velocity of 
another red-shift outflow candidate (300.504-0.176).  

% 340.78 remains as a possible red-shifted outflow

Recognition of blue-shifted outflows as a distinct characteristic 
began as a qualitative impression from perusing many spectra over 
many years.  Since the objects are quite rare, it was difficult to 
put this recognition on a rigorous quantitative 
statistical footing.  In the case of water masers, this has now been 
achieved (Caswell \& Breen 2010), with extensive evidence for an 
unusual class of water masers that show high velocity outflows 
favouring blue-shifts (Caswell 
\& Phillips 2008; Caswell \& Breen 2010; Caswell et al. 2011a;   
Motogi et al. 2011; 2013);  these primarily occur where there is 
neither OH emission nor readily detected uc\HII\ continuum emission, 
consistent with an interpretation that they are confined to an early 
evolutionary phase of massive star formation.  
However further work is needed on this matter, and similar 
investigations are required for the OH blue-shifted outflow sources.

\subsection{Variability of OH masers}

Details of variability are given in the source notes of Section 3.3.  
Among the 155 southern sites with observations in both 2004 and 2005 
presented here, only 285.263-0.050 and 323.740-0.263 show changes by a 
factor of two in their strongest feature, and two additional sources, 
318.050+0.087 and 339.884-1.259, show changes by nearly a factor of two.  
In the complementary northern region covered by Paper I, among the 99 
sources observed in both 2004 and 2005, five showed similar prominent  
variability.  Combining data from Papers I and II yields nine prominent 
variables among a sample of 254 sources, i.e. between three and 
four per cent.  Since this statistic indicates the variability of 
the strongest feature, it is a useful indicator that a single 
epoch survey is unlikely to miss 
more than a few per cent of sources that might be detectable in a 
similar survey conducted one year earlier or later.  

Our spectra also provide a valuable reference for longer term 
variability.  From the combined northern and southern regions, past 
variability can be investigated for at least 187 sites which have 
suitable archival data over several decades (notably those with earlier 
published Parkes spectra, identified with a `c' in the Refpol column of 
Table 1).   This is the largest source sample to date for 
variability studies over a long timebase.  
From observations over several decades, the sample has shown high 
variability (intensity changes by factors of 4 or more in some prominent 
features) for at least 10 per cent of the 187 masers.  Six examples  
were discussed in Paper I.  A further 12 examples from this 
paper are:  312.598+0.045, 316.811-0.057, 318.050+0.087, 318.948-0.196, 
320.232-0.284,  322.158+0.636,  327.291-0.578,  331.278-0.188,  
337.405-0.402,  337.916-0.477,  338.925+0.557, and  339.884-1.259 (see 
details  in the notes of Section 3.3).  
Although the majority of sites possess many spectral features that 
are persistently detectable over these several-decade-long intervals, 
they mostly show some variability, evident from the changing relative 
intensities of different features within a source.  Remarkably few 
of the sites (less than 10 per cent)  show no reliable evidence of 
variability.  Recognition of variability is hampered by calibration 
uncertainties, noise, slight differences in spectral resolution and 
the sparseness of data points.  With dedicated monitoring and stable 
observing procedures, it seems likely that all of the masers vary 
at the 5 per cent level over several decades.  It will be of special 
interest to monitor those sites that have not yet shown variability to 
assess whether this is so.  Apparently stable sources  from the 
present paper include 328.809+0.633, 333.135-0.431, 336.358-0.137,  
341.218-0.212, 347.870+0.014 and  349.067-0.017.  

Our data set has not only allowed preliminary characterisation of past 
variability at each maser site, but is an excellent yardstick for 
recognising future variability.  A comparable data set of spectra 
obtained 2010-2012 is being prepared from the MAGMO project, following 
the MAGMO pilot study (Green et al. 2012b).  
The combination of the present data set with MAGMO, together with 
variability studies of 
methanol masers, will allow selection of candidates that merit more 
intensive future monitoring.  Since the early discovery of OH maser 
variability many decades ago, variability studies have languished.  
Such studies  now appear  ripe for re-investigation, following the clear 
recognition of 
periodicity in some of the methanol masers that accompany them, and the 
first indication of an OH maser with periodicity at 1665 and 1667 MHz,  
matching the methanol (Green et al. 2012a).  Future in-depth studies of 
variability will require the tracking of intensity for all individual 
features, and although single dish measurements are sufficient for 
simple sources, more complex sources with many spectral features benefit 
from higher spatial resolution to distinguish components that 
overlap in velocity.  Monitoring programs at VLBI resolution are not yet 
practical for large numbers of sources, but the wide-field mapping 
capability of future telescopes such as ASKAP will provide 
simultaneous observations of a large ensemble of sources at modest 
spatial resolution, and make frequent monitoring  feasible  (Dickey et 
al. 2013).

It seems likely that several distinct variability mechanisms are 
at play.  Most readily spotted are the `flares', followed by decay, of 
single features amongst other more quiescent features.  More easily 
overlooked are amplitude changes that are smaller, but are 
simultaneously present in the majority of features;  their study 
requires a higher level of calibration precision.  The latter variables 
may be more indicative of periodic effects.  
With recognition of at least two different mechanisms in variability, 
and their interplay, we may hope to make better progress in studying 
them.  It may be that the most promising candidates for detecting 
periodic variability can best be recognised amongst sources with  
persistent overall similarity in spectral shape, with only modest 
amplitude changes, and that the periodicity is most clearly manifested 
as periodic decreases rather than periodic flares (Green et al. 2012a).

\section[]{Circular polarization of the OH emission}

When large uniformly studied samples of sources are available, almost 
all sites display some circularly polarized features, in some cases 
essentially 100 per cent polarized (e.g. Szymczak \& Gerard 2009). The  
same is true of the sources of Paper I and the present sample.  The 
Zeeman effect in magnetic fields of several mG is responsible for the 
circular polarization, and the presence of a Zeeman pair of the 
frequency-shifted $\sigma$ components of a Zeeman pattern is sometimes  
strongly indicated even in single dish spectra such as ours.  
Many of the present OH masers will soon be the subject of further study 
with the additional spatial resolution of the ATCA (see Green et al. 
2012b) and  thus, in most cases, we defer interpretation of circular 
polarization until the new observations are accessible.  However, in the 
three special cases of Zeeman triplets, where we find $\sigma$ 
components associated with linearly polarized  $\pi$ components, we 
discuss them in the following section concerning linear polarization.

\section[]{Linear polarization of the OH emission}

The Zeeman effect that satisfactorily accounts for the observed 
circular polarization is expected to also give rise to a linearly 
polarized  $\pi$ component, unshifted in frequency (velocity) by 
the magnetic field, midway between the velocity-shifted Zeeman 
$\sigma$ component pairs that are predominantly circularly polarized.  
The relative prominence of $\sigma$ and $\pi$  components  
depends on the magnetic field orientation with respect to the line of 
sight.  The expectation is that all $\pi$ components will be 100 per 
cent linearly polarized, whereas all  $\sigma$ components will have some 
linear polarization accompanying the circular (i.e. net elliptical) 
except when the field is precisely aligned along the line of sight 
between observer and source.   
Comparing these expectations with measurements to date, we first note 
that full Zeeman triplets  have proved to be remarkably elusive, with 
the sole commonly accepted candidate, at 1665 MHz, in 81.871+0.781 
i.e. W75N (Hutawarakorn, Cohen \& Brebner 2002;  Fish \& Reid 2006).  
More generally, the expectation that the majority of features will 
display some linear polarization is not borne out.  Quantitatively, the  
prevalence of linear polarization partly depends on the statistic used, 
as is clear from the study of 18 fields using the high spatial 
resolution of the VLBA by Fish \& Reid (2006).  On the one hand, 
they found only one field with an absence of linear polarization (no 
significant linear polarization amongst nearly 200 features of 
49.488-0.387).  But when considering polarization of the 1000 individual 
features distinguishable in all 18 fields, they found the majority  
(two-thirds) with  no significant linear polarization.   Of 
the  one-third that do have significant non-zero linear 
polarization, they identify some  as the $\sigma$ components of 
their 184 listed Zeeman pairs, and speculate that the remainder are a 
mixture of $\sigma$ and $\pi$ components.   In general they were unable 
to reliably distinguish  $\sigma$ and $\pi$ components except for a 
localized group of features in 81.871+0.781 (the site 
which is also host to the sole example of a Zeeman triplet).  
One motivation for our present study was to search for possible Zeeman 
$\pi$ components among the much larger sample of 200 distinct maser 
source sites - an order of magnitude larger than has hitherto been 
studied -  in the hope of establishing whether $\pi$ components are 
indeed intrinsically rare.   This is important because there have been 
suggestions that the apparent rarity of  $\pi$ components is a real 
deficiency, and could be caused by magnetic beaming which may 
preferentially favour maser emission along the magnetic field lines 
rather than transverse to them (Gray \& Field 1994, 1995).

We first consider the general statistics on the occurrence of linear 
polarization, as revealed in single-dish measurements (of low spatial 
resolution).  A major study was conducted in a sample of nearly 100 
sources  by Szymczak \&  Gerard (2009), who found that some linear 
polarization was detectable in 80 per cent of 1665-MHz spectra  and 62 
per cent of 1667-MHz spectra;  the smaller fraction at 1667 MHz is most 
likely due to the relatively weaker overall emission and fewer features, 
thus limiting polarization detectability to a smaller fraction of 
sources.   In our sample of Paper I, we found linear polarization 
statistics similar to Szymczak \& Gerard (2009).  

Here we perform the same investigation for sources in the present 
sample, which can then be combined with Paper I statistics to provide a 
larger sample and more significant statistics.  

We find detectable linear polarization at 1665 MHz in 93 of 138 
sources (excluding from the sample the sources with lowest signal to 
noise ratio, typically those with no feature above 0.5 Jy, for which 
our polarization sensitivity was inadequate, as noted in Paper I).  At 
1667 MHz, we found detectable linear polarization in 51 of 87 sources 
(similarly excluding the weakest sources).  Combining the statistics 
with those from Paper I (74 out of 89 at 1665 MHz and 39 out of 70 at 
1667 MHz), our totals are 167 out of 227 (73.6 per cent) at 1665 MHz; 
and  90 out of 157 (57.3 per cent) at 1667 MHz.  

These results establish unequivocally that the presence of linear 
polarization is common, even in data of low spatial resolution 
single dish measurements, despite these being subject to `beam 
depolarization'.  In detail, when combined within the large beam, 
maser spots at the same velocity which would be spatially 
distinct at higher resolution, will simply add in total intensity, 
but their linear polarization will partially cancel 
unless they have identical ppa.  For example, for equal linearly 
polarized signals differing in ppa by 90$^\circ$, the net linear 
polarization  will fully cancel;  if differing by 60$^\circ$, the 
net linear signal will be halved.  
While this reduces the number of features likely to exhibit high linear 
polarization in single dish observations of complex sources, the 
argument can usefully be inverted to infer that, when high linear 
polarization is seen, the emission arises predominantly from a small 
region with well-defined ppa, rather than being a blend from many 
unrelated spatially separate regions.

Thus  those sources where we find at least one feature with strong 
linear polarization are of special interest.  Retaining the criterion 
used in Paper I, in our present sample we find features that are at 
least 50 per cent linearly polarized at 1665 MHz in 36 out of 138 
sources, and at 1667 MHz in 15 out of 87 sources (10 of the 15 are also 
within the group of 36 highly polarized at 1665 MHz so the total number 
of sources with high linear polarization is 41).  
Combined with the data of Paper I, we find high linear polarization at 
1665 MHz in 68 out of 227 sources (30 per cent) and, at 1667 MHz, in 32 
out of 157 (20.4 per cent).  These fractions are similar to those found 
from the subset of sources in Paper I alone and, likewise, similar 
to those of Szymczak \& Gerard (2009).

\subsection{Individually interesting sources with linear 
polarization of their OH emission}

Amongst the 41 individual instances of sources with high linear 
polarization at 
either 1665 or 1667 MHz, we single out 16 special examples.  They fall 
into three sub-groups (Sections 6.1.1, 6.1.2, 6.1.3), the first 
associated with blue-shifted outflows with high variability (a 
category noted in Paper I with two examples, to which 
we add one new one).  The second sub-group comprises three sources, each 
with a linearly polarized feature that we identify as a $\pi$ component 
of newly recognized Zeeman triplets.  The third sub-group is the 
largest, where sources display  a linearly polarized feature  of 1667 
and of 1665-MHz emission at the same velocity, and the degree of linear 
polarization is high in at least one transition.  
These have no prominent nearby emission that could represent the 
complementary $\sigma$ components of Zeeman triplets, and thus seem to 
be prime candidates for isolated $\pi$ components;  they will   
occur where the magnetic field is predominantly in the plane of the sky, 
a field configuration where  corresponding velocity-shifted $\sigma$ 
components will be the weaker components, with low circular polarization 
and likely to be unrecognizable.  The implication of the velocity 
coincidence of $\pi$ components at the two transitions is that they are 
extremely close,  and effectively co-propagating.  We discuss these 
matters further in Section 6.3.  

We now summarize our data for each of the 16 sources that display 
linearly polarized features of special interest, and reassess data 
for similar sources from Paper I.  
The two  sub-groups associated with Zeeman patterns will then be 
discussed extensively  in Sections 6.2 and 6.3.

\subsubsection{Highly linearly polarized features associated with 
blue-shifted outflows of high variability.}

~~~~~~323.740-0.263 displays more than 70 per cent linear polarization 
in a 1665-MHz feature near velocity  -80 \kms.  The source also has two 
other (probably related) remarkable properties as already discussed in 
4.3 and 4.4, namely, the strongly polarized feature is in an outflow 
that is highly blue-shifted with respect to the systemic 
velocity near -50 \kms;  and the feature has shown remarkable flaring, 
especially from 2004 to 2005 (section 4.4), becoming the strongest  
feature in the whole source in 2005,  as is evident from the spectra 
that we have shown for both epochs.    
This remarkable source should be grouped with two other sources that 
were reported in Paper I: 12.908-0.260 which shows  even higher linear 
polarization in a blue-shifted feature that flared (becoming the 
strongest feature, exceeding 100 Jy, in 2005),  in this case at 1667 
MHz; and the source 24.329+0.145, the 
archetypal OH site dominated by an outflow at a number of blue-shifted 
velocities, variable,  with high linear polarization, and in this case 
with the outflow detected only at 1667 MHz.  

\subsubsection{Highly linearly polarized features  interpreted 
as Zeeman $\pi$ components of  three newly recognized Zeeman triplets.}
   
~~~~~~297.660-0.973.  Linear polarization is 90 per cent for the 
strongest 1665-MHz feature, at +27.6 \kms, stable  in ppa (close to 
145$^\circ$ based on the negative value of U and the weaker, but 
positive, value of Q) from 2004 to 2005.  Past evidence showing no 
net circular polarization is  consistent with the high linear fraction 
being a persistent long-term property (at least since 1982).  The 
feature lies midway between weaker features of LHCP emission at +25.6 
\kms\ and RHCP at +29.6 \kms, a separation that would imply a magnetic 
field of +6.8 mG.  If the linearly polarized feature is a $\pi$ 
component, between matching $\sigma$ components, its dominant flux 
density would suggest a magnetic  field almost in the plane of the sky.  
Furthermore, there is a hint of linear polarization in the stronger 
putative $\sigma$ component at +25.6 \kms\ with a ppa near 45$^\circ$ 
(positive U), approximately orthogonal to the ppa of the $\pi$ 
component, as expected for a true Zeeman triplet.

309.921+0.479.   The outstanding feature at this site is at -61.8 \kms\ 
where the strongest 1667-MHz emission occurs, and is essentially 100 per 
cent linearly polarized, and thus a prime candidate for a Zeeman $\pi$ 
component. Q is positive and U is of similar magnitude but negative (and 
thus the ppa is approximately 157.5$^\circ$).     At more negative 
velocity, the only 1667-MHz emission is weak RHCP at -63.1 \kms, offset 
-1.3 \kms\ from the linearly polarized feature.   Two prominent  
features are at less negative velocity, most notably the LHCP feature at 
-60.5 \kms, offset +1.3 \kms\ from the putative Zeeman $\pi$ component.  
The pattern thus appears to be a Zeeman triplet, in a magnetic field 
of -7.3 mG (assuming a splitting factor of 0.354 
${\rm{km~s^{-1}~mG^{-1}}}$).

330.953-0.182.  With regard to high linear polarization, there is a 
single 1665-MHz feature, at -88.2 \kms, with total intensity 
of nearly 10 Jy and linear polarization 7 Jy.    
This site has been studied at high spatial resolution (0.2 arcsec) with 
the LBA (Caswell, Kramer, Reynolds \& Sukom 2010a) which shows 1665-MHz 
features near -89.5 (RHCP), -87.2 (LHCP) and -88.2 \kms\ (strong at 
both R and LHCP), all arising from the same site, isolated and quite 
distinct from the majority of OH features which arise from a more 
extended nearby complex 330.954-0.182.   
The RHCP and LHCP features were interpreted as a Zeeman pair, Z1,  by 
Caswell et al. (2010a), with magnetic field -3.7 mG.  A pair of 
features, LHCP at -87.68 and RHCP at -89.80 \kms\ are another 
likely Zeeman pair in the same region, (but not remarked on by Caswell 
et al. 2010a).  At the LBA resolution, there are thus 5  
features distinguishable in position or velocity.  The 
LBA data did not record linear polarization, but the equal intensities 
of RHCP and LHCP at -88.2 \kms\ prompted Caswell et al. (2010a) to suggest 
that this feature might be linearly polarized, and indeed we now see it 
to be so, 70 per cent linear polarization, with ppa close to 
+22.5$^\circ$ (U and Q positive and approximately equal).  We suggest 
that it is the  $\pi$ component of Z1, closely matching the mean 
velocity, -88.4 \kms\ of Z1, although with a formal offset in 
position of 0.2 arcsec (close to the LBA beamwidth to half-power).  We 
interpret it as a Zeeman triplet.  

\subsubsection{Linearly polarized features  of  1665 and 
1667-MHz transitions coincident  in velocity;  interpreted as $\pi$ 
components of a Zeeman pattern, where the two transitions approximately 
co-propagate.  }

~~~~~324.200+0.121.  In this source, the strongest feature at both 
1665  and 1667 MHz is at the same velocity, -91.6 \kms, and displays 
close to 50 per cent polarization at both transitions.  

327.120+0.511.   Linear polarization of more than 50 per cent is 
present at 1665 MHz in three features, one of which, at -84.80 \kms, 
shows nearly 50 per cent linear polarization at 1667 MHz.  

327.402+0.444.   Linear polarization of more than 50 per cent 
is present in a weak feature at -82.5 \kms\ with peak flux density 
near 0.5 Jy at both 1665 and 1667 MHz.  

331.342-0.346.  High linear polarization of 90 per cent is present at the 
strongest 1667-MHz feature;  prominent linear polarization is also 
present at the strongest 1665-MHz feature which, notably, is at the same
velocity, -66.95 \kms.

335.060-0.427.   
1667 and 1665-MHz emission features at the same velocity, 
-42.7 \kms, both have a high degree of linear polarization.

335.585-0.285.  The feature at velocity -48.0 \kms\ has linear 
polarization near 100 per cent at 1665 and 1667 MHz.

338.280+0.542.  1667-MHz emission is currently seen as a single feature 
at -61.6 \kms\ with 60 per cent linear polarization and 50 per cent RHCP.  
Near this velocity, 1665-MHz features also 
show high linear polarization. The  1665 feature at -63 \kms\ 
has even more striking linear polarization, essentially 100 per cent, 
but has no 1667-MHz counterpart. 

339.282+0.136.  The high linear polarization of a 1667-MHz feature 
at -73.5 \kms\ is matched by similar polarization of a weaker 
1665-MHz feature at this velocity.  

339.622-0.121.  The high linear polarization of the strongest 
1667-MHz feature at -36.3 \kms\ is matched by significant linear 
polarization at 1665 MHz.  Other features also show high linear 
polarization.

345.437-0.074. High linear polarization is present at both 1665 and 
1667 MHz near -24.2 \kms.  

345.504+0.348.  There is high linear polarization at both 1665 and 
1667 MHz near -19.5 \kms.  

347.628+0.148.  Linear polarization exceeds 50 per cent at -97 \kms\ 
for both 1665 and 1667 MHz.  More generally, there is prominent 
linear polarization displayed by many features which has  persisted 
over nearly 1 year from 2004 and 2005.

In Paper I, eight examples similar to the twelve listed above 
can be recognised in the Table of source properties, but were not 
explicitly discussed in Paper I.  
Here we list these further examples from Paper I so that the whole 
sample of 20 from Papers I and II can be considered together in the 
subsequent discussion of Zeeman $\pi$ components in Section 6.3.  

356.662-0.264.  The dominant features are at -54.0 \kms, with 
LINP 1.6 Jy at 1665 MHz and 0.4 Jy at 1667 MHz compared with 
total intensities of 2.0 and 0.5 Jy respectively.  

8.683-0.368.  A 1667-MHz feature at +40.8 \kms\ has LINP 
flux density 0.8 Jy (80 per cent of the total intensity) and at 
the same velocity is the highest LINP of 1665-MHz emission, 
1.3 Jy (65 per cent of the total intensity).  

9.621+0.196 (sometimes referred to as 9.62 with  the suffix `e', or 
`north').   Our data show a 1667-MHz feature with the 
strongest LINP of 6 Jy (55 per cent of the total intensity) at 
+1.5 \kms; at this velocity is a matching 1665-MHz feature with 
LINP 3.5 Jy, but its total intensity  is clearly blended with 
other features.  This source was 
mapped with the VLBA in 2001 (Fish et al. 2005) and their 1667-MHz 
emission corroborates our 1667-MHz measurement of LINP, with the same 
ppa as our measurement (Q negative and U small positive), 80$^\circ$.  
The VLBA measurement of 1665-MHz emission shows a feature at essentially 
the same velocity, and with LINP 3.19 Jy;  within a fraction of 
the VLBA beamsize (36 x 22 milliarcsec) it coincides with the 1667-MHz 
feature, and the VLBA is able to spatially distinguish it from a nearby 
feature (stronger in total intensity but with lower LINP), that overlaps 
in velocity (at +1.77 \kms).   To VLBA precision we conclude that the 
linearly polarized features (at +1.5 \kms) at the two transitions are 
spatially coincident, and stronger at  1667 MHz than at 1665 MHz.  

10.444-0.018.  The feature at +75.4 \kms\ is quite distinct at 1667 MHz 
with LINP 1 Jy and total intensity 2.5 Jy.  Its counterpart at 1665 MHz 
is weaker, with total intensity 0.5 Jy and LINP 0.3 Jy (evident from Q 
and U spectra).  

10.473+0.027.  The 1667-MHz feature at +65.2 \kms\ has LINP 0.7 Jy 
and total intensity 1.2 Jy.  Its counterpart at 1665 MHz 
is confused by stronger features at slightly higher velocity, but we 
estimate its LINP to be 0.3 Jy (evident from Q and U spectra), and total 
intensity 0.6 Jy.  

12.889+0.489.  Spectra at both 1665 and 1667 MHz show a 
feature at  velocity +31.6 \kms\ which has high linear 
polarization (Green et al. 2012a; and Paper I). The 
velocity is strikingly the same at both transitions, and the position 
angles of linear polarization are similar.  It is an isolated feature 
spatially and spectrally.   Observations in 2004 and 2005 (Paper I), as 
well as a series of observations in 2010 and 2011 (Green et al. 2012a), 
and the measurements in 2003 of Szymczak \& Gerard (2009)
confirm the persistence of these features over more than 8 years.  
We interpret them as Zeeman $\pi$ features, with 1665-MHz intensity 
of nearly 1 Jy and 1667-MHz slightly weaker.  

30.703-0.069.  The strongest feature at both 1665 and 1667 MHz 
appears to be a blend, with peak at velocity +91.1 \kms\ predominantly 
LHCP.  The components at slightly lower velocity, +91 \kms, have a 
higher level of linear polarization, with LINP 4.6 Jy at 1667 MHz and 
3.9 Jy at 1665 MHz, and we interpret them as Zeeman $\pi$ features, 
with percentage linear polarization greater than 50 per cent 
but difficult to measure in view of the blending.  

35.197-0.743. In this source, the 1667-MHz main feature with total 
intensity peak of 
1.0 Jy at +29.1 \kms\ has linearly polarized emission of 0.8 Jy (80 
per cent linearly polarized).  At 1665 MHz, a clear counterpart matches 
in velocity with a linearly polarized emission peak of 1.8 Jy, but 
corresponding to only 25 per cent linear polarization, and 
RHCP emission stronger than LHCP.  From MERLIN studies of (only) 
1665-MHz emission in 1993 with 0.2 arcsec spatial resolution,  
Hutawarakorn \& Cohen (1999) interpret a 
feature at this velocity  as the RHCP component of their Zeeman 
pair Z3, although they measure higher linear polarization (62.5 per 
cent) than circular (58.1 per cent).  We suggest this is blended in the 
MERLIN beam, where it contains both the RHCP $\sigma$  component of 
Z3, and also the highly linearly polarized counterpart to our 1667-MHz 
feature.  With this interpretation, the linearly polarized emission is 
the $\pi$ component of a different maser spot that has no detectable 
$\sigma$ counterparts, partly because the magnetic field lies close 
to the plane of the sky.  It is spatially near Z3 in projection, 
but offset from the `de-magnetized' velocity of Z3 by 0.9 \kms.  
The corresponding 1667-MHz $\pi$ feature in our measurements is 
nearly half the intensity of the 1665-MHz $\pi$ feature.  

More detailed consideration is now given separately to the two 
classes of candidate $\pi$ components.

\subsection{Candidate Zeeman $\pi$ components in triplets}

We first summarize additional data on each of the three new triplets, 
and then consider generalizations from the class as a whole.

297.660-0.973.  We noted in Section 6.1.2 that the 1665-MHz data 
showed clear evidence for a Zeeman triplet, centred at +27.6 \kms,  
with magnetic field +6.8 mG.  The dominant $\pi$ component indicates 
a field close to the plane of the sky.  
There are no matching features at 1667 MHz near this velocity, 
but centred at +24.8 \kms\ there is a convincing  Zeeman pair 
(detectable in 2004, 2005 and, with hindsight, persistent from 1982), 
separation 2 \kms\ with LHCP more positive, and thus in a  field 
almost as large, but of opposite sign, -5.65 mG. 
Since the Zeeman splitting is a measurement of the full field strength 
(rather than line-of sight component), the similarity of the field 
magnitudes further supports the idea that the field in the source is 
mostly close to the plane of the sky, and it is very plausible that 
small variations in field direction would lead to some locations showing 
fields of opposite sign with respect to the line-of-sight.  
An associated  water maser is present but no methanol.  
The nearby continuum shows an extended \HII\ region of several arcmin 
diameter with measured recombination line emission centred at +28 \kms\ 
(Caswell \& Haynes 1987b).  
The large positive velocity indicates a location outside the solar 
circle in a distant portion of the Carina-Sagittarius arm, at a distance 
of about 10 kpc.  No high resolution continuum image is available 
to see whether the old extended \HII\ region is accompanied by 
compact features indicative of more recent star formation.  The 
absence of any nearby methanol maser suggests the site may be  
relatively evolved, and the presence of an associated water maser 
(Breen et al. 2010b) is compatible with this, since the water masers 
often survive into the later evolutionary stages of a high mass SFR.

309.921+0.479.  We noted that the 1667-MHz spectrum shows a Zeeman 
triplet centred at -61.8 \kms, with splitting corresponding to a 
magnetic field of -7.3 mG.  

%${\rm{km~s^{-1}~mG^{-1}}}$.  

The 1665-MHz emission is more confused, and there is no 
clear evidence for a matching Zeeman triplet.  
However, excited-state OH measurements (Caswell 2003, 2004b) clearly 
demonstrate several Zeeman pairs that reveal a 
mixture of positive and negative magnetic fields, 
indicating that there must be some locations where the field is in 
the plane of the sky.  This is corroborated by another long-known 
property of the 6035-MHz emission, namely, the strong 
linear polarization of a feature at -59.55 \kms\ (Knowles et al. 
1976).  Irrespective of whether this is a $\pi$ or $\sigma$ component, 
the linear polarization is further evidence for a magnetic field 
close to the plane of the sky in at least some portions of this 
source.  Thus the presence of a triplet is not unexpected.  

More directly related to the present 1667-MHz triplet are specific 
features in the excited-state OH spectra, revealing likely Zeeman 
pairs centred near the velocity of the Zeeman 1667-MHz triplet.  At 
6035 MHz, RHCP is at more negative velocity than a 
LHCP feature, separation 0.2 \kms, and at 6030 MHz, RHCP is at 
more negative velocity than a LHCP feature, separation 0.25 \kms.   
Adopting the usual Zeeman splitting factors of 0.056 and 0.079 
${\rm km~s^{-1}~mG^{-1}}$ respectively, gives  
magnetic field estimates of $-3.6$ and $-3.2$ mG. The features at each 
polarization both appear to be broadened in velocity, suggesting the 
overlap of several 
maser spots, all with similar magnetic field.   Overall, the  
excited-state and the 1667-MHz data concur that there is a negative 
magnetic field near this velocity, but small offsets of mean velocity 
indicate that the different transitions are not strictly co-propagating, 
which  readily accounts for the different magnitude of the field 
estimate.  There is a strong, 
0.5-Jy, uc\HII\ region at this site, suggestive of a 
quite late evolutionary stage.

%${\rm km~s^{-1}~mG^{-1}}$ this form is also OK
%$\rm{km\,s^{-1}}$

Sources between Galactic longitudes 284.5 and  295$^\circ$ show 
positive fields (Green et al. 2012b), all of them located in the 
Carina arm, which is viewed along its length in this longitude range.  
The fact that 309.921+0.479, with its internal field `reversals', 
and a likely predominant field orientation close to the plane of 
the sky, clearly does not fit 
this pattern, is not surprising, since it does not lie in 
the Carina arm, but in the Crux-Scutum arm, near its tangent point 
as viewed from the Sun, at a likely distance of 4.8 kpc.  

%JAG***[this para not for publication]*** 
%Green et al. 6035 polarimetry shows multiple Zeeman pairs, several 
%differing from those apparent on the earlier spectra;  but they confirm 
%the presence of magnetic fields both positive and negative, with 
%magnitude of several mG. No claimed triplet in the current text, 
%although one looks plausible.  So maybe modify 6035 paper.     

330.953-0.182.  The Zeeman triplet at this site yielded a 
magnetic field of -3.7 mG, and has a flux density in the $\pi$ component 
similar to or larger than the $\sigma$ components, and thus the 
field orientation has a significant (perhaps dominant) component in the 
plane of the sky rather than along  the line of sight.  Other Zeeman 
pairs in the adjacent complex 330.954-0.182 indicate fields of 
similar magnitude and the same negative sign;  however, a field of 
opposite (positive) sign is shown by a loosely related 1720-MHz 
maser site (offset by several arcsec and  probably in the same 
star forming complex).  A field reversal suggests that the true 
field may be mostly in the plane of the sky, perhaps  
over the whole region.  
As summarized by Caswell et al. (2010a), continuum emission in the 
vicinity appears to be resolved into an extended complex of 
several arcsec extent encompassing the nearby maser features of 
330.954-0.182 and, at the location of the maser features of 
330.953-0.182 discussed here, a more compact uc\HII\ region of about 40 
mJy.  The distance was estimated to be at least 5.6 kpc.
Excited-state OH emission at 6035 MHz  and methanol maser emission also 
arise from 330.953-0.182 (Caswell 2003) but has not been detected  at 
the offset, more extended site, 330.954-0.182.

%JAG***[this para not for publication]*** 
%Green et al. 6035 polarimetry shows 3 Z pairs;  the strongest 
%centred at -87.94 indicates a magnetic field of -3.4 mG (LHCP at more 
%positive velocity) and in fact is a Zeeman triplet;  the other two, 
%centred at -88.8 and -89.8 \kms\ are weaker;  they also indicate a 
%field of -3.4 mG.  We also note that the ppa of the $\pi$ component is 
%close to +22.5$^\circ$ (U and Q both positive), similar to that of the 
%1665-MHz $\pi$ component.    

The four known Zeeman triplets display a range of properties which 
we now summarize.   
Three are at 1665 MHz and one at 1667 MHz, thus both transitions are 
represented.  
Distances vary from 1.3 kpc (81.871+0.781)  to approximately 10 kpc, 
with the 1667-MHz example lying at intermediate distance, 4.8 kpc.  All 
except 297.660-0.973 (the most distant)  have an associated methanol 
maser, and associated continuum emission likely to be a uc\HII\ region 
(the weakest being the nearest, 81.871+0.781, only 10 mJy).  
297.660-0.973, with no associated methanol, may be more evolved, and 
lies towards an extended \HII\ region observed only at low spatial 
resolution;  however, in order to distinguish any possible closely 
associated uc\HII\ region, we will require observations of   
high resolution combined with high sensitivity, since it lies at a 
greater distance than any of the others. 
%w98 have 20 mJy ucHII at 12 04 07.89, -63 21 30.6 which is offset by 
%1s ie -7.5" and +5.4" from the OH position.  

\subsubsection{The precision of spatial coincidence in Zeeman triplets 
or pairs}

The tolerance in spatial coincidence needed to identify Zeeman pairs has 
been discussed by Fish \& Reid (2006).  Many Zeeman pairs fail the 
criterion of coincidence to within the measurement precision when that  
precision has been measured from the longest baselines of VLBI 
observations, and yet the  derived magnetic field 
appears satisfactory.  This is attributed to spatial 
clusters of maser spots that possess the same velocity and a common 
value of magnetic field, and thus the components of an apparent  Zeeman 
`pair' often arise from different spots in the 
same cluster, with no major impact on the validity of the field 
measurement.   Fish \& Reid (2006) suggest a loose distinction 
between `true' Zeeman pairs and Zeeman `cousins' of slightly larger 
spatial separation,  Overall, too tight a criterion of coincidence 
would not be helpful, since it would  remove many excellent determinations 
of magnetic field.  Indeed, Fish \& Reid's (2006) measurements of the 
Zeeman triplet 81.871+0.781 show deviations from coincidence much 
greater than the position uncertainties (several times larger than 
the beam width) and yet it is most useful to treat it as a true triplet.  
% also disagreement on the ppa of the linear polarization and  
% stated discrepancy of positions in text differs from Table.  
Thus an understanding of the clustering has been an important step in 
evaluating the reliability of magnetic field determinations.  Spatial 
resolution just sufficient to distinguish separate clusters is usually 
adequate. With this in mind, we regard the position discrepancies 
(greater than nominal errors) between components of  our proposed triplet 
330.953-0.182 as acceptable.

\subsection{Candidate Zeeman $\pi$ components with no triplet 
association}

Our criterion for identifying these $\pi$ components (in the absence of a 
recognisable Zeeman triplet) is the velocity coincidence of the linearly 
polarized feature at 1665 and 1667 MHz (noting that $\sigma$ components 
of Zeeman patterns arising from the same physical region at the two 
transitions would be differently shifted in velocity;  only matching  
$\pi$ components can remain at the same matching velocity).   
We are not aware of any earlier use of this criterion to identify 
$\pi$ components.  The above results show the criterion to be 
spectacularly successful, with 40 suggested $\pi$ features (20 at 
each transition) whereas previously the only confidently 
claimed $\pi$ was the one example in a Zeeman triplet.

The two transitions arising from the same physical region 
implies near co-propagation.  Although the coincidence of the two 
transitions in velocity has been established, the spatial 
coincidence remains to be tested.  Fortunately, in one example, 
9.621+0.196, the high precision spatial measurements at both 
transitions already exist.  As previously discussed in detail 
(Section 6.1.3), they indeed support our suggestion 
of spatial coincidence, as well 
as corroborating the linear polarization measurements (despite a 
difference in measurement epoch of 3 or 4 years).  
Similar observations are needed for the other candidates to assess the 
precision of coincidence between 1665 and 1667-MHz features.  

Other distinctive properties might be associated with the suggested 
co-propagation, and here we explicitly explore intensity ratios of the 
transitions for the suggested $\pi$ components.  
Summarizing the information in section 6.1.3:  of the eight candidates 
from Paper I, 
9.621+0.196, 10.444-0.018, 10.473+0.027 and 30.703-0.069 are all 
stronger at 1667 MHz, by small factors, and 356.662-0.264, 8.683-0.368, 
12.889+0.489 and 35.197-0.743 are all stronger at 1665 MHz, again by 
small factors.  
Of the twelve instances in Paper II, in two (339.282+0.136 and 
339.622+0.121), the 1667-MHz intensity exceeds the 1665-MHz intensity, 
in five others (324.200+0.121, 327.402+0.444, 331.342-0.346, 
335.060-0.427 and 338.280+0.542) they are approximately equal, 
in three cases (345.437-0.074, 345.504+0.348 and 347.628+0.148) 
the 1665-MHz intensity is greater by a factor near two, in 
327.120+0.511 a factor of nearly four, and in only one case 
(335.585-0.285) is the 1665-MHz intensity much greater than at 1667 MHz.  

Thus the typical 1667 and 1665-MHz intensities in the putative 
co-propagating features are nearly equal (i.e. 
the ratio of 1665 to 1667-MHz intensity has a median value close to   
one), whereas the ratio for the vast majority of features 
is 4, as estimated in Section 4.1.  We also noted in Section 4.1 that 
Wright et al. (2004b) found, in their detailed study of the W3(OH) 
site, that most of the stronger features are at 1665 MHz rather than 
1667 MHz, and for the sum of all features, the flux density at 1665 MHz 
is 15 times larger than at 1667 MHz; in contrast, they found that, 
at six locations where the two transitions appear to co-propagate, 
the 1667-MHz intensity in 
most cases nearly equals or surpasses the 1665-MHz intensity.  
Thus our result appears to support the finding by Wright et al. (2004b), 
with regard to a distinctive intensity ratio for the rare 
co-propagating regions, compared to the average for all features.   
However it is not clear whether this might be a trivial result, since 
regions selected by apparent co-propagation necessarily have moderately 
similar intensities at the two transitions.   
We finally draw attention to a difference between our sample of likely 
co-propagating features and those of Wright et al. (2004b):  the 
features that they consider are, in all cases,  Zeeman $\sigma$ 
components; thus although the spatial coincidence of the two transitions 
is a direct measurement (at milliarcsecond precision), the  velocity 
`coincidence'  is necessarily  indirect,  since the two transitions 
have different splitting factors and are differently offset 
from the inferred `demagnetized' velocity.  

It is only with full observations of high resolution that can 
spatially distinguish all individual features in every source 
that we will be able to thoroughly study the statistics of 
co-propagation (either precise co-propagation, or near co-propagation) 
and explore whether this can be used as a tool to ascertain the 
predominant combinations of temperature and density in the masing 
region.  

\section[]{Polarization patterns}

We first ask, why are Zeeman triplets so rarely seen?  This appears 
to be largely related to the long-known inequality seen between 
the amplitudes of  $\sigma$ components in Zeeman pairs, and quantified 
for a large sample of Zeeman pairs by Fish \& Reid (2006).  A 
commonly accepted explanation is the `Cook effect' (Cook 1966), 
with magnetic field gradients and velocity gradients present 
which will favour the propagation at the emitted frequency of 
one of the Zeeman components, and attenuate at the other.  This 
explanation will equally apply to Zeeman triplets, with one component 
favoured more than the other two.  In some situations it can be 
the $\pi$ component that is favoured, and both $\sigma$ components 
attenuated.  

A second puzzle that relates to Zeeman triplets, and to $\pi$
components more generally, is that the $\pi$ components
are generated as 100 per cent linear, yet most features discovered 
to exhibit high linear polarization have been found on closer inspection 
to also have significant circular polarization (net elliptical). 
Consequently, it has been common to discard these candidates 
as potential $\pi$ components, leading to the extreme conclusion 
that $\pi$ components are completely absent.  
An explanation suggested by Fish \& Reid (2006) 
is that $\pi$ components usually acquire some circular polarization 
during subsequent passage through weak spots of circularly 
polarized $\sigma$ components of other Zeeman patterns.  Then,  
with a weaker constraint on the purity of linear polarization for 
$\pi$ candidates, we retain many more candidates, and are no longer 
forced to the  unlikely conclusion 
that $\pi$ components  are completely absent.  The puzzle may be 
solved, but an unfortunate implication is that the difficulty of 
distinguishing $\pi$ components from $\sigma$ components has been 
aggravated.   

The  Fish \& Reid (2006)  
explanation can also contribute to the absence of another flavour of 
Zeeman triplet - `triple treats' - where a central linearly polarized 
feature is expected to be bracketed by a pair of predominantly linearly 
polarized features of half its intensity, and polarized  orthogonally to 
the central feature, if the maser arises in a large magnetic field close 
to the plane of the sky.  
The `Fish \& Reid' effect suggests to us that even where all three 
components of a `triple treat' are present (perhaps with relative 
intensities modified by the `Cook effect'), it is unlikely that all 
of them will retain their full original linear polarization, 
unscathed by passage through unrelated masing regions of 
different polarization.  

We now turn to a further puzzle.  
Many of the Zeeman pair $\sigma$ components display some linear 
polarization, which will naturally occur whenever the magnetic 
field is not precisely in the line of sight to the observer and, 
in that case, a weak $\pi$ component midway between the pair is 
expected.  Various arguments have been made that the $\pi$ component 
may not be recognisable as a linearly polarized feature if it has 
acquired some circular polarization during propagation.  But that does 
not account for the usual total absence of any feature at the frequency 
midway between the $\sigma$ components.  The observational puzzle is 
most starkly evident in the study of W3(OH) with the VLBA (Wright et 
al. 2004a, 2004b) where candidate $\pi$ components are absent from 
all but two of the 83 Zeeman pairs recognised at 1665 MHz, and even 
these two were rejected when investigated in more detail.  Indeed, 
Wright et al. (2004a) find no linear polarization greater than 15 
per cent in any 1665-MHz feature stronger than 0.35 Jy.    
A likely explanation for W3(OH) is that its magnetic field is very 
uniform in direction, and closely aligned 
with the line of sight, by chance.  The other source with very little 
linear polarization at VLBA  resolution, W51 (Fish \& Reid 2006) may 
be accounted for similarly.  
The rarity of triplets appears to extend to other sources, but there are 
relatively few sources where the statistics are adequate to confirm that 
this remains a serious puzzle.   

Encouraged by the greatly increased  number of Zeeman $\pi$ 
components that are now recognized from the present study, we now 
ask whether there remains a puzzle in which there are fewer 
$\pi$ components than expected.  
We first recall the importance of investigating a large sample 
of sources.  There are certainly some sources where $\pi$ 
components appear to be absent, such as W3(OH) at Galactic 
coordinates 133.947+1.064 (Wright et al. 2004b), and W51e1 and e2,
at Galactic coordinates 49.488-0.388 and 49.491-0.387 (Fish \& 
Reid 2006).  As we suggested earlier, these are likely to be sites 
where the predominant field direction is almost along the line of 
sight.  We note that Fish \& Reid (2006) prefer to attribute 
sites with low linear polarization to stronger internal Faraday 
rotation, but our suggestion that it is due to the gross 
orientation of magnetic field within the site can more simply 
account for it.
There are also some sources where linear polarization is especially 
prevalent e.g. 81.871-0.781 (Fish \& Reid 2006), and these are 
likely to be sites where the predominant field direction is almost 
in the plane of the sky.

Our presentation of 43 new candidate $\pi$ components 
(including three in Zeeman triplets) has hugely increased 
their number, compared to previous assessments of a near absence.  
However, the total number remains small compared to the number 
of sources searched, and does not exclude the possibility that 
$\pi$ components are discriminated against.    
Of course, we would expect the number that we identified from 
their approximate co-propagation at 1665 and 1667 MHz 
to be only a small fraction of the underlying number of $\pi$ 
components that are likely to be present at only one transition;  
but distinguishing these from $\sigma$ components 
(which can also have quite high linear polarization) is 
difficult, except in a small number of 
cases where linear polarization greatly exceeds circular 
polarization.  Extending high spatial resolution VLBI studies to many 
more sources will clarify this.  The resulting improved 
statistics will then indicate whether there is a true 
depletion of $\pi$ components relative to expectations.  

Past indications of a near absence of $\pi$ components  
appeared to be accounted for by the `magnetic beaming' effect 
(Gray and Field 1994, 1995).  Although it need no longer be 
invoked for the near total absence of $\pi$  
components, the `magnetic beaming' effect may, none the less,  
be needed to account for their 
possible depleted rate of occurrence.  

With the large body of full polarization data now accumulating for 
the ground-state main-line transitions, and the prospect of similar 
data for excited-state transitions (especially at 6035 and 6030 MHz), 
there may be scope for also distinguishing some 
propagation-related influences on the linear polarization 
of $\pi$ components, by comparing the statistics for 
transitions at greatly different frequencies, where effects such as 
Faraday rotation are grossly different.

\section[]{Conclusion}

Characterising the full polarization spectra of the complete sample 
of known OH masers has had the special value of revealing 
rare phenomena (including Zeeman $\pi$ components and blue-shifted 
outflows) that were absent or not recognised in smaller samples.  

We have identified three new Zeeman triplets.  We have also used 
the apparent co-propagation at 1665 and 1667 MHz of features 
with high linear polarization to identify a further 40 features  
as Zeeman $\pi$ components.  Not only is this the first large sample of 
$\pi$ components, but it is also the largest sample of features 
exhibiting near co-propagation at two transitions in the OH 
ground-state.  They lend support to a tentative earlier finding that the 
1667-MHz intensity is comparable to the 1665-MHz intensity in such 
conditions, whereas elsewhere it is, on average, only one-quarter the 
intensity.

In Paper I, we singled out rare sources with wide velocity spread, and 
noted two distinct varieties.  Here we have continued this 
investigation using the larger sample, and have focused on ten sites 
that may be the result of collimated outflows.  These are 
similar to those from water masers, with indications of a 
preponderance of blue shifts.  Future VLBI measurements are 
needed to validate these suggestions, but extensive  MAGMO results 
currently in progress using the ATCA may allow preliminary 
corroboration.  
Further follow-up will include searching for similar kinematics in 
possibly accompanying methanol and water masers, and testing for 
continuum emission.  

Combining the variability statistics from the 104 sources in Paper I 
and the 157 sources of the present paper, we find that between 2004 and 
2005, only nine show variations of more than a factor of two in their 
strongest feature.  
On a longer timescale, for sources where comparisons were possible 
with earlier data over several decades, dramatic changes, mostly 
strong `flares', were recognised amongst 18 sources, and  quite high 
stability was evident at only a handful of sites (10).  
Further investigation of 
variability will be possible from the MAGMO project combined with the 
present survey.  This will allow better selection of candidates that 
merit more intensive monitoring, and enhance prospects for 
identifying periodic variable sites, where there is unique  potential 
for probing the history and more fully investigating the physical 
properties.  

\section*{Acknowledgments}

%Note numbering of section suppressed by the use of section*.  

We thank Warwick Wilson for implementing the correlator enhancements, 
John Reynolds and Parkes Observatory staff for enabling the 
non-standard observing procedures, and Malte Marquarding for 
providing within the ATNF spectral analysis package ({\sc asap}), 
the spectropolarimetric reduction features developed 
for the present data.

\bsp

\begin{figure*}
 \centering
\includegraphics[width=15.5cm]{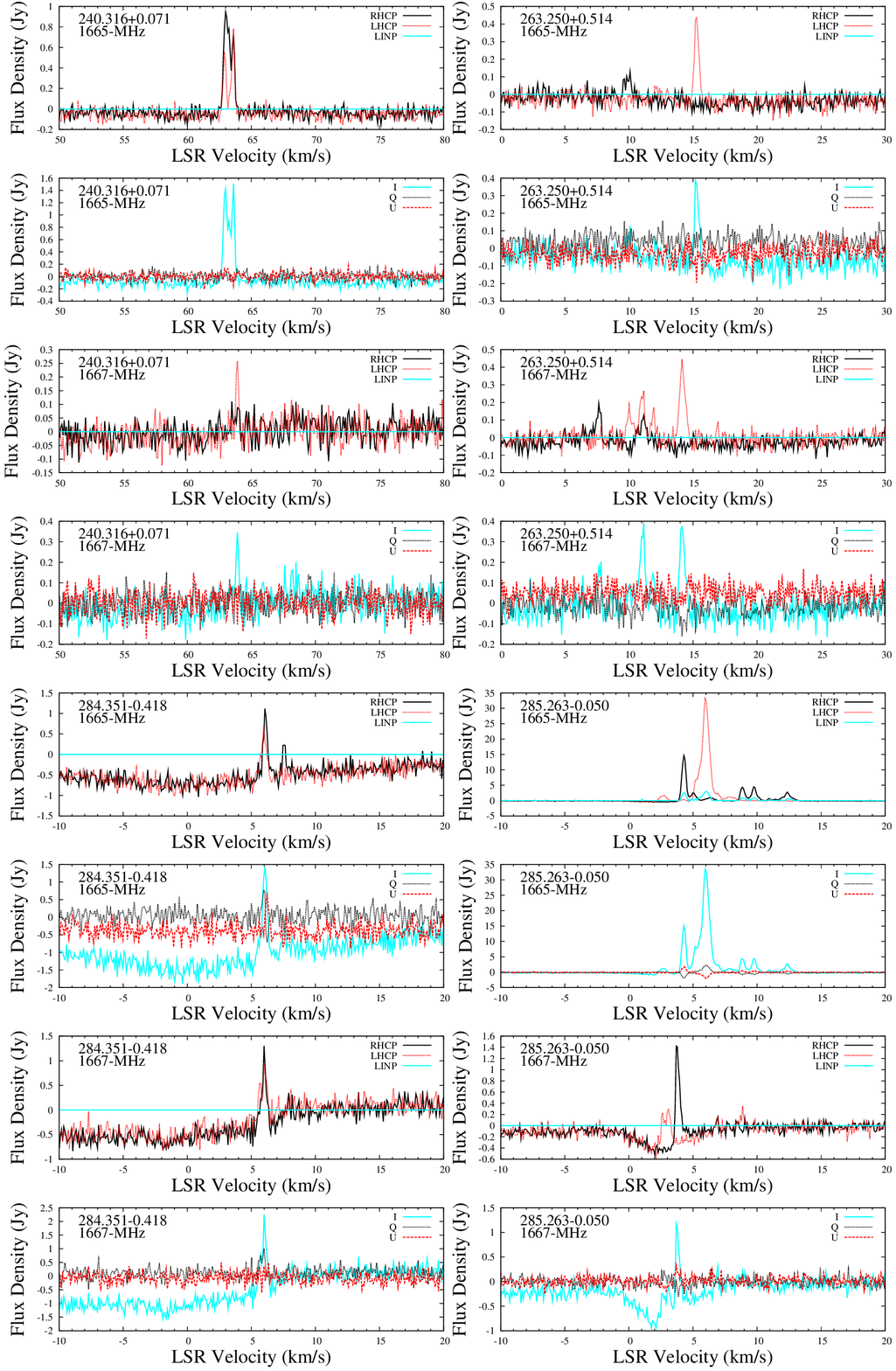}
\caption{Spectra of OH masers at 1665 and 1667 MHz. Within each 
panel, the source name and transition are given, and the  
polarization parameters are plotted as: overlaid spectra of  
RHCP and LHCP with LINP (linearly polarized flux density);  
and overlaid spectra of Q and U with I. }   

\label{fig1 part 1}

\end{figure*}

\begin{figure*}
 \centering

\addtocounter{figure}{-1}

\includegraphics[width=15.5cm]{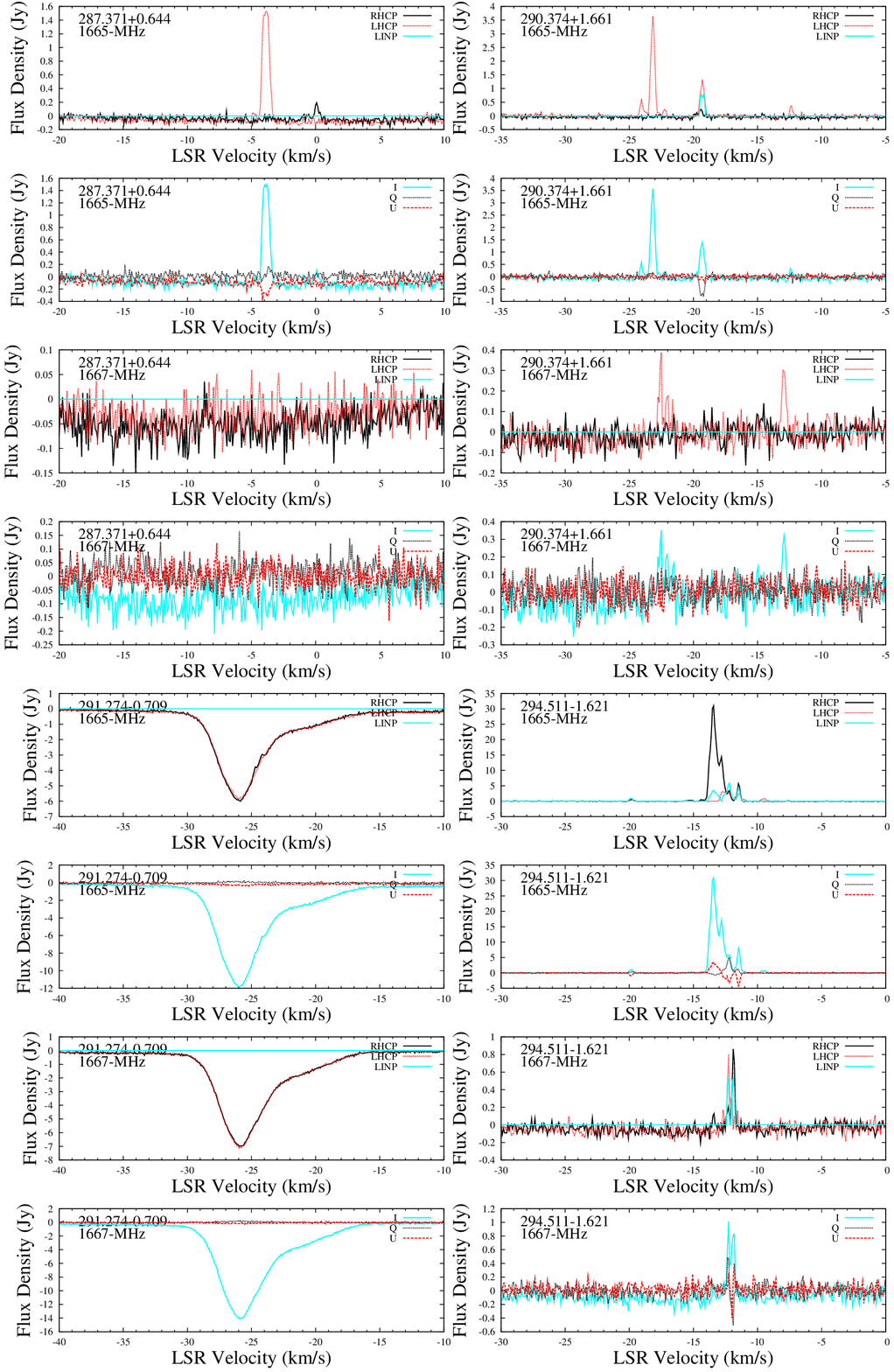}

\caption{\textit{- continued p2 of 35}}

\label{fig1p2} 

\end{figure*}

\begin{figure*}
 \centering

\addtocounter{figure}{-1}

\includegraphics[width=15.5cm]{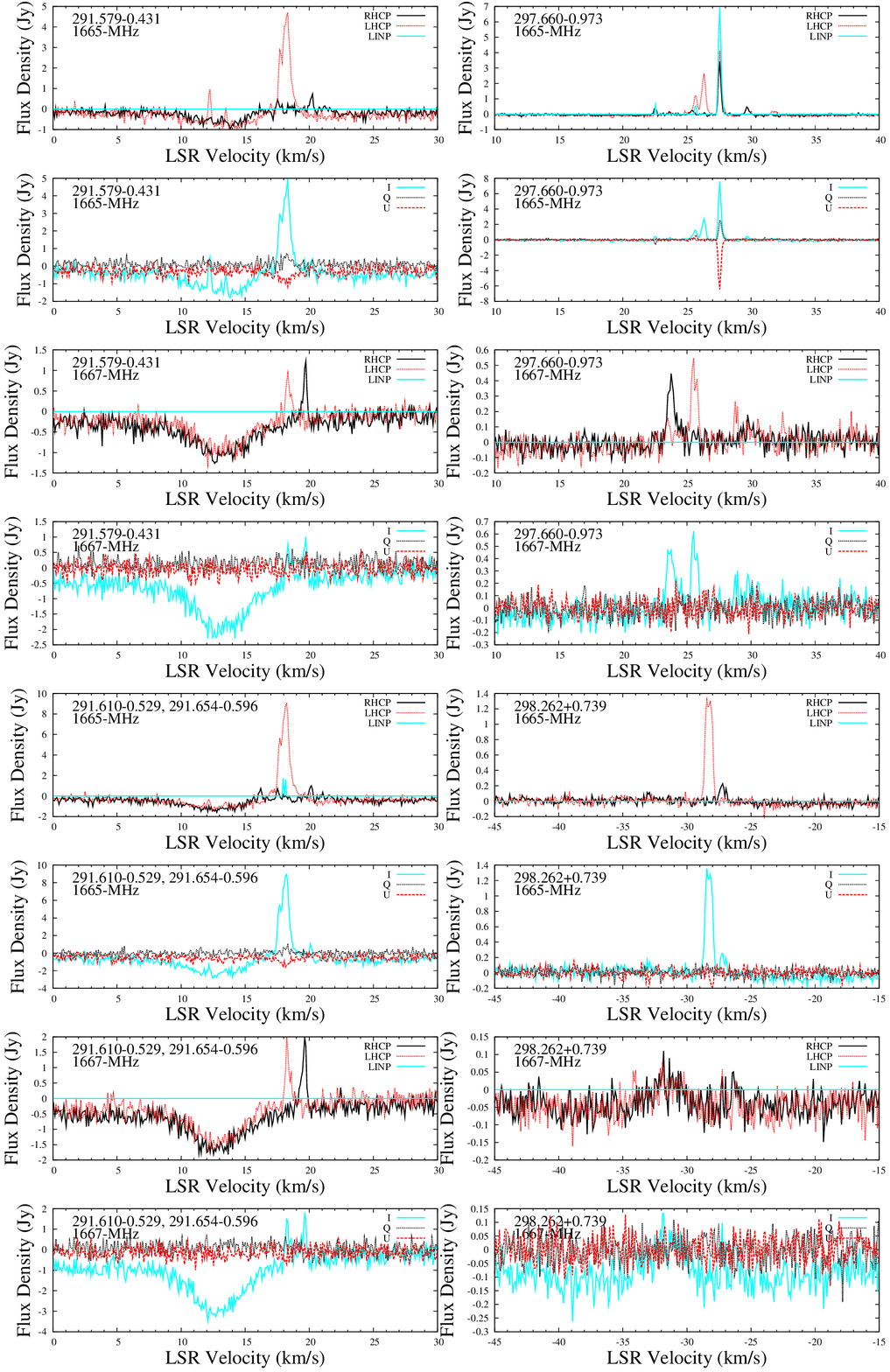}

\caption{\textit{- continued p3 of 35}}

\label{fig1p3} 

\end{figure*}

\begin{figure*}
 \centering

\addtocounter{figure}{-1}

\includegraphics[width=15.5cm]{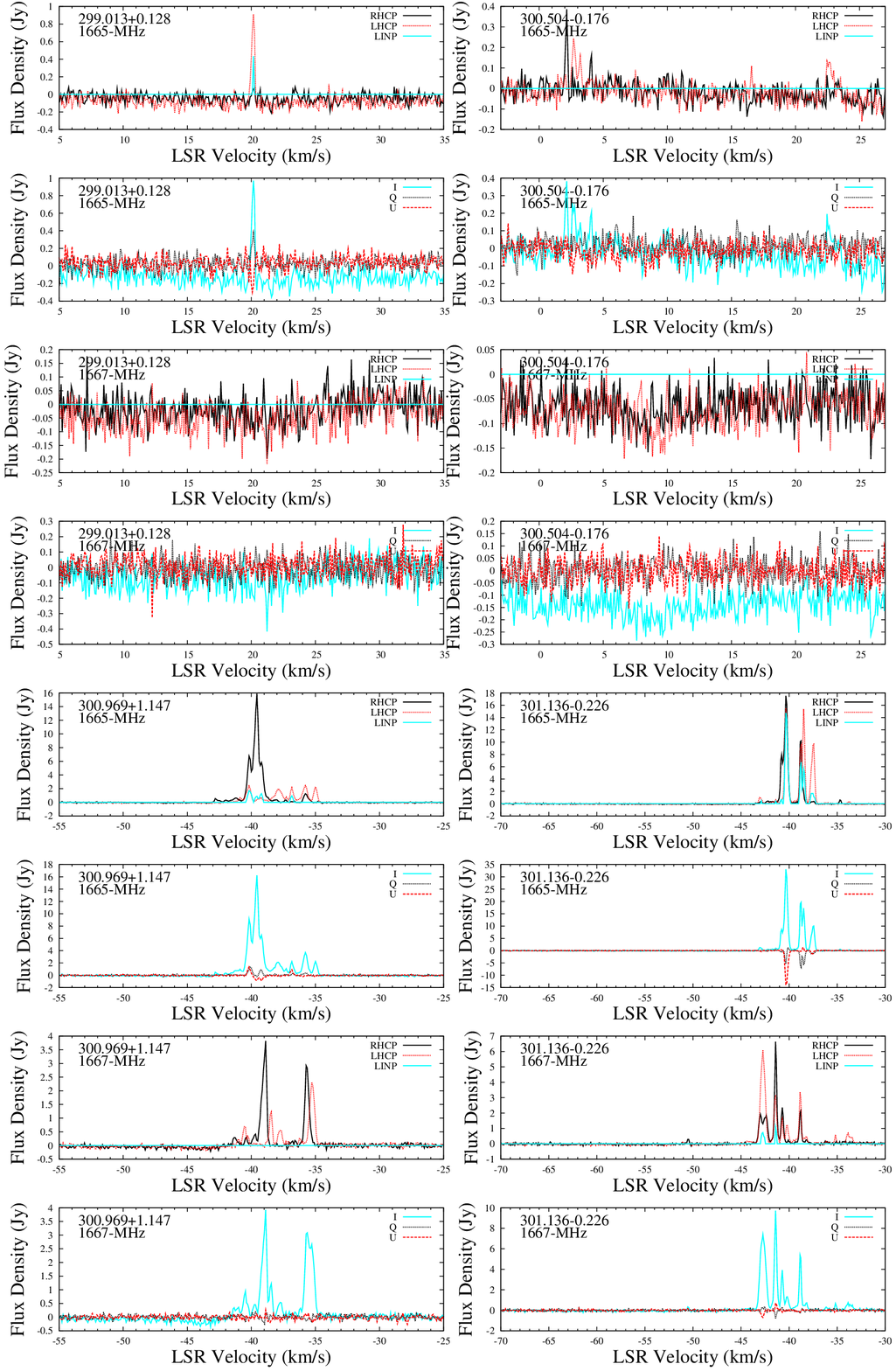}

\caption{\textit{- continued p4 of 35}}

\label{fig1p4} 

\end{figure*}

\begin{figure*}
 \centering

\addtocounter{figure}{-1}

\includegraphics[width=15.5cm]{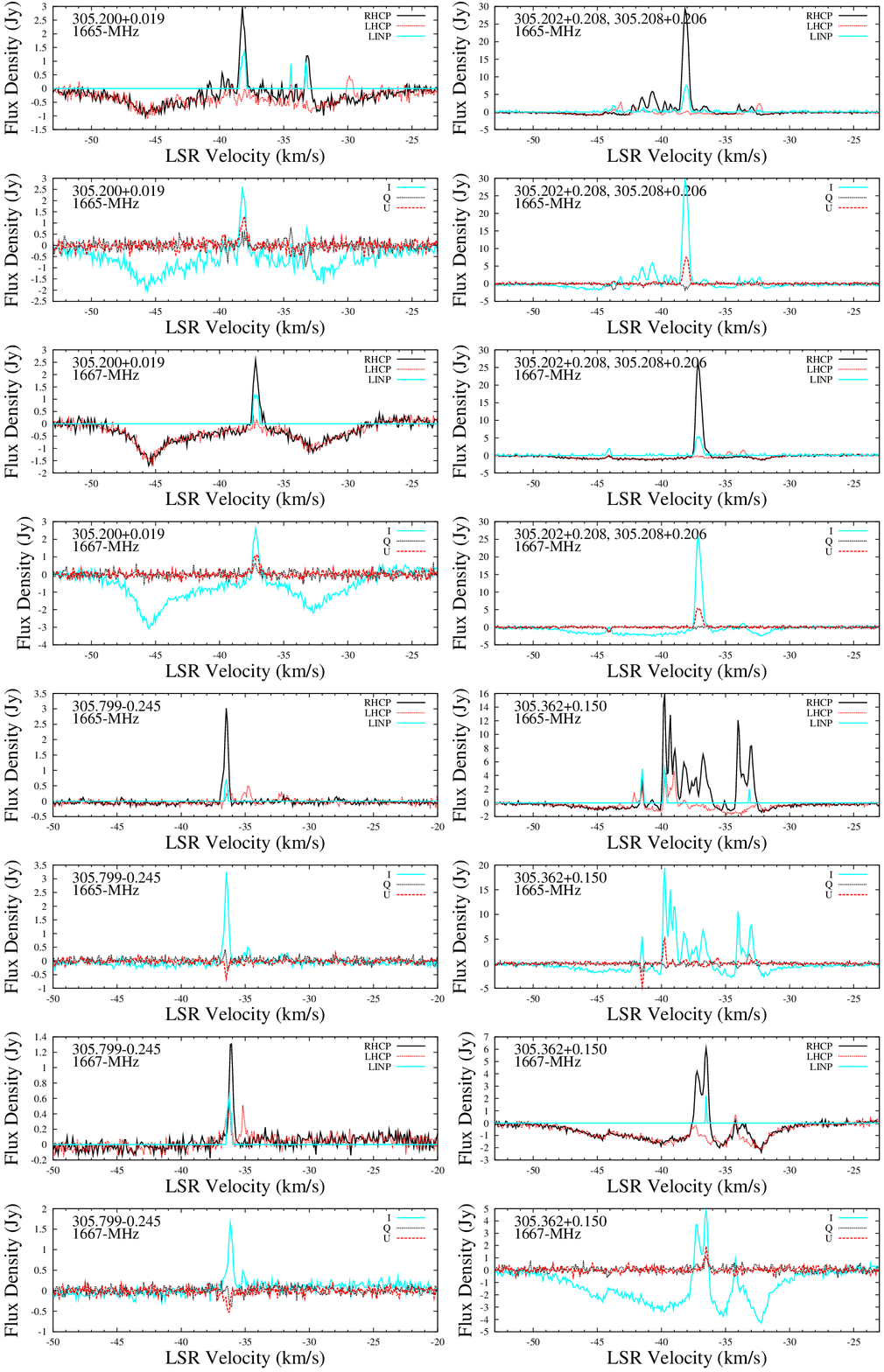}

\caption{\textit{- continued p5 of 35}}

\label{fig1p5} 

\end{figure*}

\begin{figure*}
 \centering

\addtocounter{figure}{-1}

\includegraphics[width=15.5cm]{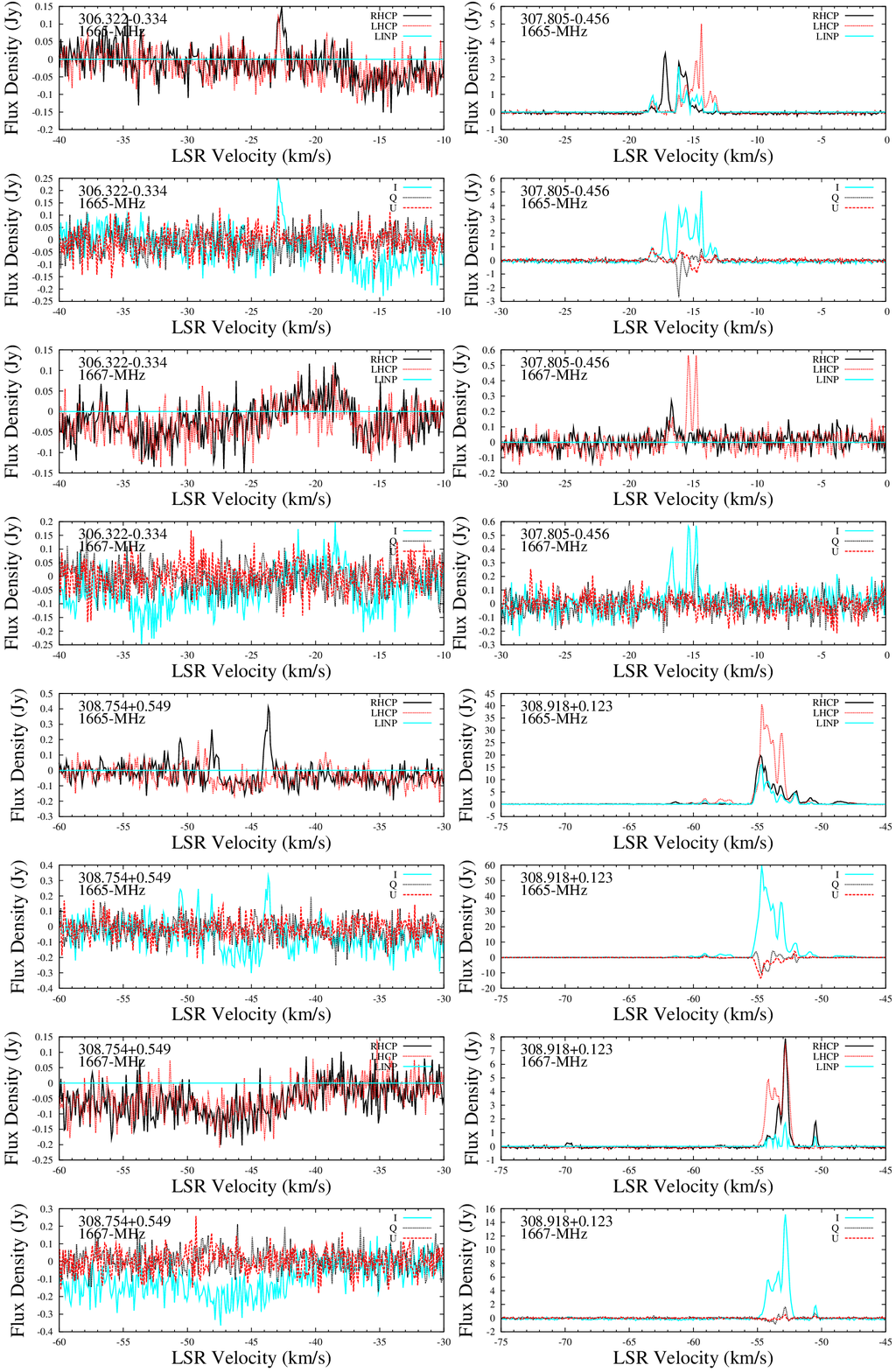}

\caption{\textit{- continued p6 of 35}}

\label{fig1p6} 

\end{figure*}

\begin{figure*}
 \centering

\addtocounter{figure}{-1}

\includegraphics[width=15.5cm]{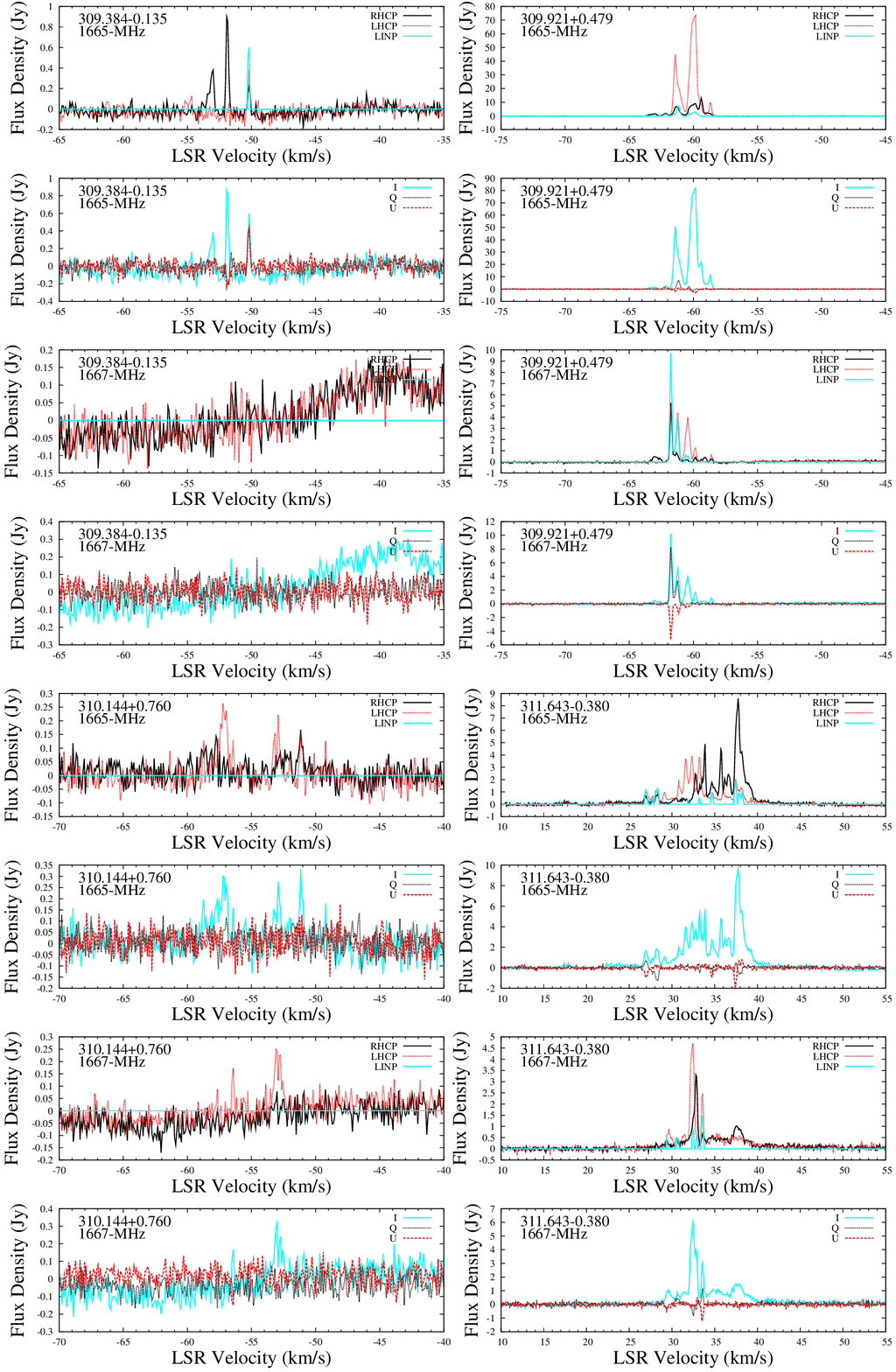}

\caption{\textit{- continued p7 of 35}}

\label{fig1p7}

\end{figure*}

\begin{figure*}
 \centering

\addtocounter{figure}{-1}

\includegraphics[width=15.5cm]{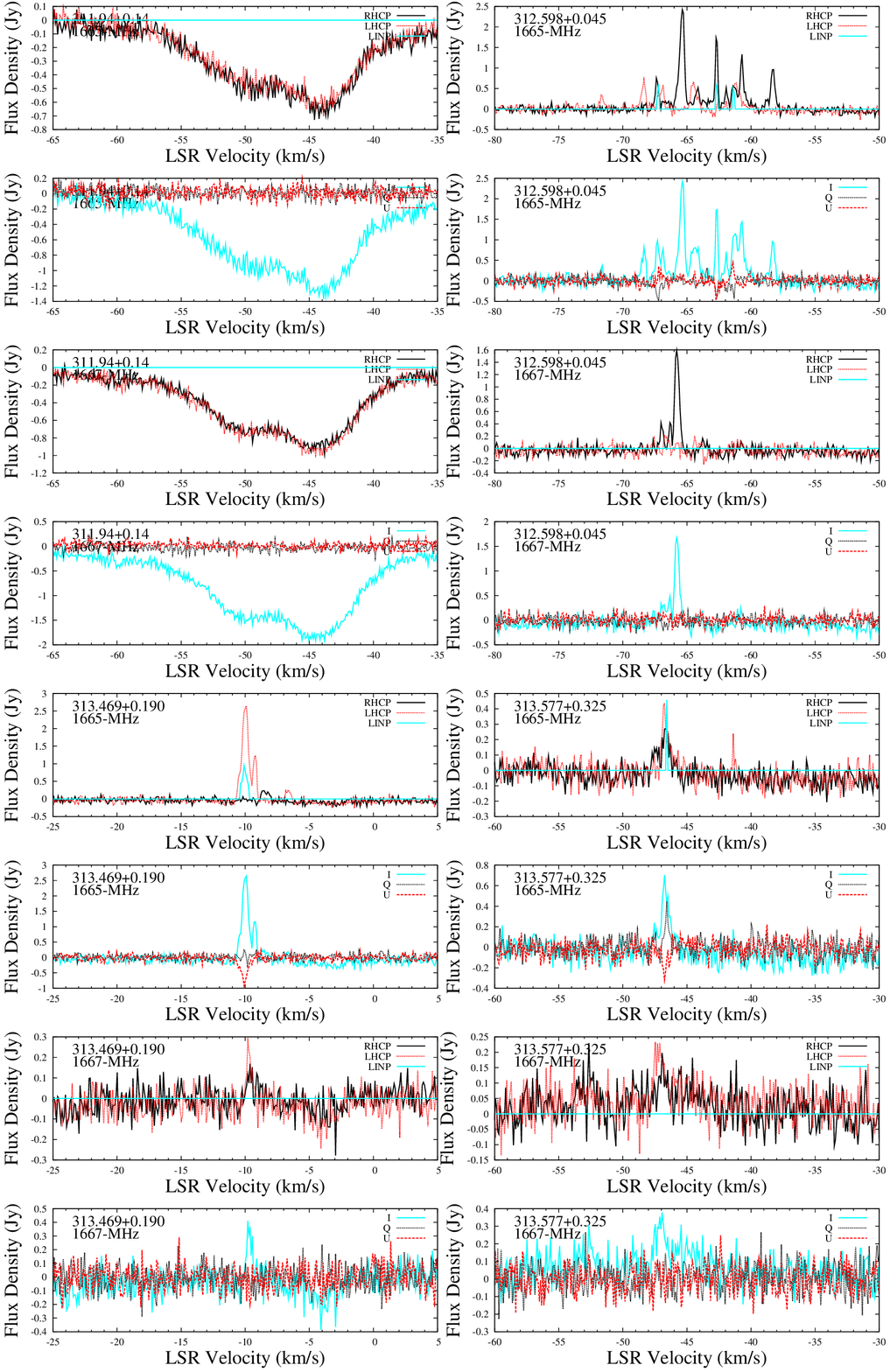}

\caption{\textit{- continued p8 of 35}}

\label{fig1p8}

\end{figure*}

\begin{figure*}
 \centering

\addtocounter{figure}{-1}

\includegraphics[width=15.5cm]{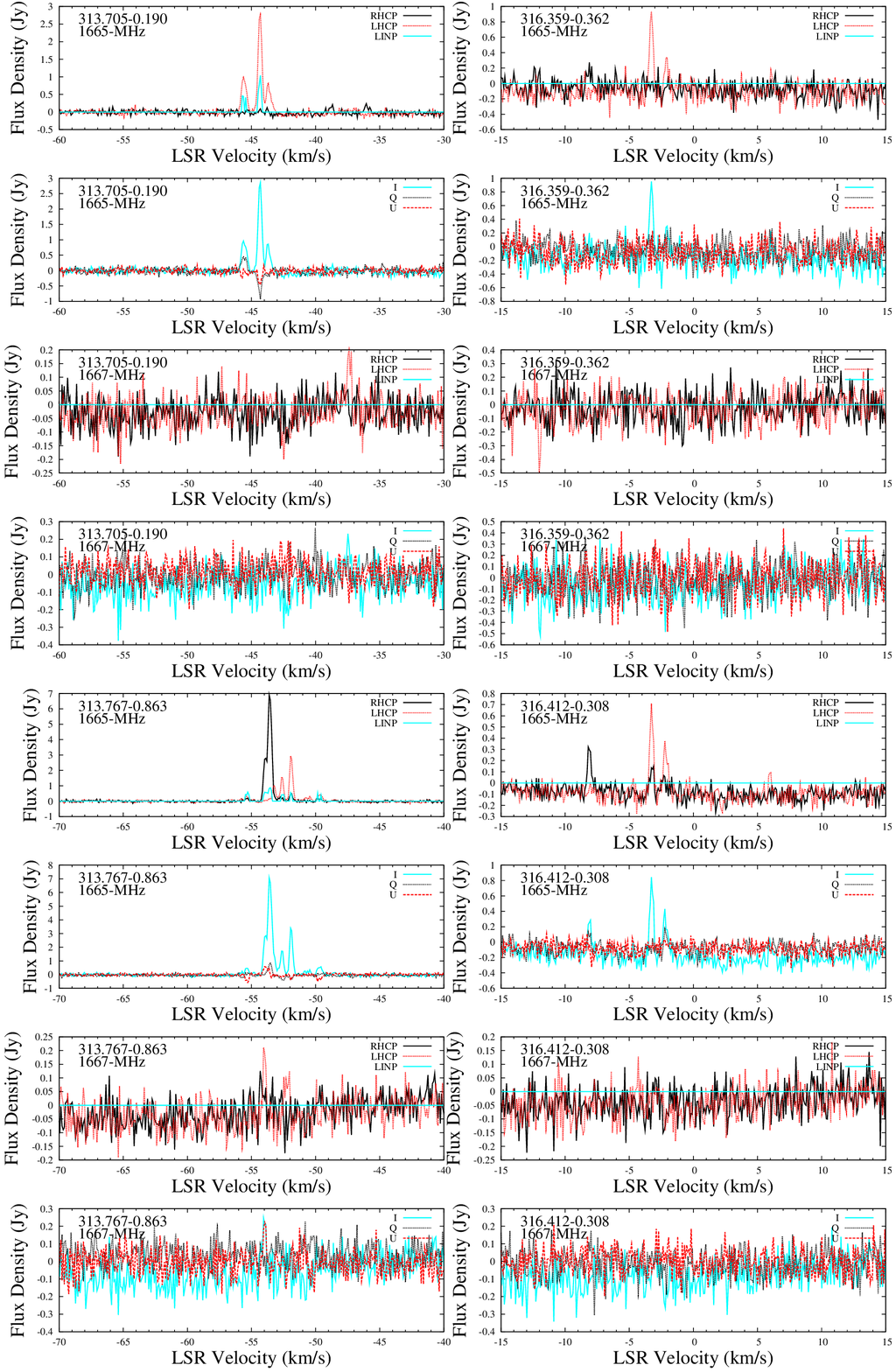}

\caption{\textit{- continued p9 of 35}}

\label{fig1p9}

\end{figure*}

\begin{figure*}
 \centering

\addtocounter{figure}{-1}

\includegraphics[width=15.5cm]{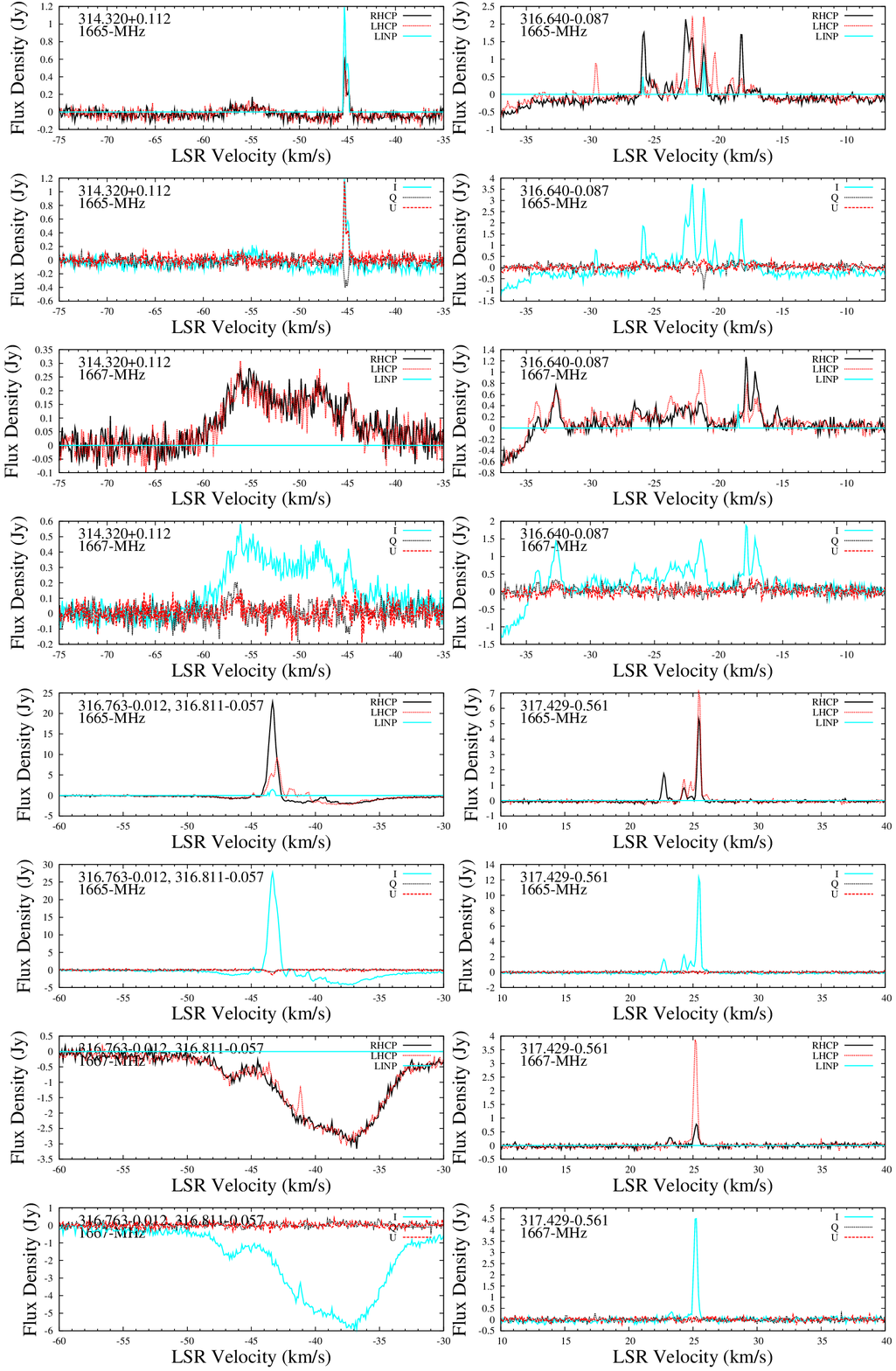}

\caption{\textit{- continued p10 of 35}}

\label{fig1p10}

\end{figure*}

\begin{figure*}
 \centering

\addtocounter{figure}{-1}

\includegraphics[width=15.5cm]{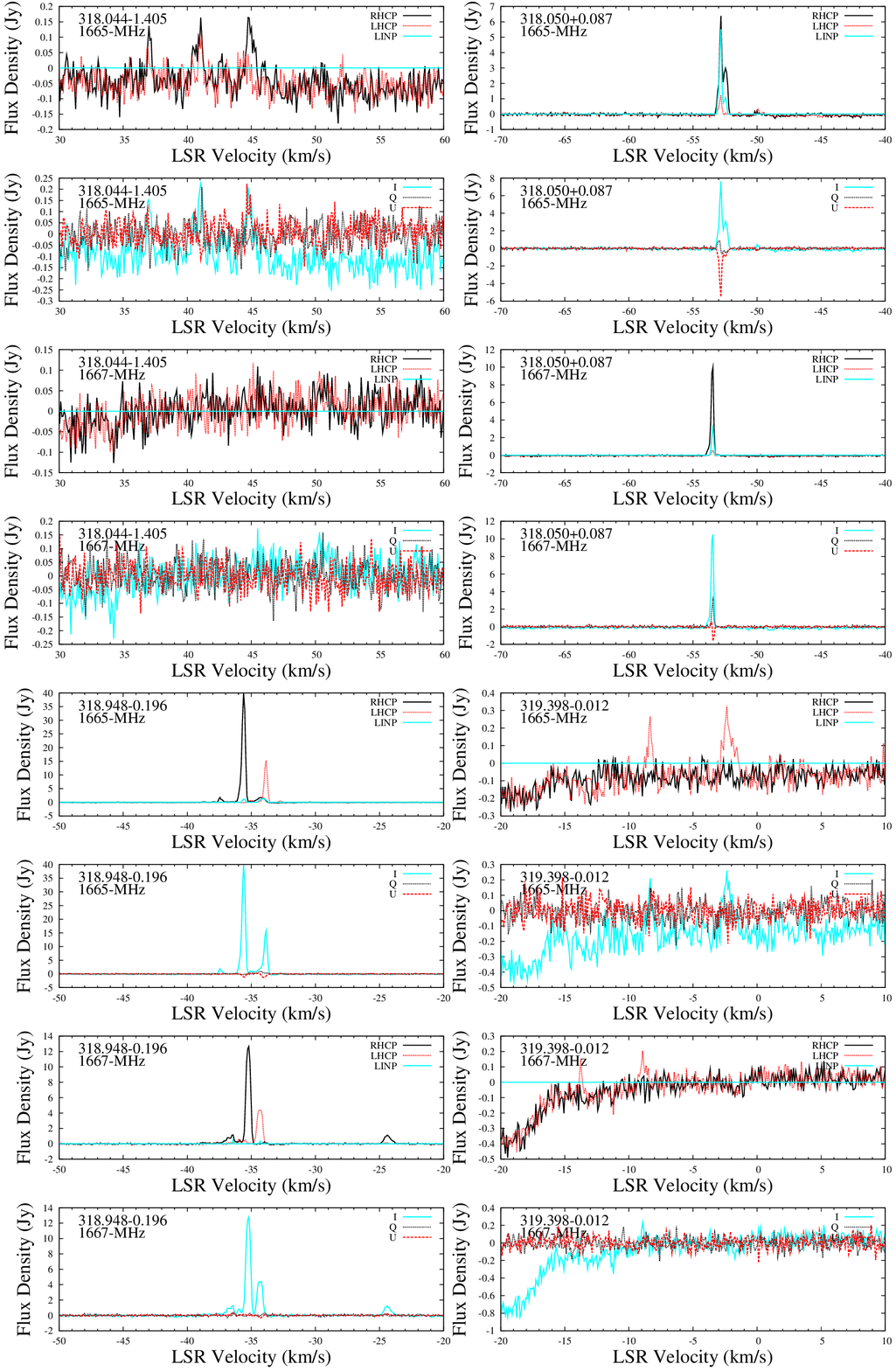}

\caption{\textit{- continued p11 of 35}}

\label{fig1p11}

\end{figure*}

\begin{figure*}
 \centering

\addtocounter{figure}{-1}

\includegraphics[width=15.5cm]{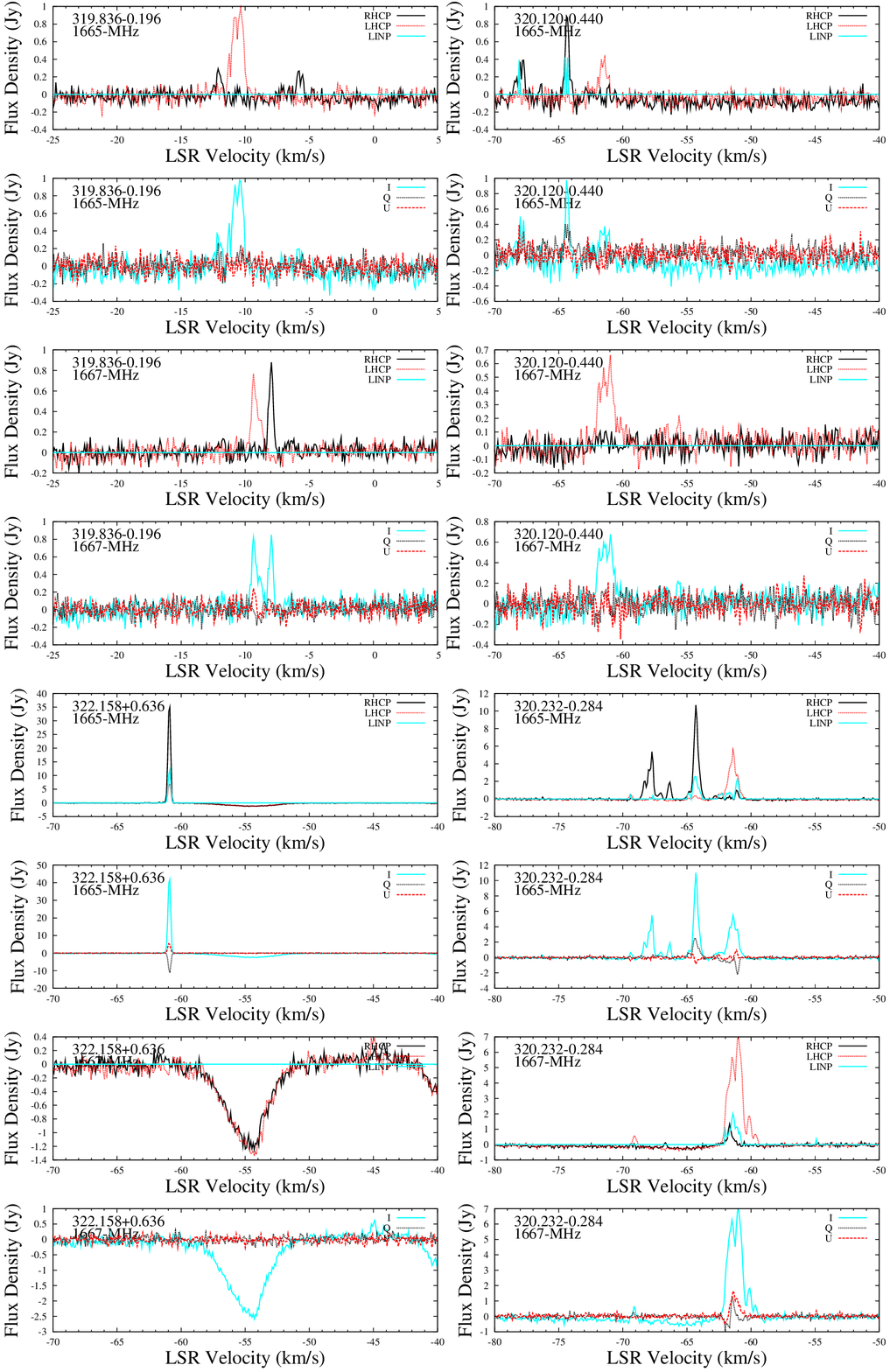}

\caption{\textit{- continued p12 of 35}}

\label{fig1p12}

\end{figure*}

\begin{figure*}
 \centering

\addtocounter{figure}{-1}

\includegraphics[width=15.5cm]{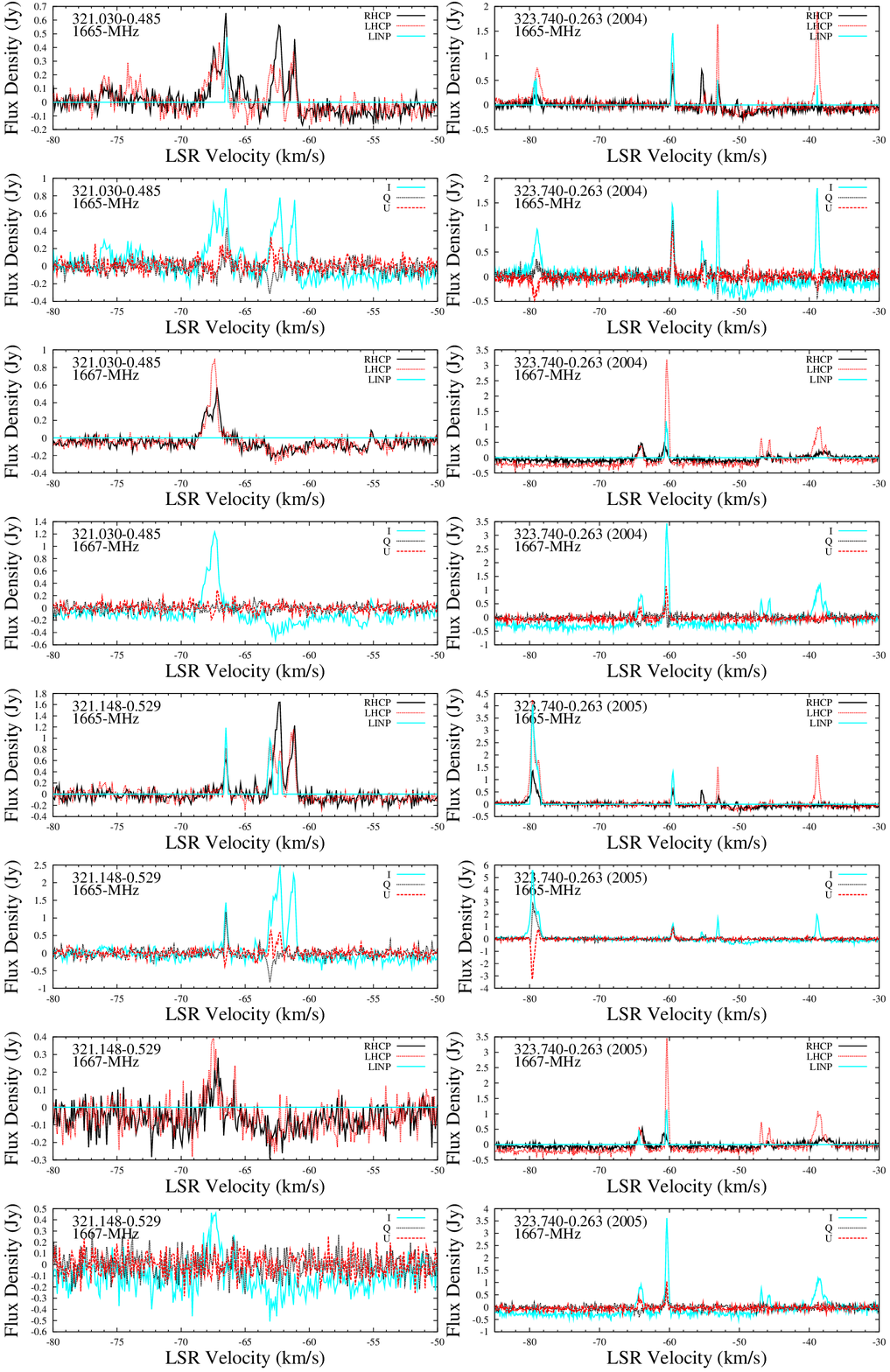}

\caption{\textit{- continued p13 of 35}}

\label{fig1p13}

\end{figure*}

\begin{figure*}
 \centering

\addtocounter{figure}{-1}

\includegraphics[width=15.5cm]{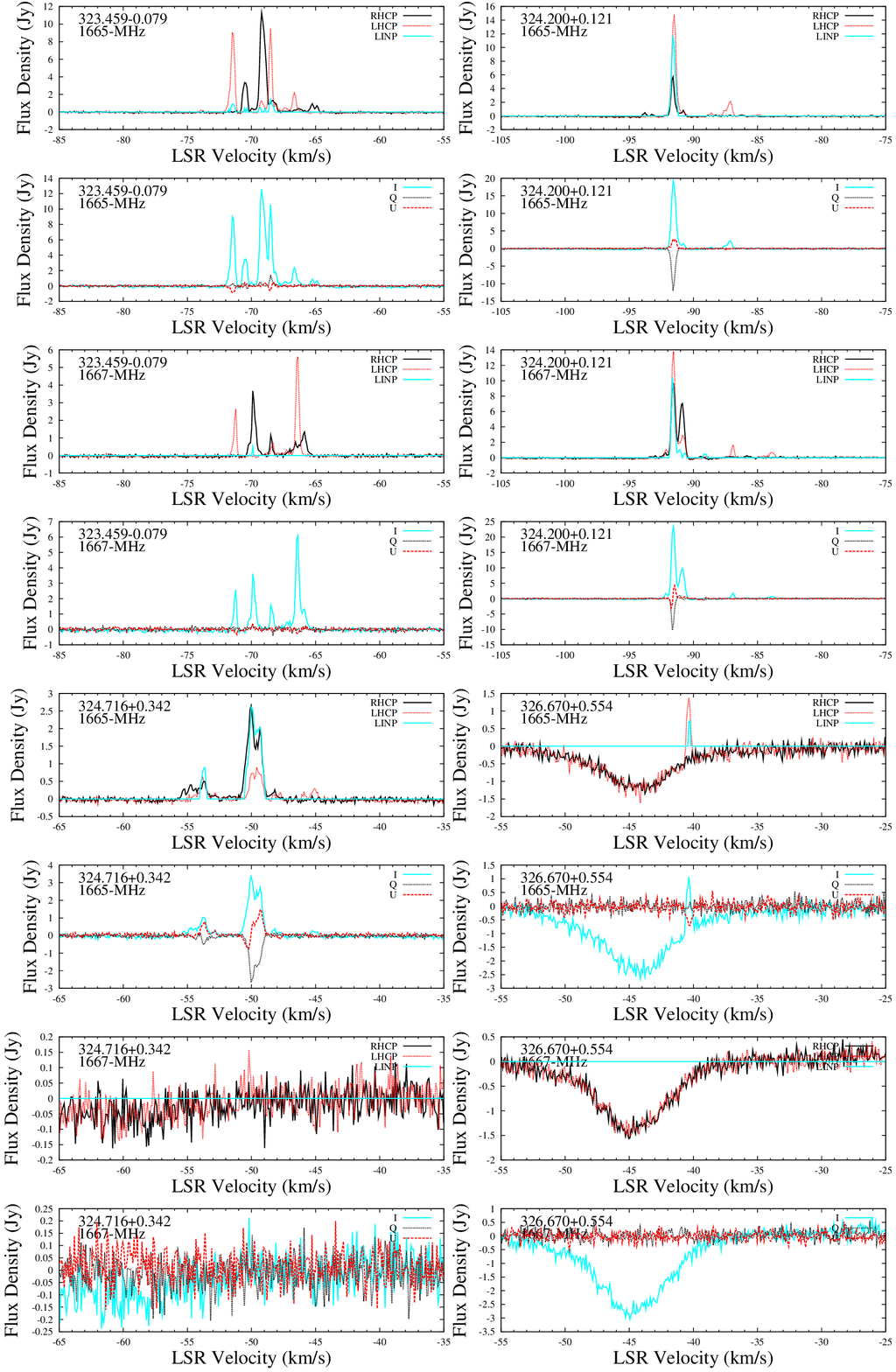}

\caption{\textit{- continued p14 of 35}}

\label{fig1p14}

\end{figure*}

\begin{figure*}
 \centering

\addtocounter{figure}{-1}

\includegraphics[width=15.5cm]{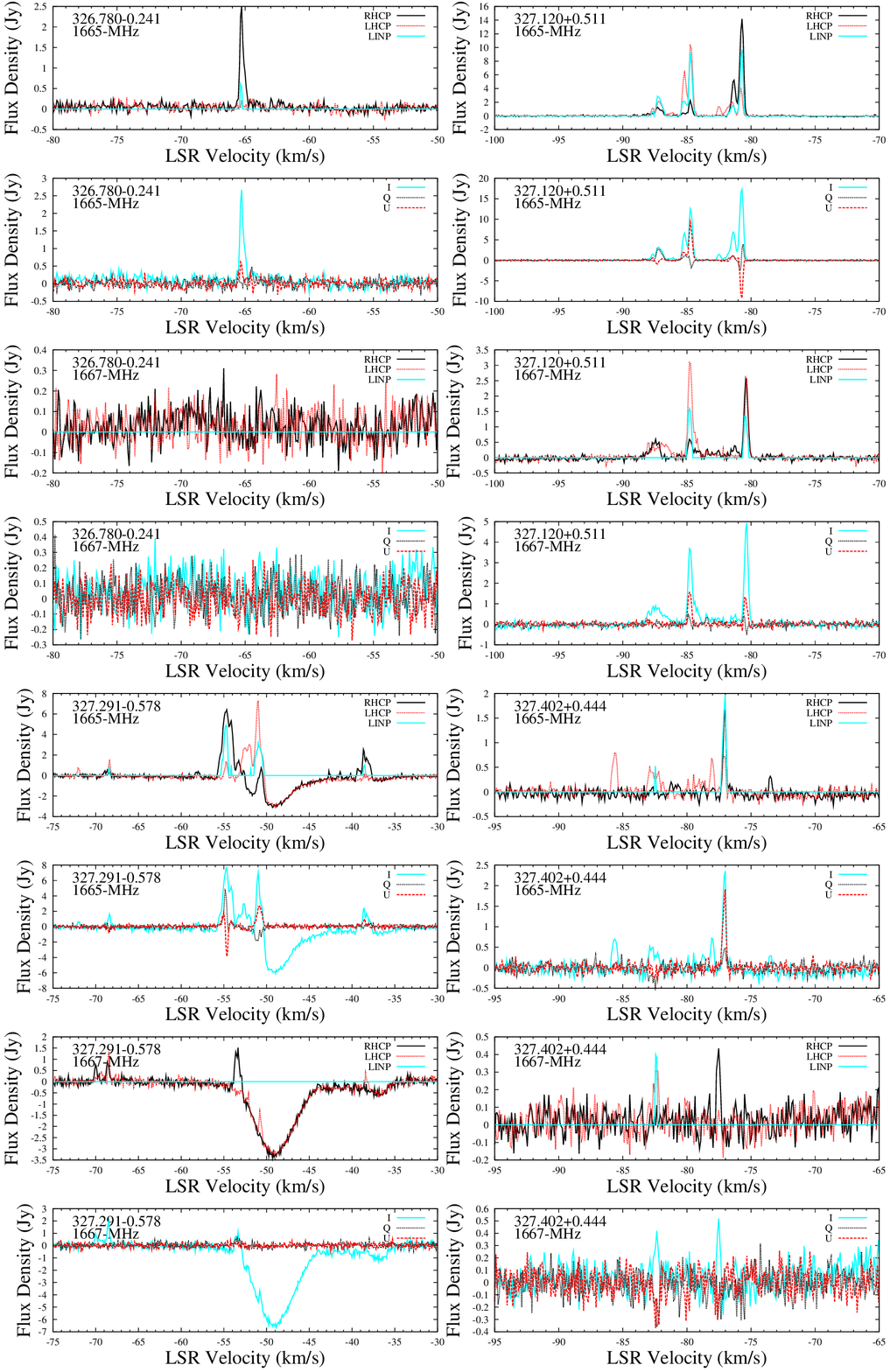}

\caption{\textit{- continued p15 of 35}}

\label{fig1p15}

\end{figure*}

\begin{figure*}
 \centering

\addtocounter{figure}{-1}

\includegraphics[width=15.5cm]{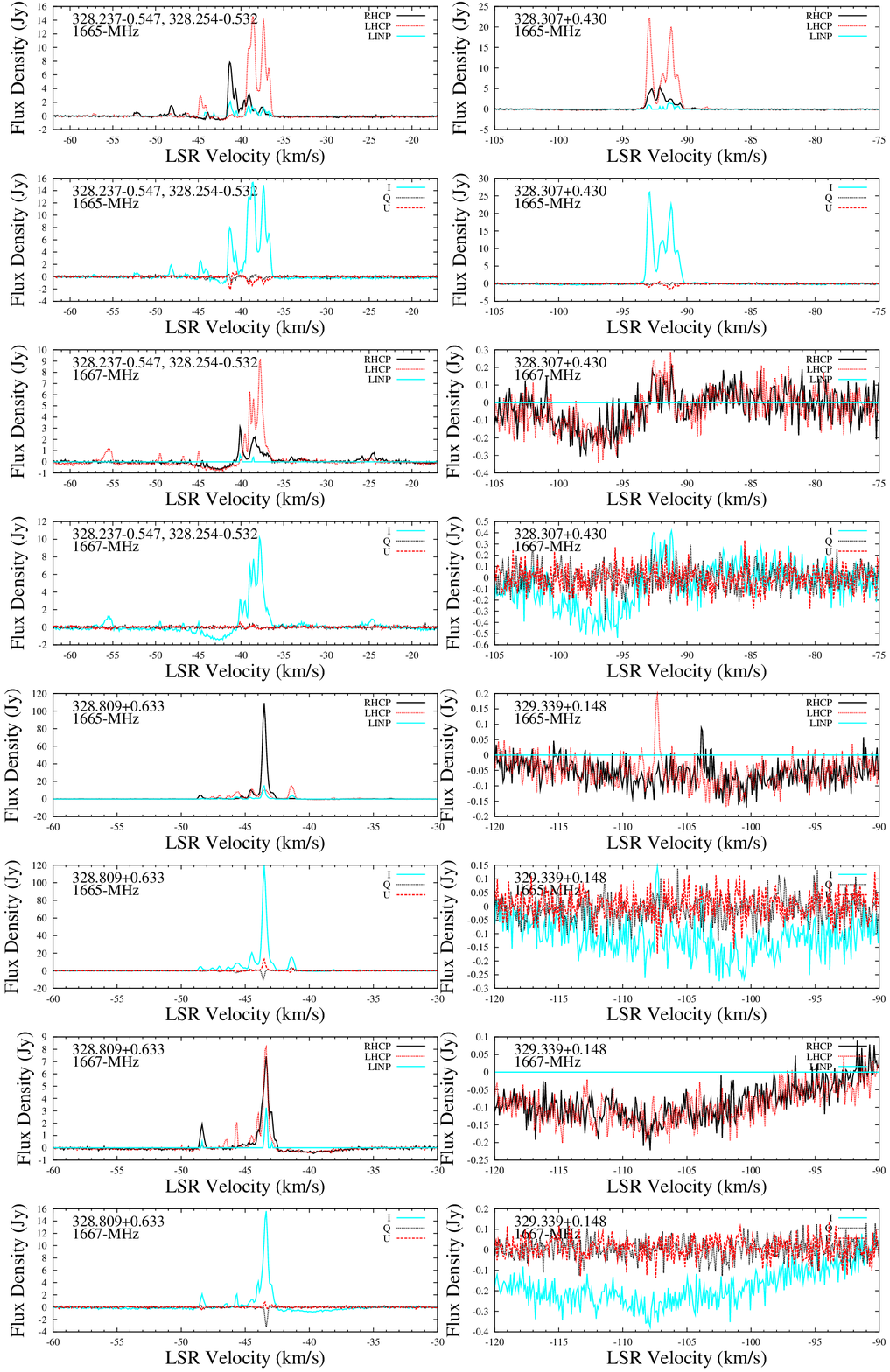}

\caption{\textit{- continued p16 of 35}}

\label{fig1p16}

\end{figure*}

\begin{figure*}
 \centering

\addtocounter{figure}{-1}

\includegraphics[width=15.5cm]{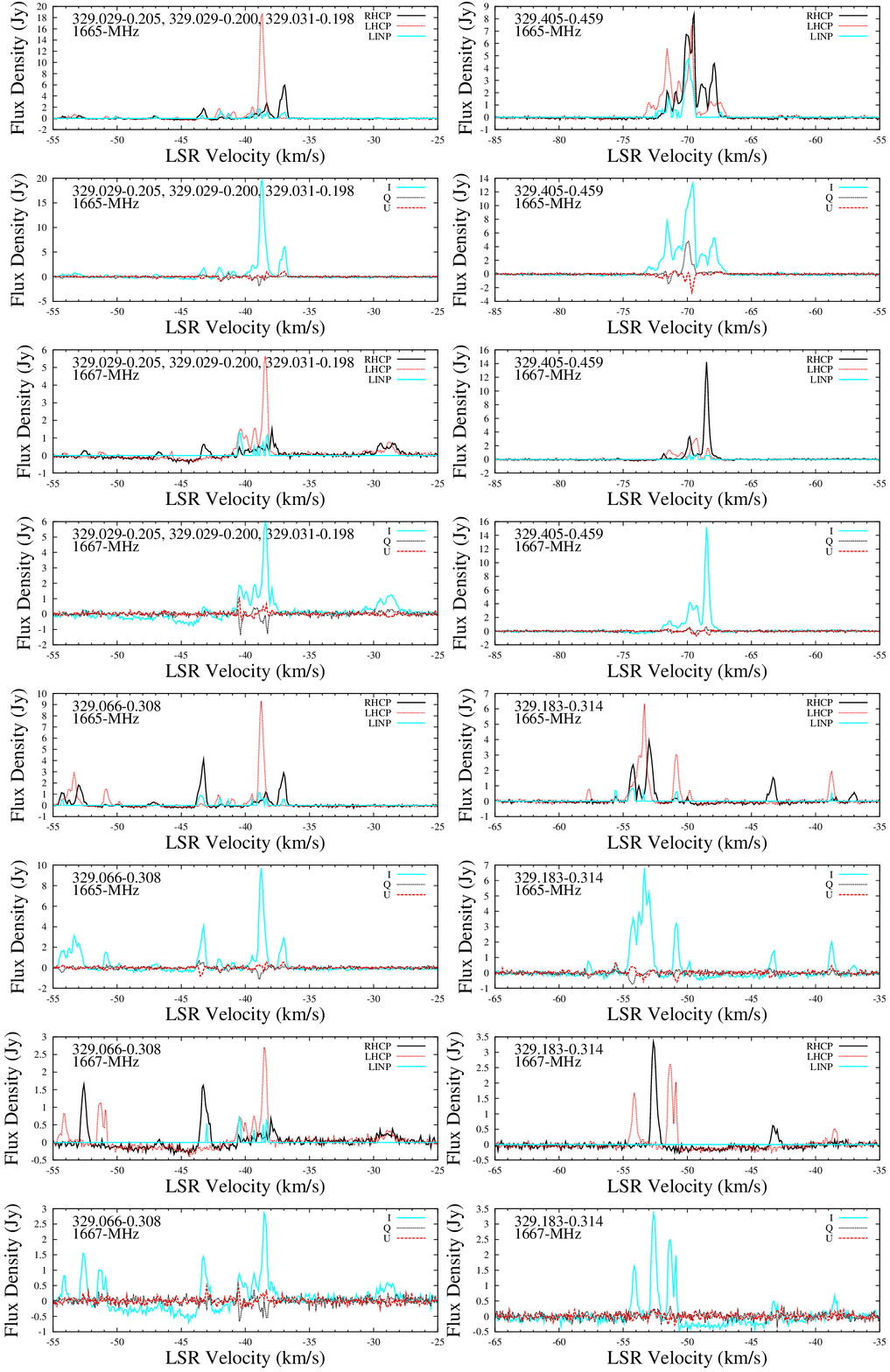}

\caption{\textit{- continued p17 of 35}}

\label{fig1p17}

\end{figure*}

%******break up the figs

\clearpage

\clearpage

\begin{figure*}
 \centering

\addtocounter{figure}{-1}

\includegraphics[width=15.5cm]{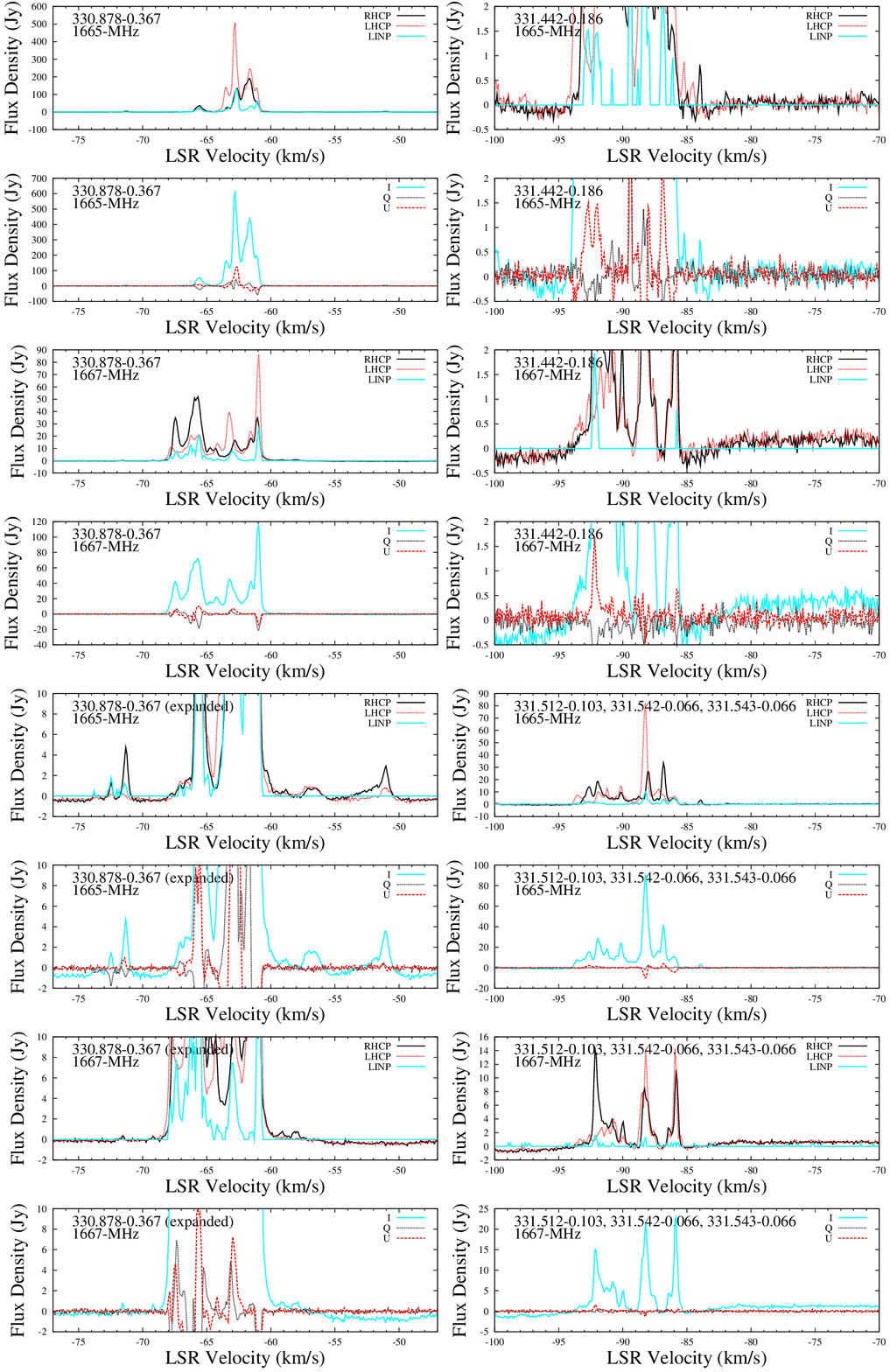}

\caption{\textit{- continued p18 of 35}}

\label{fig1p18}

\end{figure*}

\begin{figure*}
 \centering

\addtocounter{figure}{-1}

\includegraphics[width=15.5cm]{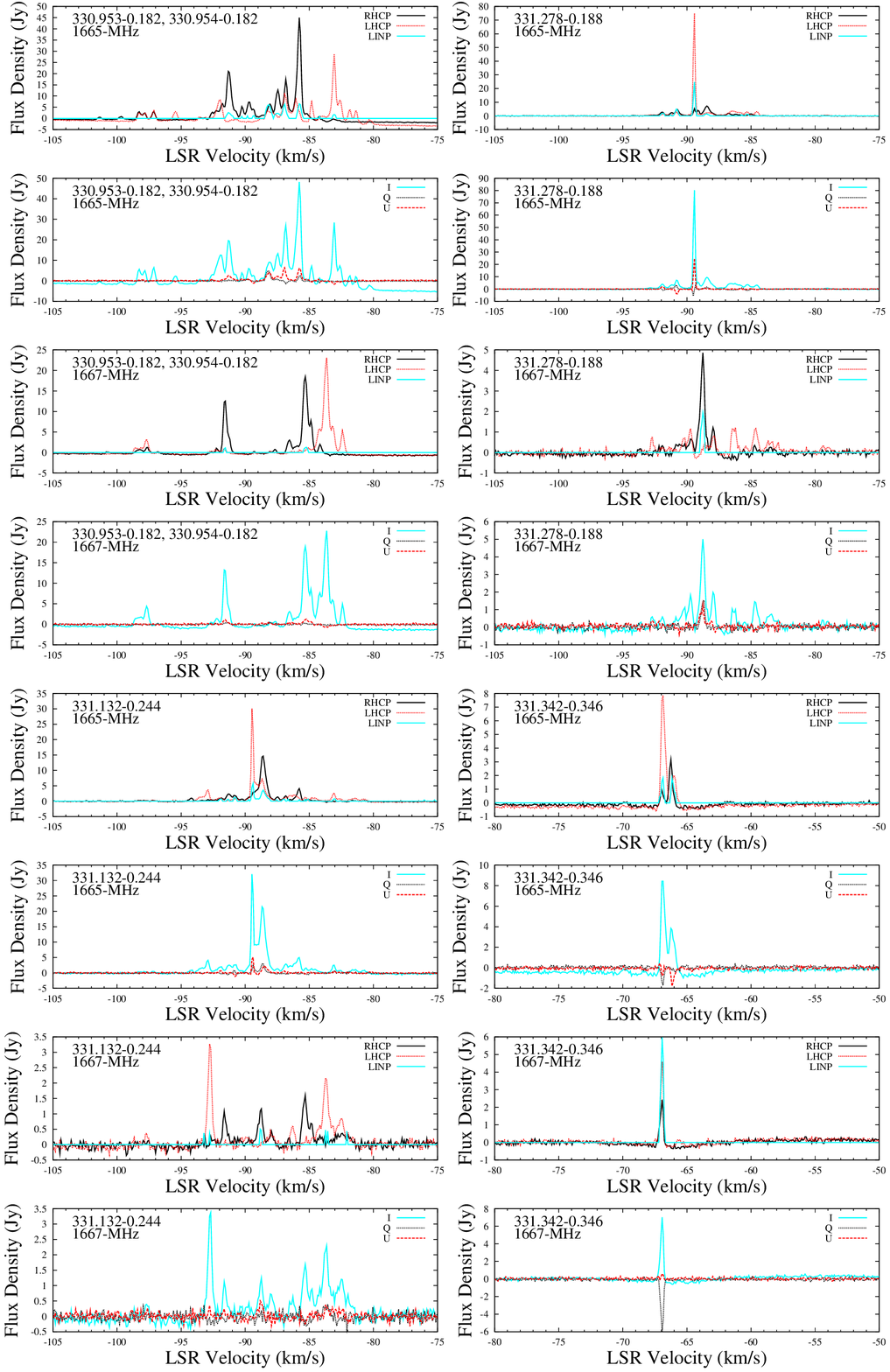}

\caption{\textit{- continued p19 of 35}}

\label{fig1p19}

\end{figure*}

\begin{figure*}
 \centering

\addtocounter{figure}{-1}

\includegraphics[width=15.5cm]{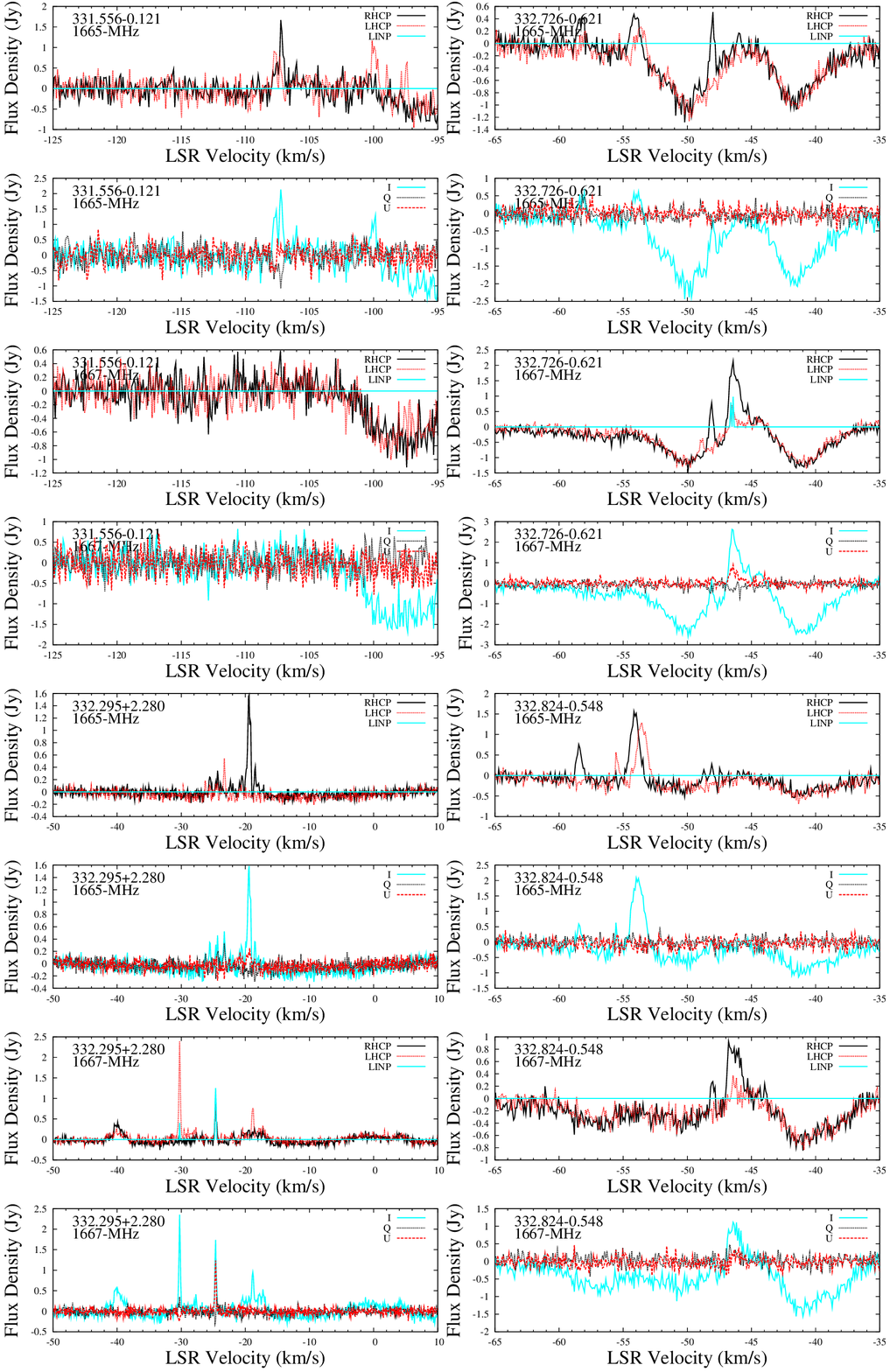}

\caption{\textit{- continued p20 of 35}}

\label{fig1p20}

\end{figure*}

\begin{figure*}
 \centering

\addtocounter{figure}{-1}

\includegraphics[width=15.5cm]{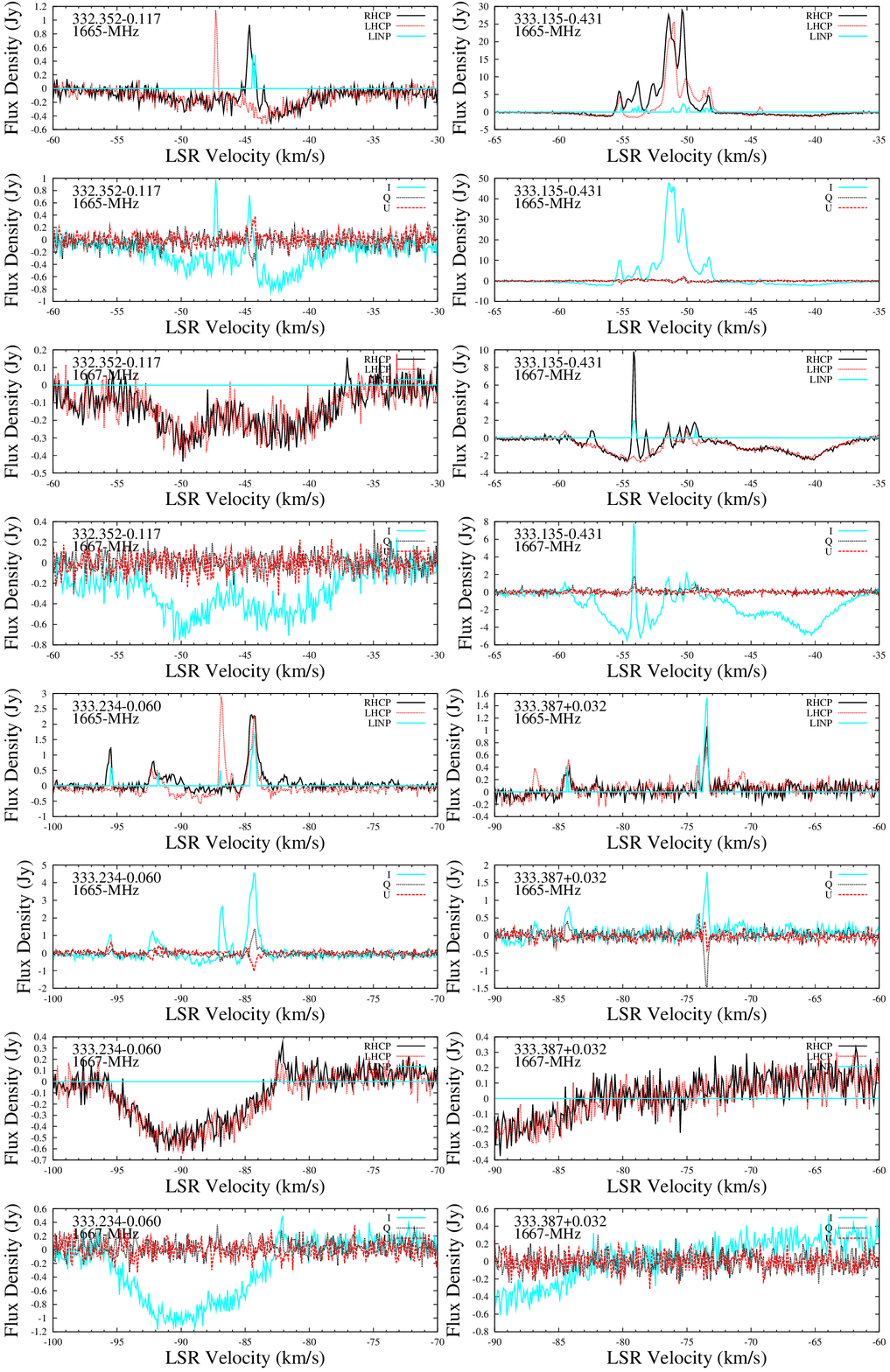}

\caption{\textit{- continued p21 of 35}}

\label{fig1p21}

\end{figure*}

\begin{figure*}
 \centering

\addtocounter{figure}{-1}

\includegraphics[width=15.5cm]{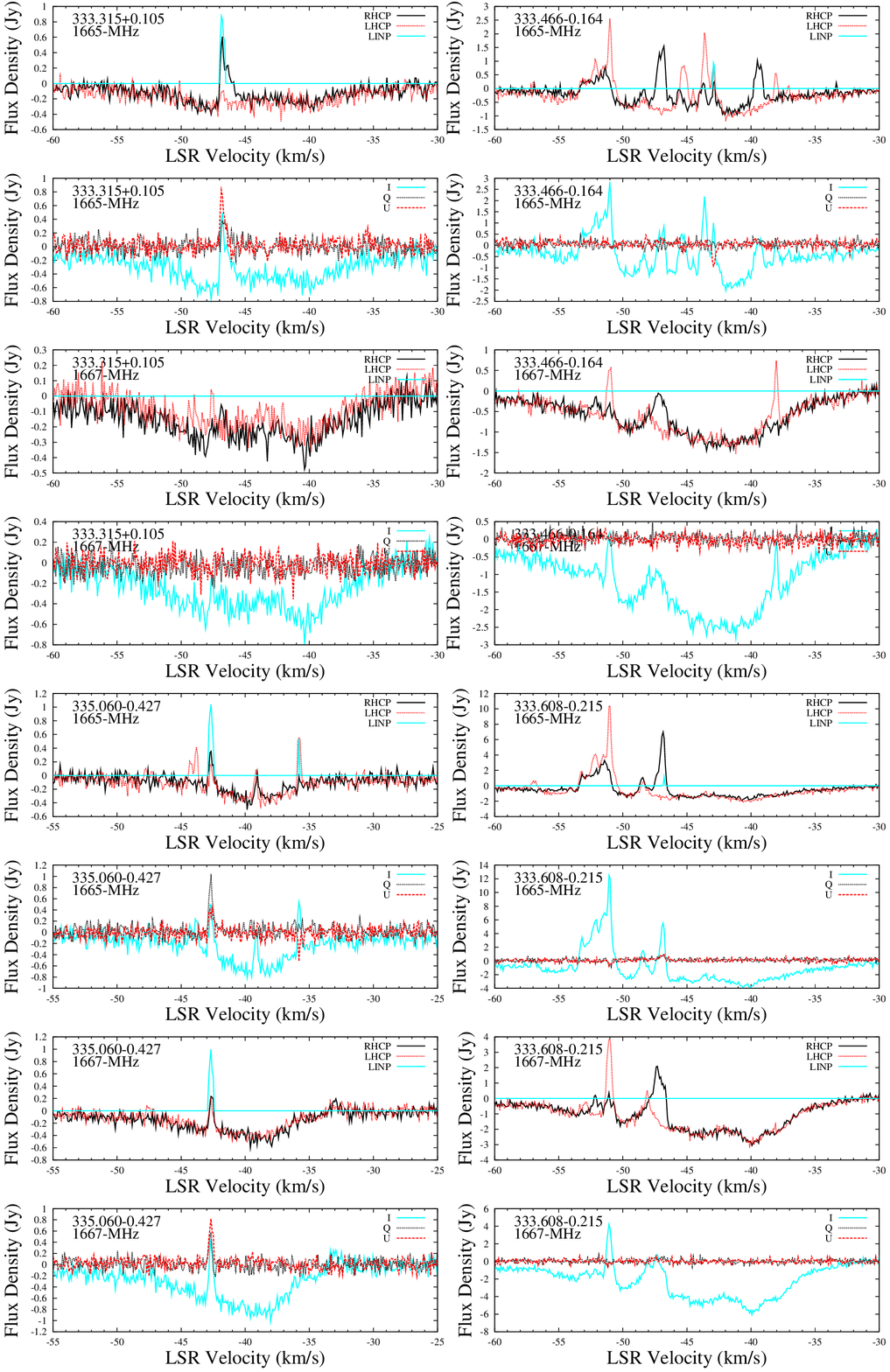}

\caption{\textit{- continued p22 of 35}}

\label{fig1p22}

\end{figure*}

\begin{figure*}
 \centering

\addtocounter{figure}{-1}

\includegraphics[width=15.5cm]{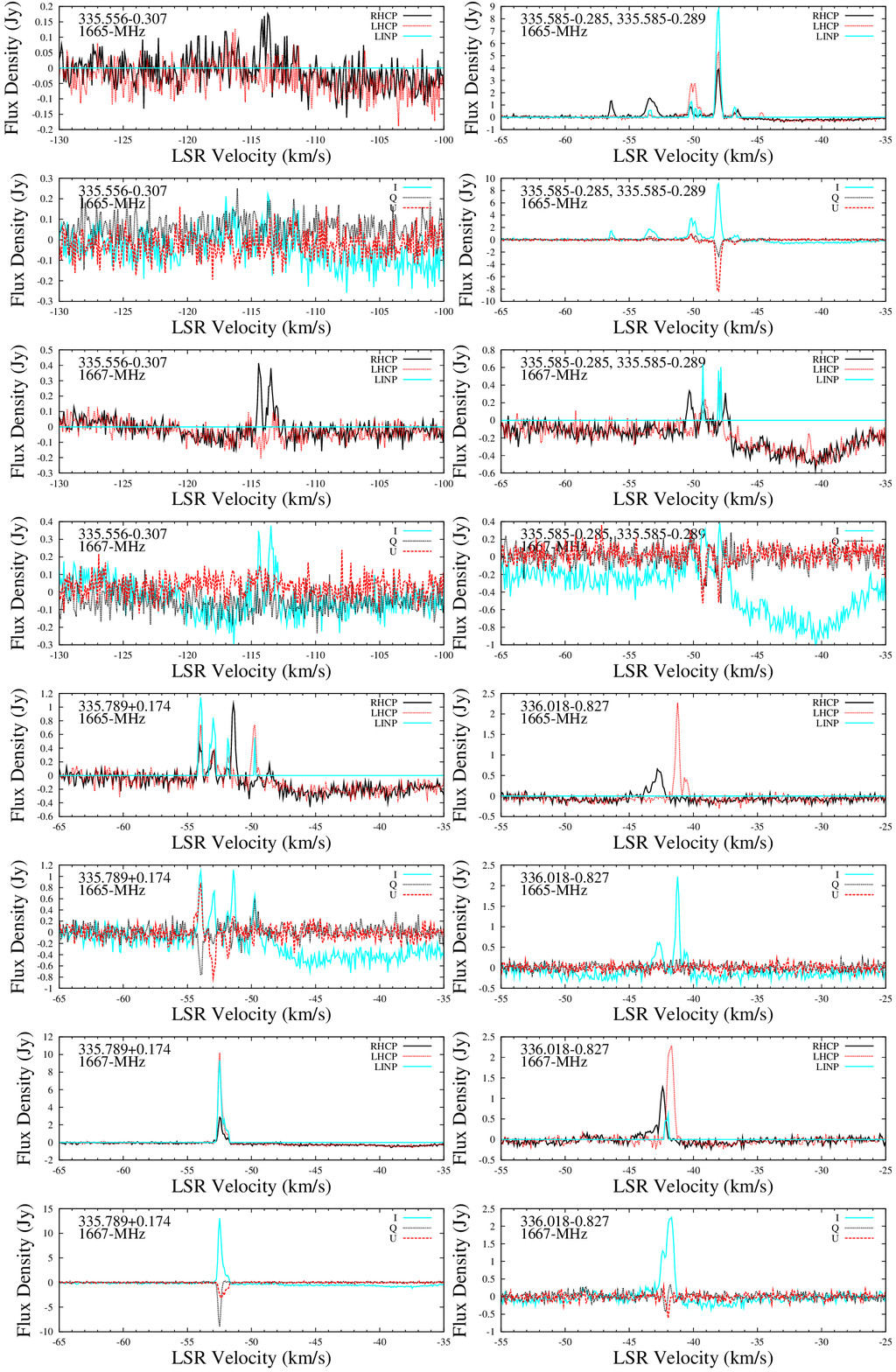}

\caption{\textit{- continued p23 of 35}}

\label{fig1p23}

\end{figure*}

\begin{figure*}
 \centering

\addtocounter{figure}{-1}

\includegraphics[width=15.5cm]{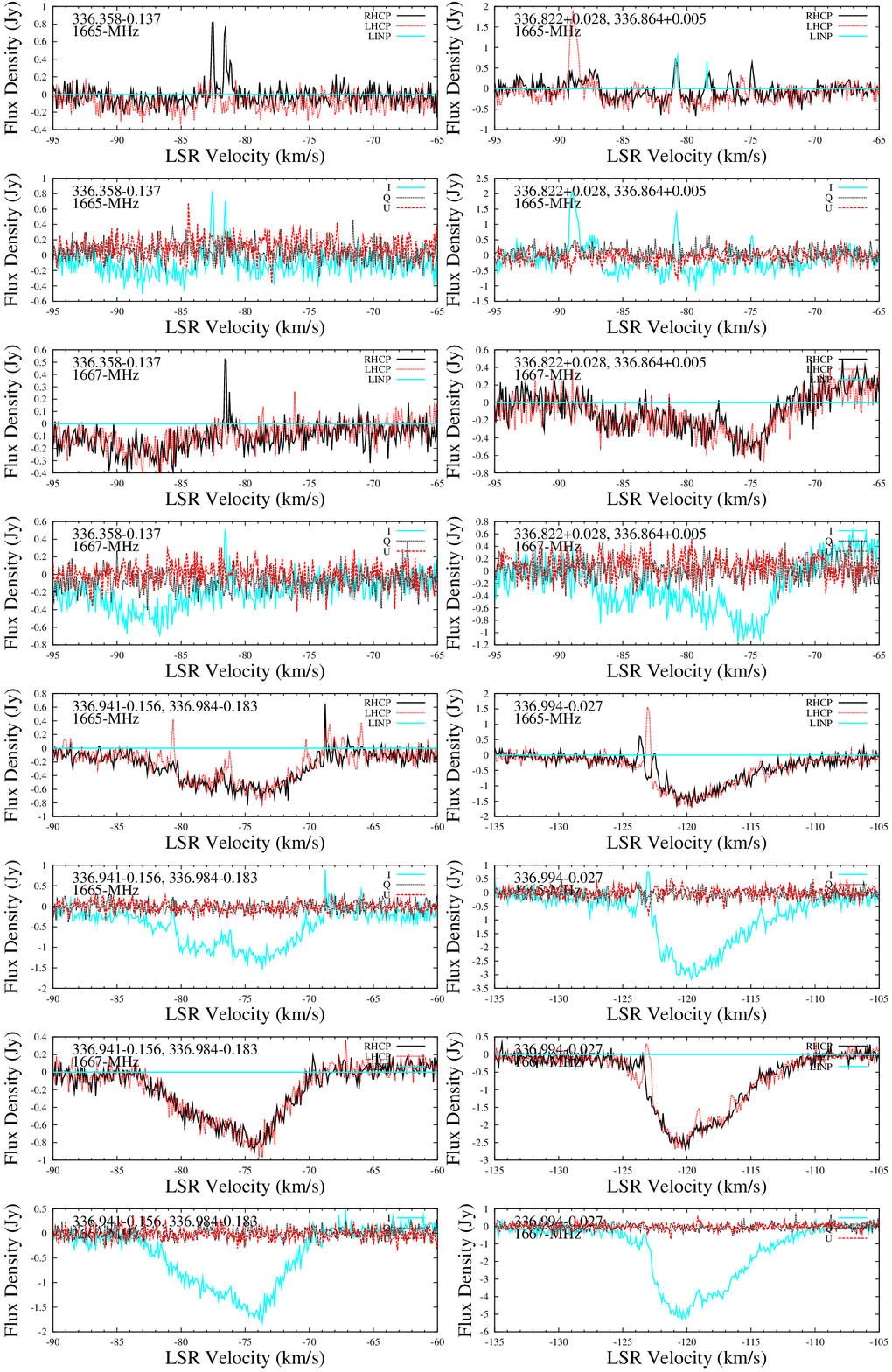}

\caption{\textit{- continued p24 of 35}}

\label{fig1p24}

\end{figure*}

\begin{figure*}
 \centering

\addtocounter{figure}{-1}

\includegraphics[width=15.5cm]{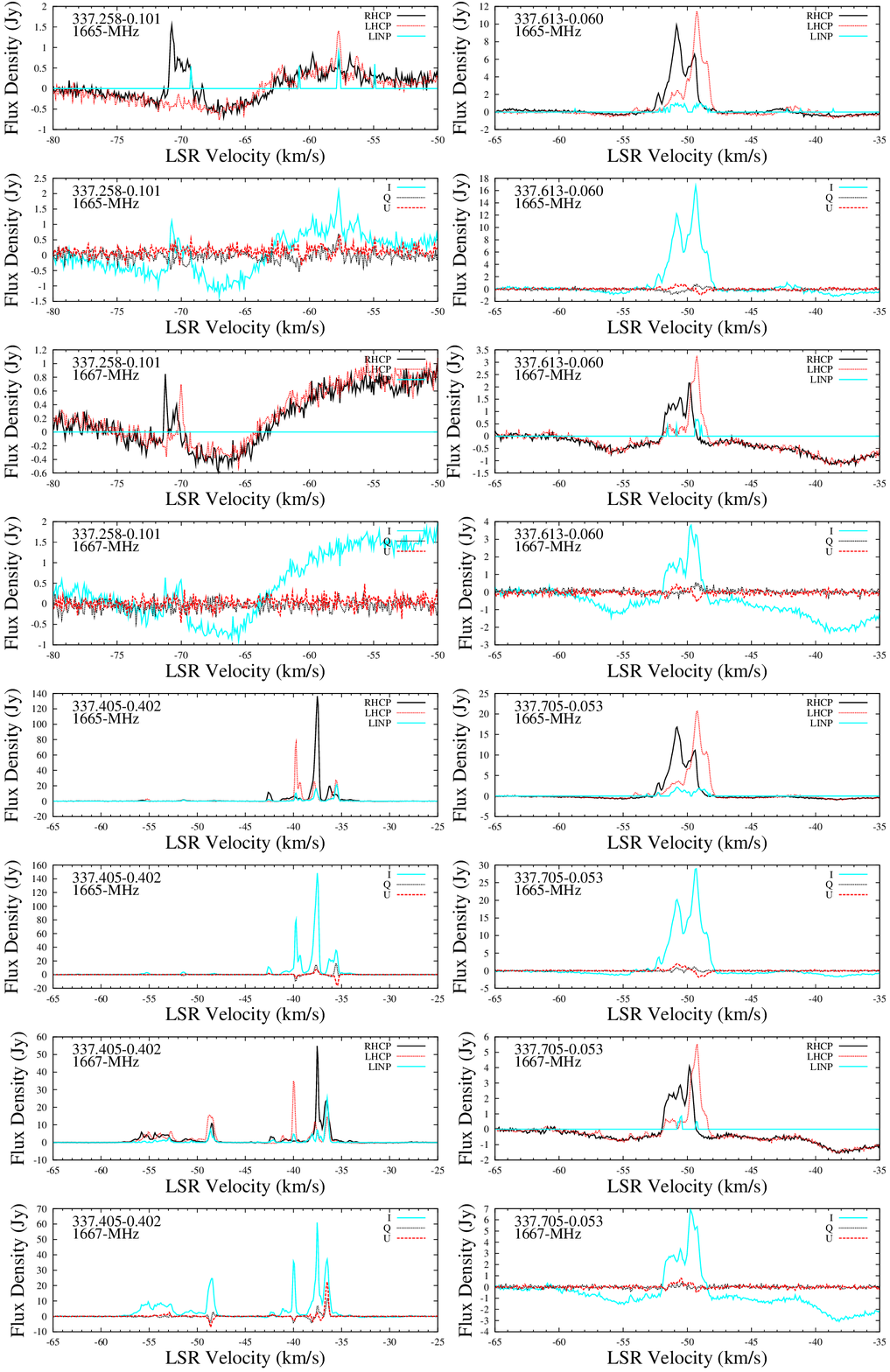}

\caption{\textit{- continued p25 of 35}}

\label{fig1p25}

\end{figure*}

\begin{figure*}
 \centering

\addtocounter{figure}{-1}

\includegraphics[width=15.5cm]{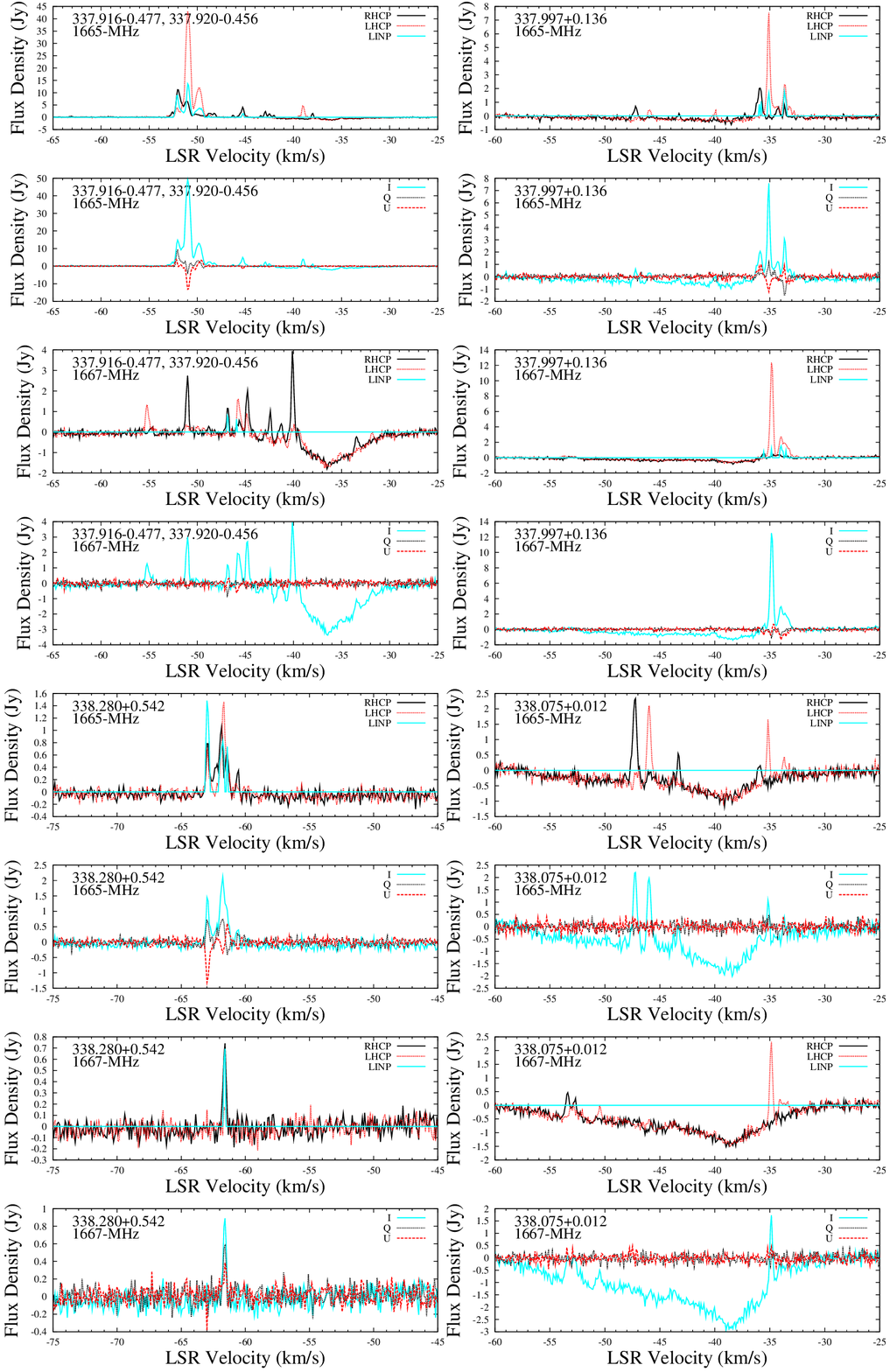}

\caption{\textit{- continued p26 of 35}}

\label{fig1p26}

\end{figure*}

\begin{figure*}
 \centering

\addtocounter{figure}{-1}

\includegraphics[width=15.5cm]{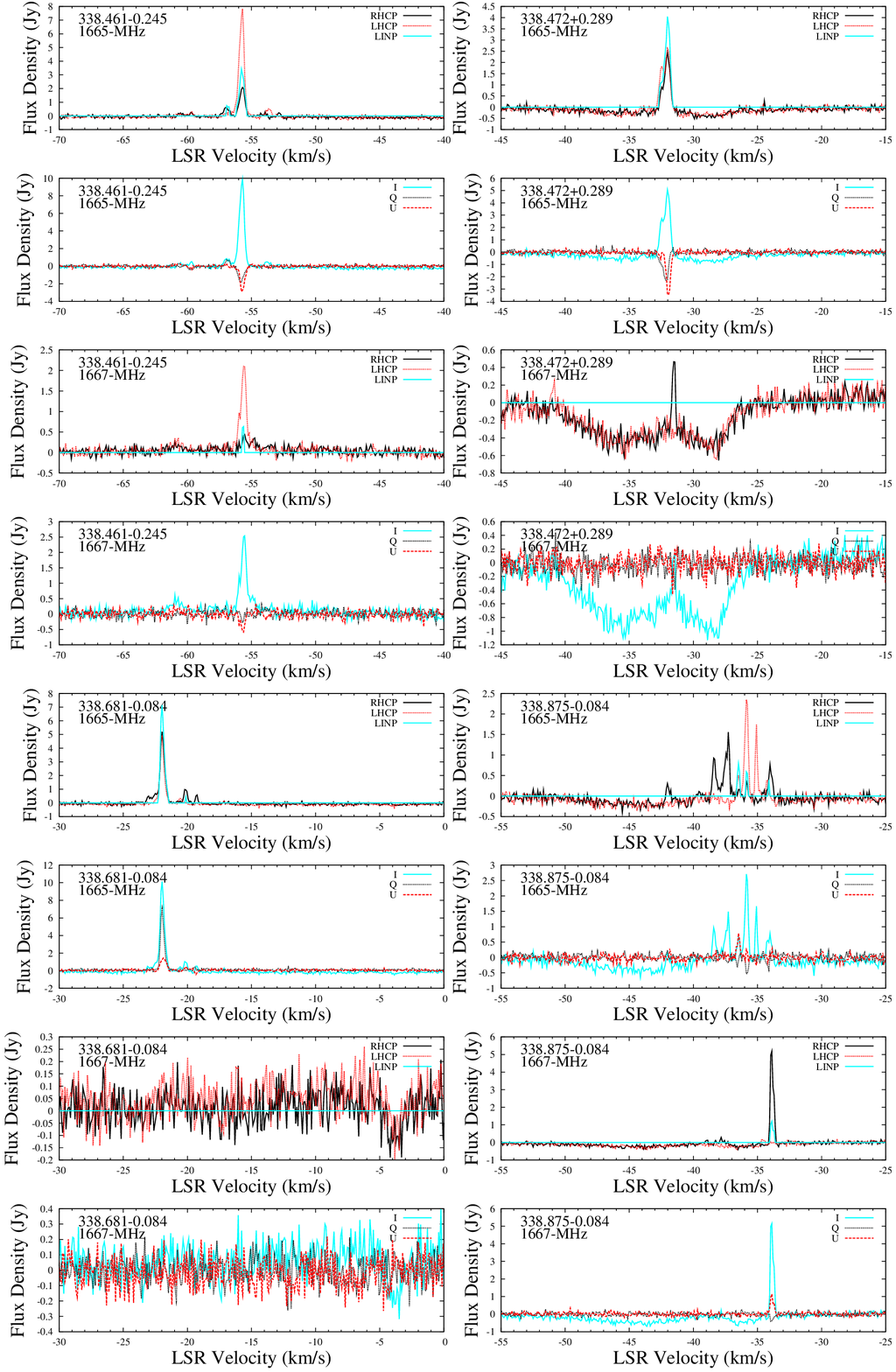}

\caption{\textit{- continued p26 of 35}}

\label{fig1p26}

\end{figure*}

\begin{figure*}
 \centering

\addtocounter{figure}{-1}

\includegraphics[width=15.5cm]{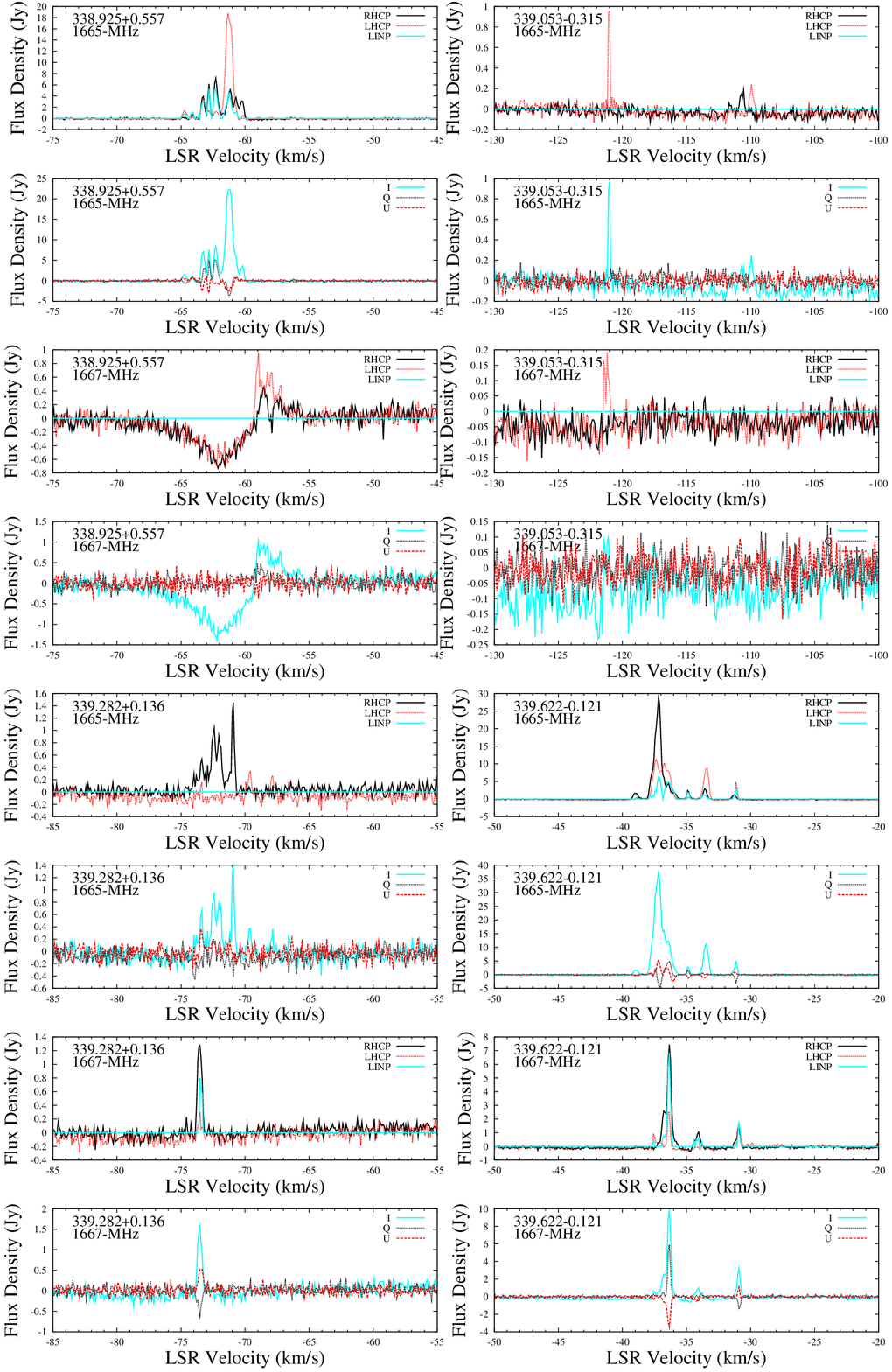}

\caption{\textit{- continued p28 of 35}}

\label{fig1p28}

\end{figure*}

\begin{figure*}
 \centering

\addtocounter{figure}{-1}

\includegraphics[width=15.5cm]{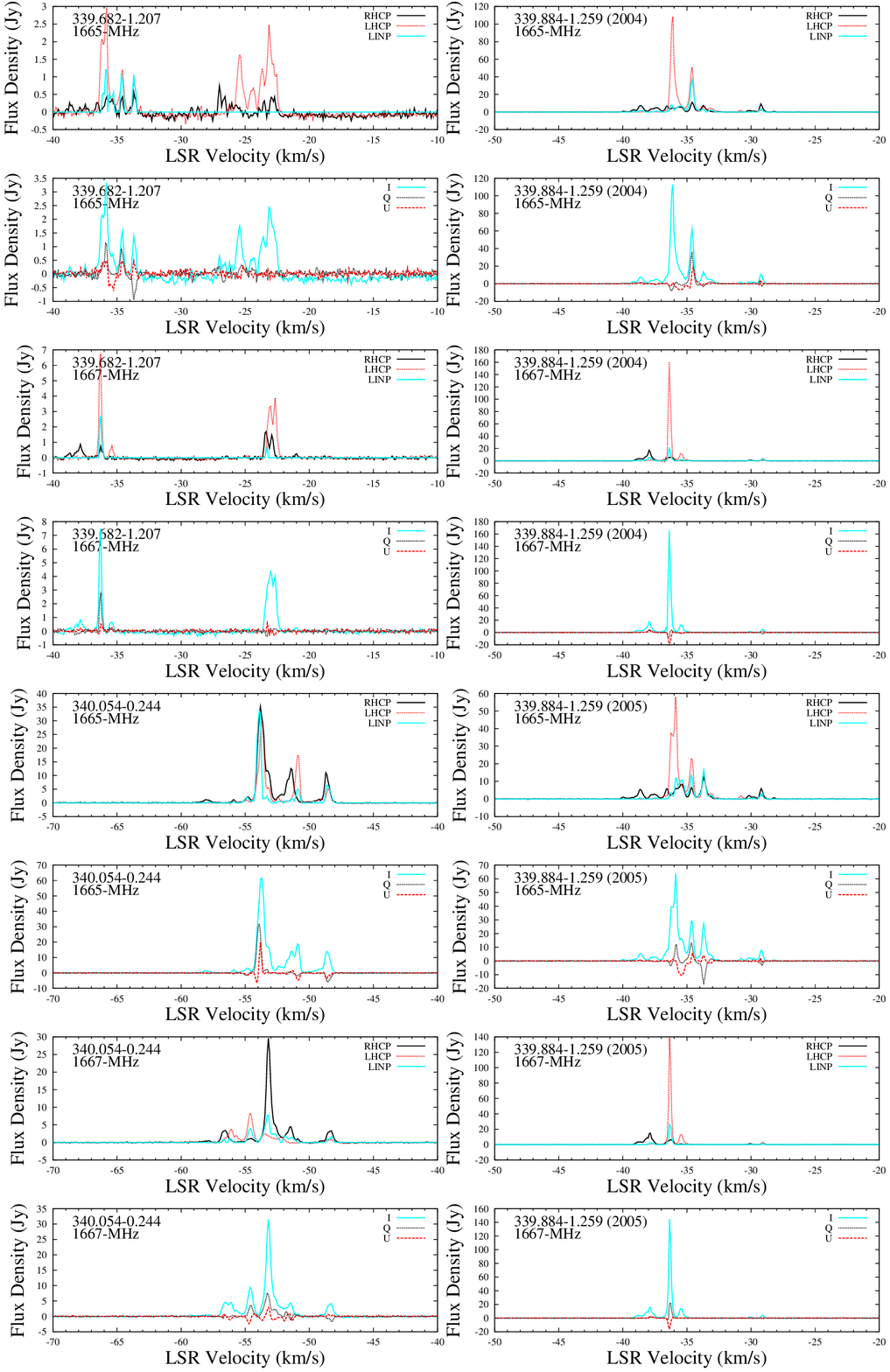}

\caption{\textit{- continued p29 of 35}}

\label{fig1p29}

\end{figure*}

\begin{figure*}
 \centering

\addtocounter{figure}{-1}

\includegraphics[width=15.5cm]{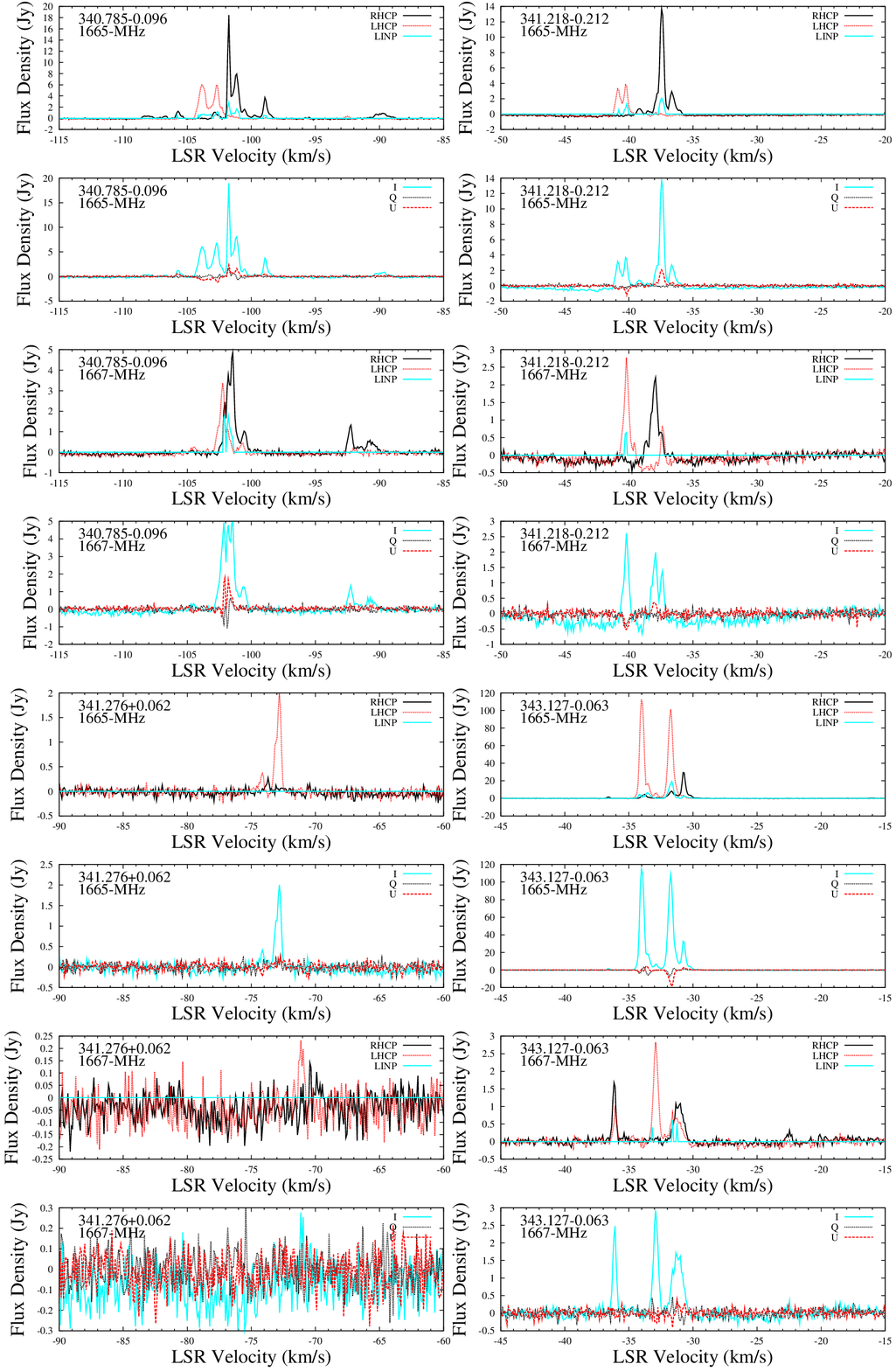}

\caption{\textit{- continued p30 of 35}}

\label{fig1p30}

\end{figure*}

\begin{figure*}
 \centering

\addtocounter{figure}{-1}

\includegraphics[width=15.5cm]{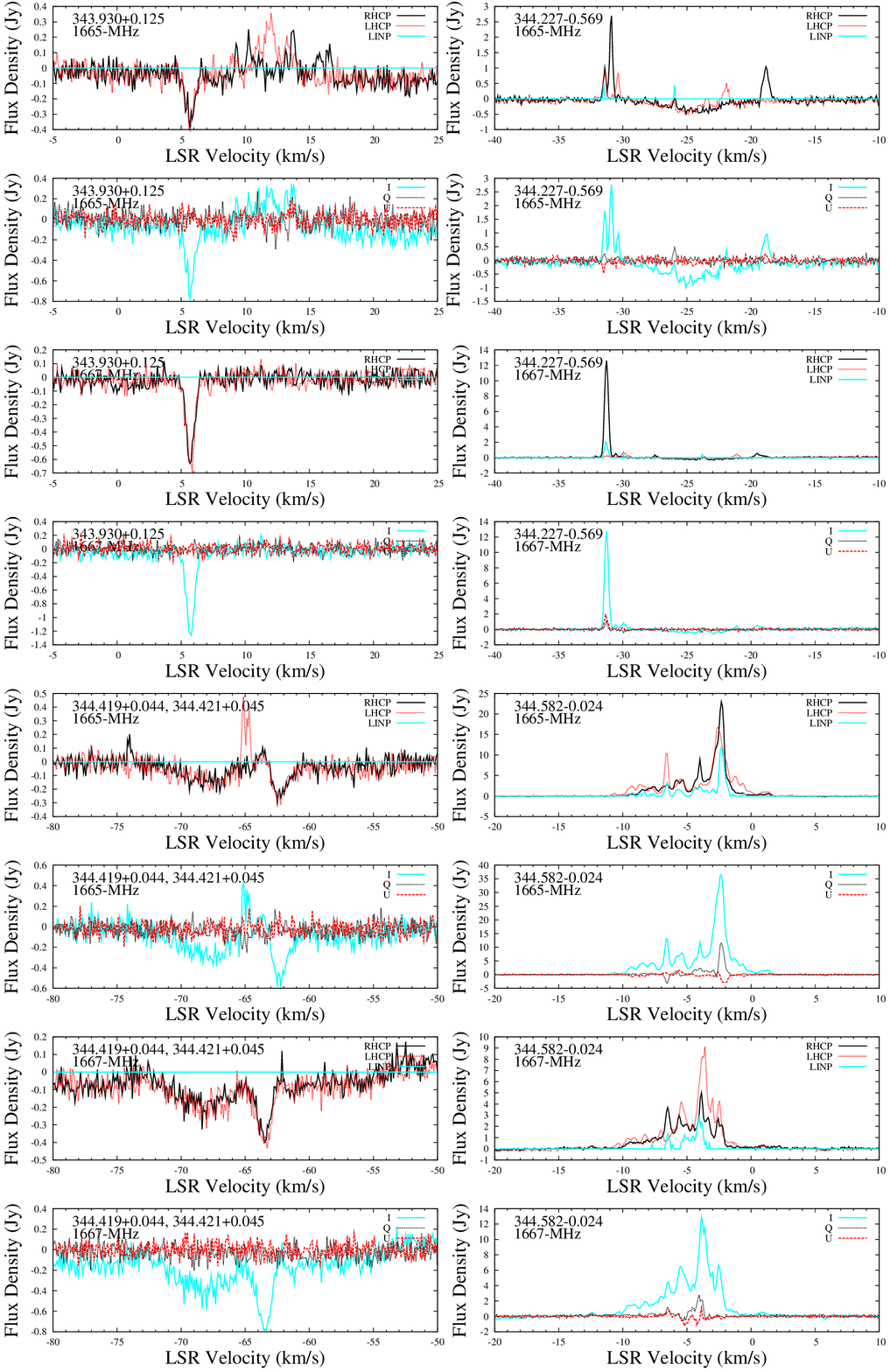}

\caption{\textit{- continued p31 of 35}}

\label{fig1p31}

\end{figure*}

\begin{figure*}
 \centering

\addtocounter{figure}{-1}

\includegraphics[width=15.5cm]{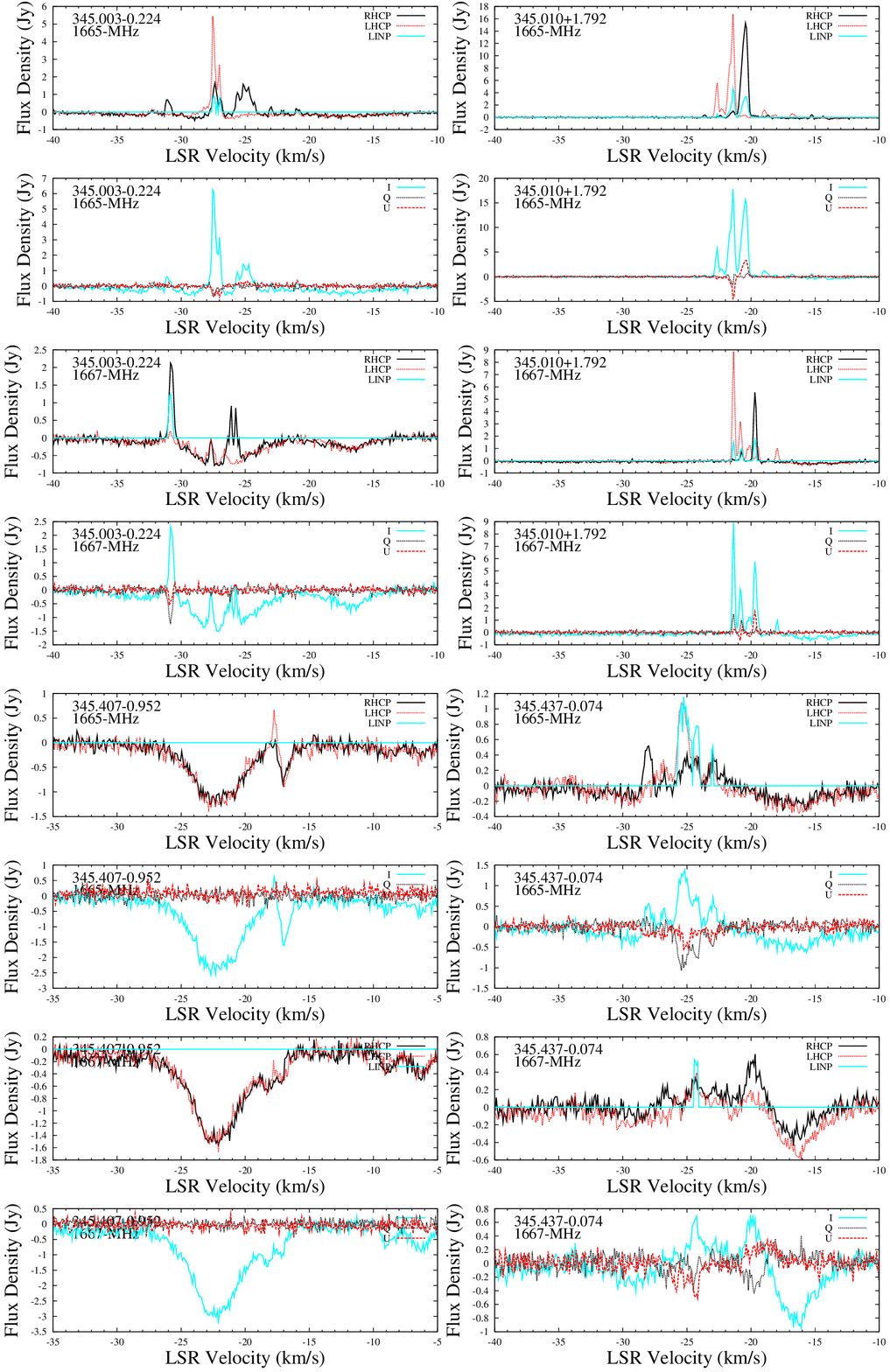}

\caption{\textit{- continued p32 of 35}}

\label{fig1p32}

\end{figure*}

\begin{figure*}
 \centering

\addtocounter{figure}{-1}

\includegraphics[width=15.5cm]{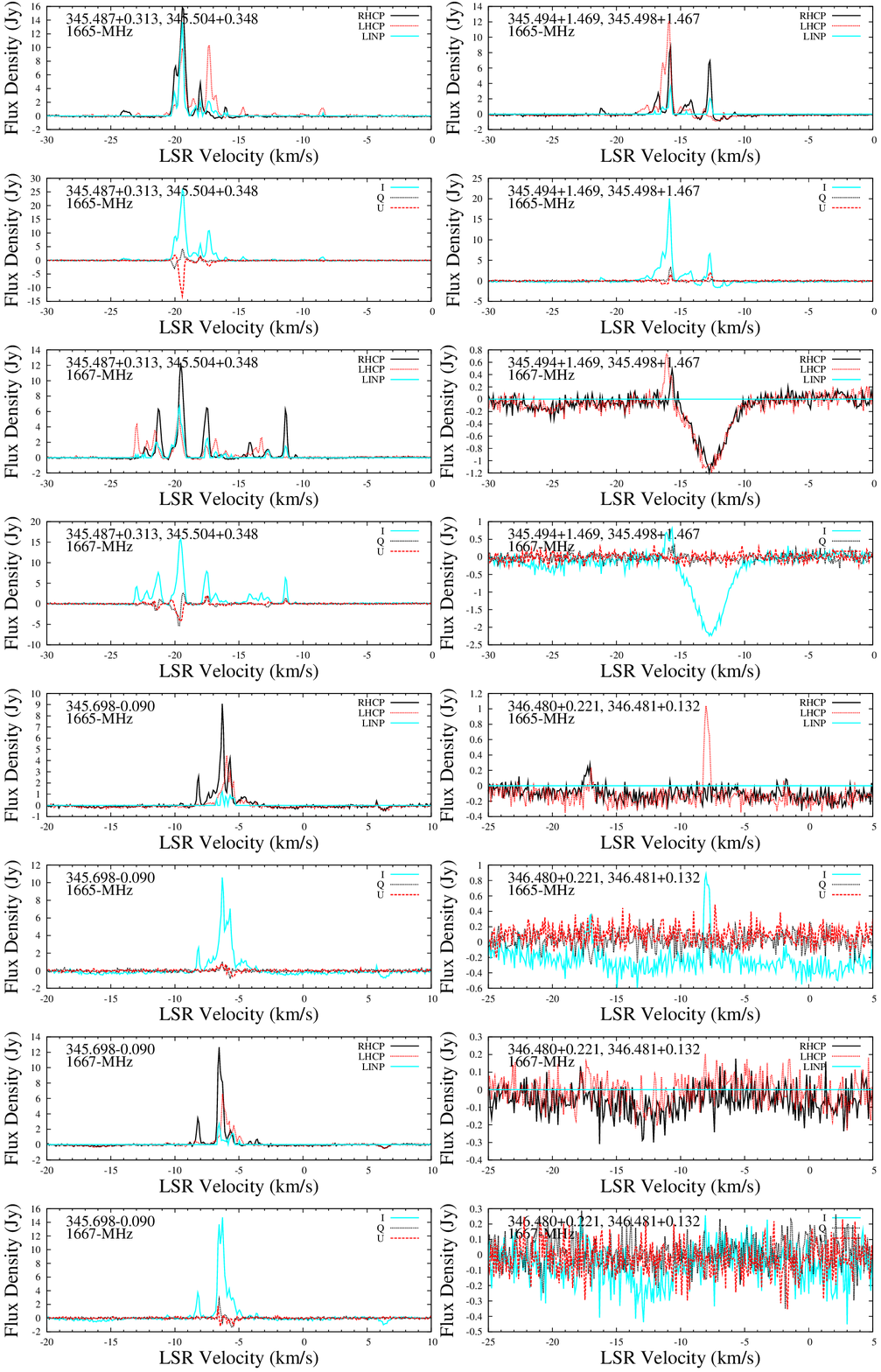}

\caption{\textit{- continued p33 of 35}}

\label{fig1p33}

\end{figure*}

\begin{figure*}
 \centering

\addtocounter{figure}{-1}

\includegraphics[width=15.5cm]{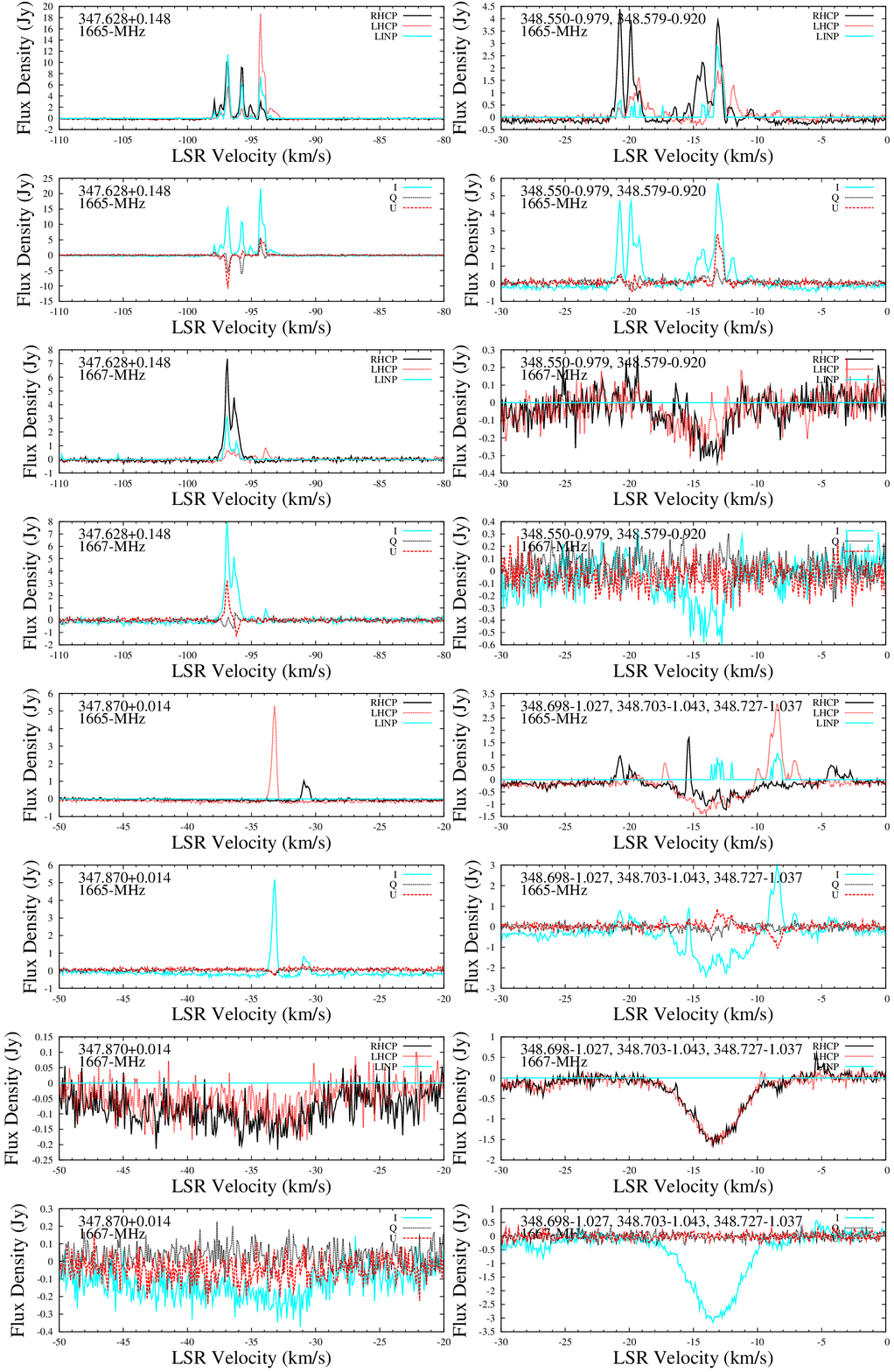}

\caption{\textit{- continued p34 of 35}}

\label{fig1p34}

\end{figure*}

\begin{figure*}
 \centering

\addtocounter{figure}{-1}

\includegraphics[width=15.5cm]{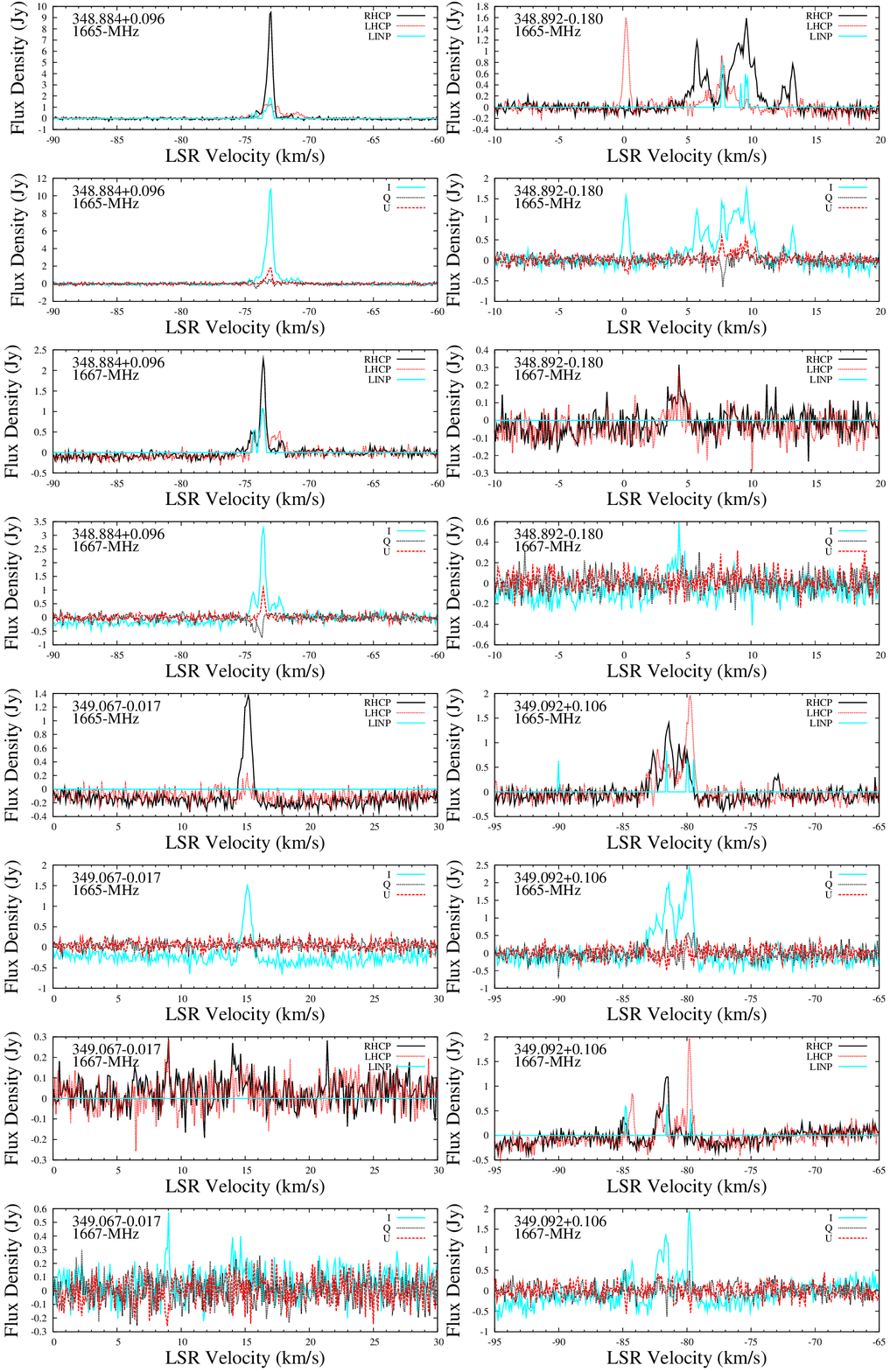}

\caption{\textit{- continued p35 of 35}}

\label{fig1p35}

\end{figure*}

\label{lastpage}

\end{document}